\documentclass[stslayout, noinfoline]{imsart}
\usepackage{geometry} 
\usepackage{subfigure}
\usepackage{graphicx}	
\usepackage{amssymb, amsmath}
\usepackage{color}
\usepackage{natbib}
\usepackage[colorlinks,citecolor=blue,urlcolor=blue]{hyperref}

\usepackage{tikz}
\usetikzlibrary{arrows.meta, math}

\definecolor{light}{RGB}{220, 188, 188}
\definecolor{mid}{RGB}{185, 124, 124}
\definecolor{dark}{RGB}{143, 39, 39}
\definecolor{highlight}{RGB}{180, 31, 180}
\definecolor{light_teal}{RGB}{107, 142, 142}
\definecolor{mid_teal}{RGB}{72, 117, 117}
\definecolor{dark_teal}{RGB}{29, 79, 79}
\definecolor{gray10}{gray}{0.1}
\definecolor{gray20}{gray}{0.2}
\definecolor{gray30}{gray}{0.3}
\definecolor{gray40}{gray}{0.4}
\definecolor{gray60}{gray}{0.6}
\definecolor{gray70}{gray}{0.7}
\definecolor{gray80}{gray}{0.8}
\definecolor{gray90}{gray}{0.9}
\definecolor{gray60}{gray}{0.95}

\tikzmath{
  function soft(\x) {
    if \x > 0 then {
      return \x + ln(1 + exp(-\x));
    } else {
      return ln(1 + exp(\x));
    };
  };
  function weight(\x) {
    if \x > 1 then {
      return 1 / ( (1 / \x) * (1 / \x) * (1 / \x) * (1 / \x) * (1 / \x) + 1);
    } else {
      return \x * \x * \x * \x * \x / (1 + \x * \x * \x * \x * \x);
    };
  };
}

\begin{document}

\begin{frontmatter}

\title{Modifying Survival Models To Accommodate Thresholding Behavior}
\runtitle{A Threshold Survival Model}

\begin{aug}
  \author{Michael Betancourt%
  \ead[label=e1]{inquiries@symplectomorphic.com}}

  \runauthor{Betancourt}
  \address{Symplectomorphic, LLC. \printead{e1}.}
\end{aug}

\begin{abstract}
Survival models capture the relationship between an accumulating hazard 
and the occurrence of a singular event stimulated by that accumulation.
When the model for the hazard is sufficiently flexible survival models
can accommodate a wide range of behaviors.  If the hazard model is less 
flexible, for example when it is constrained by an interpretation as a 
physical stimulus, then the resulting survival model can be much too rigid.  
In this paper I introduce a modified survival model that generalizes the 
relationship between accumulating hazard and event occurrence with 
particular emphasis on capturing thresholding behavior.  Finally I 
demonstrate the utility of this approach on a physiological application.
\end{abstract}

\end{frontmatter}

Survival modeling is often motivated from a probabilistic perspective,
with the occurrence of an event moderated by a subtle conditional 
probability distribution known as the hazard function.  The hazard
function and its development into a survival model, however, can also 
be motivated from a more physical perspective.  This latter perspective 
suggests an immediate generalization of the survival model construction 
that can be used to enforce certain behaviors independent of the assumed 
hazard function.

This paper begins with a discussion of both of these perspectives, the 
general construction of modified survival models, and the particular 
construction of a modified survival model that enforces thresholding 
behavior.  In the second section I apply both the standard and threshold 
survival model to phenology data from winegrape physiology to demonstrate 
their relative performance.  The appendix works through a reversed 
construction of modified survival models and auxiliary calculations used 
in the winegrape phenology models.

\section{Delayed Survival}

In this section I first present the various interpretations of the 
standard survival model construction and its generalization to modified 
survival models.  Next I consider a thresholding survival model and 
compare it to some common models for thresholding behavior.

\subsection{Survival Models}

Survival modeling \citep{CoxEtAl:1984, IbrahimEtAl:2001, HosmerEtAl:2008, Betancourt:2022a} 
concerns events that occur once, and only once, over the open time interval $(t_{i}, \infty)$.  
In general the probability density function for such event times can always be decomposed 
into two terms,
\begin{align*}
\pi(t)
&\equiv \quad\;\,
\pi(\text{event occurs at }t)
\\
&= \quad \;\,
\pi(\text{event occurs at }t \mid \text{no occurrence before }t )
\\
& \quad\quad \cdot
\mathbb{P}_{\pi} [ \text{no occurrence before }t ]
\\
&\quad + \;\,
\pi(\text{event occurs at }t \mid \text{occurrence before }t )
\\
& \quad\quad \cdot
\mathbb{P}_{\pi} [ \text{occurrence before }t ].
\end{align*}

Because of the foundational assumption that the event can occur only
once we must have 
\begin{equation*}
\pi(\text{event occurs at }t \mid \text{occurrence before }t ) = 0
\end{equation*}
in which case the event probability density function becomes
\begin{align*}
\pi(t)
&= \;\,
\pi(\text{event occurs at }t \mid \text{no occurrence before }t )
\\
& \quad \cdot
\mathbb{P}_{\pi} [ \text{no occurrence before }t ]
\\
&\equiv \;\,
\lambda(t) \cdot S(t),
\end{align*}
where $\lambda(t)$ is denoted the \emph{instantaneous hazard function} 
and $S(t)$ is the complementary cumulative distribution function or 
\emph{survival function},
\begin{equation*}
S(t) 
= \int_{t}^{\infty} \mathrm{d} t \, \pi(t) 
= 1 - \int_{t_{i}}^{t} \mathrm{d} t \, \pi(t).
\end{equation*}

By definition the event probability density function is also minus the 
derivative of the survival function,
\begin{equation*}
\pi(t) = - \frac{ \mathrm{d} S }{ \mathrm{d} t }(t).
\end{equation*}
Substituting this into the above relationship gives an ordinary 
differential equation for the survival function in terms of the 
hazard function,
\begin{align*}
\pi(t) &= \lambda(t) \cdot S(t)
\\
- \frac{ \mathrm{d} S }{ \mathrm{d} t }(t)
&=
\lambda(t) \cdot S(t).
\end{align*}
Solving yields
\begin{equation*}
S(t) 
= \exp \left( - \int_{t_{i}}^{t} \mathrm{d} t' \lambda(t') \right)
\equiv \exp \left( - \Lambda(t) \right)
\end{equation*}
where $\Lambda(t)$ is the \emph{cumulative hazard function}
\begin{equation*}
\Lambda(t) = \int_{t_{i}}^{t} \mathrm{d} t' \lambda(t').
\end{equation*}

This result immediately implies two key properties of the hazard function
and its integral.  Firstly, because the survival function is monotonically 
non-increasing and the exponential function is monotonically increasing the 
cumulative hazard function must be monotonically non-decreasing in order to 
define a mathematically consistent model.  This requires in particular that 
the instantaneous hazard function be everywhere positive.  Secondly, in order 
to ensure that the event occurs before $t = \infty$ we must have $S(\infty) = 0$.  
This is true if and only if the cumulative hazard function diverges,
\begin{align*}
\Lambda(\infty)
&=
\int_{t_{i}}^{\infty} \mathrm{d} t \, \lambda(t)
\\
&=
\int_{t_{i}}^{\infty} \mathrm{d} t \, \frac{1}{S(t)} \, \pi(t)
\\
&=
- \int_{t_{i}}^{\infty} \mathrm{d} t \,
\frac{1}{S(t)} \, \frac{ \mathrm{d} S }{\mathrm{d} t}(t)
\\
&=
- \int_{t_{i}}^{\infty} \mathrm{d} t \,
\frac{ \mathrm{d} \log S }{\mathrm{d} t}(t)
\\
&=
- \Big( \log S(\infty) - \log S(0) \Big)
\\
&=
- \Big( \log(0) - \log 1 \Big)
\\
&=
\infty.
\end{align*}
Consequently $\lambda(t) = \pi(\text{event occurs at }t \mid \text{no occurrence before }t )$
is not a normalized probability density function and should be interpreted with care.

Once we have constructed the survival function from the cumulative hazard 
function we can recover the event probability density function by differentiation,
\begin{align*}
\pi(t)
&= - \frac{\mathrm{d} S}{\mathrm{d} t}(t)
\\
&= - \frac{\mathrm{d} }{\mathrm{d} t}
\exp \left( - \int_{t_{i}}^{t} \mathrm{d} t' \, \lambda(t) \right)
\\
&=
\left[ \frac{\mathrm{d} }{\mathrm{d} t}
\int_{t_{i}}^{t} \mathrm{d} t' \, \lambda(t) \right] \,
\exp \left( - \int_{t_{i}}^{t} \mathrm{d} t' \, \lambda(t) \right)
\\
&= \lambda(t) \, \exp \left( - \int_{t_{i}}^{t} \mathrm{d} t' \, \lambda(t) \right)
\\
&= \lambda(t) \, S(t).
\end{align*}

This survival model can also be derived as a model of an explicit, physical stimulus.  
For example we might use the hazard function to directly model instantaneous cell damage 
that stimulates death or instantaneous energy accumulation that stimulates a physiological 
transition in a developing organism.

If we assume that this physical stimulus $\lambda(t)$ cannot be negated or depleted in 
any way then it will be positive across time, and its integral will be non-decreasing.  
Moreover if we assume that the physical stimulus is inexhaustible then that integral 
will eventually diverge with increasing time.  Given these two assumptions the exponential 
of the cumulative hazard function defines a valid survival function,
\begin{equation*}
S(t)
=
\exp \left( - \Lambda(t) \right)
 = \exp \left( - \int_{t_{i}}^{t} \mathrm{d} t' \, \lambda(t) \right).
\end{equation*}
In other words the exponential decay of the survival function with the increasing
cumulative hazard is no longer a definitional consequence but rather a particular 
modeling assumption.

Mathematically this construction is equivalent to the previous, probabilistic 
derivation of a survival model.  The explicit interpretation of the hazard function,
however, allows us to use domain expertise about the underlying, physical stimulus 
to motivate the particular form of the instantaneous hazard function.

\subsection{Modified Survival Models}

When the hazard function is not tied to any particular stimulus it is often modeled 
heuristically with an emphasis on flexible functional forms.  For example one might 
use a piecewise linear function across small time intervals or a non-parametric 
functional model such as splines.  This flexibility allows the exponential survival 
model to accommodate a wide range of observed survival behaviors.  A hazard function 
that can saturate at zero for long times, for instance, can model long gaps between 
the observed event times. 

If the hazard function is tied to an explicit, physical phenomenon, however, then it
may not enjoy this flexibility.  In this case the exponential survival may be too 
restrictive, requiring a generalization beyond the exponential relationship between 
the cumulative hazard function and the survival function in order to adequately model 
the observed data.

Fortunately this generalization is straightforward.  Assuming that the hazard function 
is positive and integrates to infinity then not only does 
\begin{equation*}
S(t) = \exp \left( - \Lambda(t) \right)
\end{equation*}
define a valid survival function but so too does
\begin{equation*}
S(t) = \exp \left( - g(\Lambda(t)) \right)
\end{equation*}
for any monotonically non-decreasing \emph{warping} function 
$g : \mathbb{R} \rightarrow \mathbb{R}$.  This modification obstructs the probabilistic 
interpretation of the hazard function, so that we no longer have
\begin{equation*}
\lambda(t) = \pi(\text{event occurs at }t \mid \text{no occurrence before }t ),
\end{equation*}
but it does not prevent any physical interpretation of the hazard function.

The event probability density function for this modified survival model is given by
\begin{align*}
\pi(t) 
&=
- \frac{ \mathrm{d} S }{ \mathrm{d} t}(t)
\\
&=
- \frac{ \mathrm{d} }{ \mathrm{d} t} \bigg[ \exp \left( - g(\Lambda(t)) \right) \bigg]
\\
&= 
- \bigg[ - \frac{ \mathrm{d} g }{ \mathrm{d} \Lambda}(\Lambda(t)) \bigg]
\cdot \bigg[ \frac{ \mathrm{d} \Lambda }{ \mathrm{d} t}(t) \bigg]
\cdot \bigg[ \exp \left( - g(\Lambda(t)) \right) \bigg]
\\
&= 
\frac{ \mathrm{d} g }{ \mathrm{d} \Lambda}(\Lambda(t)) 
\cdot \lambda (t) 
\cdot \exp \left( - g(\Lambda(t)) \right).
\end{align*}
Taking $g$ to be the identify function with $g(\Lambda) = \Lambda$ gives
$\mathrm{d} g / \mathrm{d} \Lambda \, (\Lambda(t)) = 1$ in which case the modified event 
probability density function reduces to that of the standard survival model.

Importantly this modified survival is as straightforward to implement in practice as the 
standard survival model.  So long as we can evaluate the cumulative hazard function then 
the event probability density functions for both models can be efficiently evaluated in 
closed form.

\subsection{A Threshold Survival Model} \label{sec:threshold_survival}

One of the most common behaviors that an exponential survival model can have difficulty
accommodating is \emph{thresholding}, where events don't start to occur until the 
cumulative hazard function reaches some minimal value $\Lambda_{0}$.  In the standard 
survival model events start occurring as soon as the cumulative hazard function is 
non-zero, no matter how far that non-zero value might be from the desired threshold.

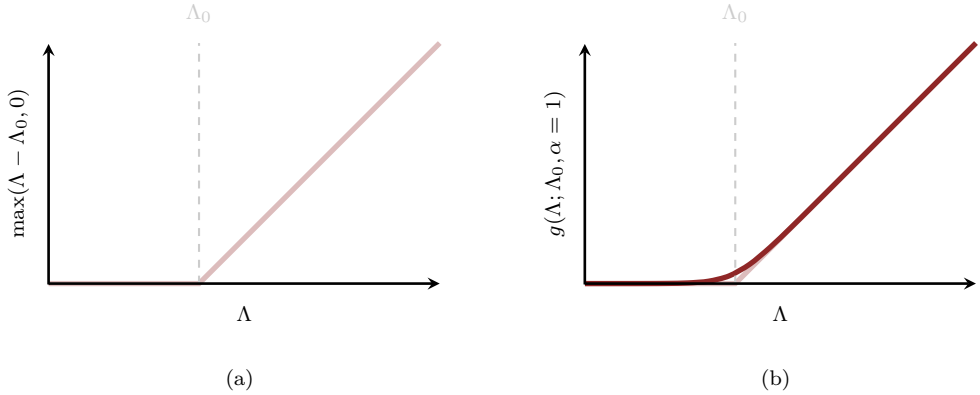
\begin{figure}
\centering
\subfigure[]{ 
\begin{tikzpicture}[scale=0.2, thick]

\draw[white] (-17, -13) rectangle (17, 13);

    \draw[light, line width=2] (-13, -8) -- (-3, -8) -- +(16, 16);

    \draw[gray80, dashed] (-3, -8) -- +(0, 16);
    \node[gray80] at (-3, 10) { $\Lambda_{0}$ };

    \draw [->, >=stealth, line width=1] (-13, -8) -- +(26, 0);
    \node[] at (0, -10) { $\Lambda$ };
        
    \draw [->, >=stealth, line width=1] (-13, -8.058) -- +(0, 16);
    \node[rotate=90] at (-15, 0) { $\max(\Lambda - \Lambda_{0}, 0)$ };

\end{tikzpicture} 
}
\subfigure[]{
\begin{tikzpicture}[scale=0.2, thick]

\draw[white] (-17, -13) rectangle (17, 13);

    \draw[light, line width=2] (-13, -8) -- (-3, -8) -- +(16, 16);

    \draw[domain={0.01:26}, smooth, samples=20, line width=0.5, 
          variable=\x, color=dark, line width=2] 
        plot ({\x - 13},{soft( 1 * (\x - 10) ) - 8});

    \draw[gray80, dashed] (-3, -8) -- +(0, 16);
    \node[gray80] at (-3, 10) { $\Lambda_{0}$ };

    \draw [->, >=stealth, line width=1] (-13, -8) -- +(26, 0);
    \node[] at (0, -10) { $\Lambda$ };
        
    \draw [->, >=stealth, line width=1] (-13, -8.058) -- +(0, 16);
    \node[rotate=90] at (-15, 0) { $g(\Lambda; \Lambda_{0}, \alpha = 1) $ };

\end{tikzpicture}
}
\caption{The (a) hinge function is non-differentiable at the threshold 
$\Lambda_{0}$ but it is well approximated by a (b) smooth thresholding function.}
\label{fig:thresholding_functions} 
\end{figure}

In order to capture this threshold behavior we need a warping function that can suppress 
the cumulative hazard function until it reaches the desired threshold value, at which 
point it can influence the survival function as in the standard model.  For example 
the hinge function,
\begin{equation*}
g(\Lambda) = \max(\Lambda - \Lambda_{0}, 0),
\end{equation*}
is identically zero below $\Lambda_{0}$ before it pivots to allow the cumulative hazard 
function to pass unperturbed (Figure \ref{fig:thresholding_functions}a).

Unfortunately there are two problems with the hinge function.  Firstly the 
cusp at $\Lambda = \Lambda_{0}$ frustrates inferential computational methods 
that rely on gradient information, such as Hamiltonian Monte Carlo.  Secondly 
for each observation $\tilde{t}_{\text{event}}$ the complete suppression below 
the threshold implicitly defines a neighborhood
\begin{equation*}
\Lambda(\tilde{t}_{\text{event}}) < \Lambda_{0}
\end{equation*}
for which the corresponding likelihood function $\pi(\tilde{t}_{\text{event}})$ 
vanishes.  In practice model configurations that fall into this unfeasible 
neighborhood can be difficult, if not impossible, to avoid.

One way to moderate these issues is to replace the hard hinge function with 
something a little softer.  For example the function
\begin{equation*}
g(\Lambda; \Lambda_{0}, \alpha) 
= 
\log \left(1 + \exp \left( \alpha \cdot (\Lambda - \Lambda_{0}) \right) \right).
\end{equation*}
smooths out the hinge cusp for any finite value of $\alpha$ 
(Figure \ref{fig:thresholding_functions}b).  Model configurations that result in 
$\Lambda(\tilde{t}_{\text{event}}) < \Lambda_{0}$ are still suppressed, but no 
longer impossible.  This allows the gradients of the likelihood function to guide 
computation towards more suitable model configurations with 
$\Lambda(\tilde{t}_{\text{event}}) > \Lambda_{0}$.

To construct a modified survival model from this warping function we will also  
need the derivative of the threshold function as well,
\begin{align*}
\frac{\mathrm{d} g}{ \mathrm{d} \Lambda}(\Lambda; \Lambda_{0}, \alpha)
&=
\frac{\mathrm{d} }{ \mathrm{d} \Lambda} 
\log \left(1 + \exp \left( \alpha \cdot (\Lambda - \Lambda_{0}) \right) \right)
\\
&=
\frac{\alpha \, \exp \left( \alpha \cdot (\Lambda - \Lambda_{0}) \right)}
{1 + \exp \left( \alpha \cdot (\Lambda - \Lambda_{0}) \right)}.
\end{align*}

Conveniently the resulting event probability density function simplifies 
considerably,
\begin{align*}
\pi(t)
&=
\lambda (t) \, 
\frac{ \mathrm{d} g }{ \mathrm{d} \Lambda} \left( \Lambda(t) \right) \,
\exp \left( - g \left( \Lambda(t) \right) \right)
\\
&=
\lambda (t) \,
\frac{1}{\alpha}  \,
\frac{\exp \left( \frac{ \Lambda(t) - \Lambda_{0} }{ \alpha } \right) }
{ 1 + \exp \left( \frac{ \Lambda(t) - \Lambda_{0} }{ \alpha } \right) } \,
\exp \left( - \log \left(1 + \exp \left( \frac{ \Lambda(t) - \Lambda_{0} }{ \alpha } \right) \right) \right)
\\
&=
\lambda (t) \,
\frac{1}{\alpha}  \,
\frac{\exp \left( \frac{ \Lambda(t) - \Lambda_{0} }{ \alpha } \right)}
{ \left(  1 + \exp \left( \frac{ \Lambda(t) - \Lambda_{0} }{ \alpha } \right) \right)^{2} }
\\
&=
\lambda (t) \, \mathrm{logistic} (\Lambda(t); \Lambda_{0}, \alpha ).
\end{align*}

Interestingly this construction can also be reversed, starting with a probability
density function for the cumulative hazard function and then deriving a modified 
survival model with a particular warping function.  I discuss this construction in 
more detail in Appendix \ref{sec:implicit_modified}.

\subsection{Scaling Non-Identifiabilities} \label{sec:scaling_degeneracies}

In applications where the hazard function models a physical stimulus it will often 
take the form of a unitful quantity $\psi(t)$ that is scaled into an instanenous 
hazard function $\lambda(t)$ with units of inverse time,
\begin{equation*}
\lambda(t) = \gamma \cdot \psi(t).
\end{equation*}
Under this assumption the unitless cumulative hazard function becomes a scaled version 
of the cumulative physical stimulus,
\begin{align*}
\Lambda(t) 
&=
\int_{t_{i}}^{t} \mathrm{d} t' \lambda(t')
\\
&=
\gamma \int_{t_{i}}^{t} \mathrm{d} t' \psi(t')
\\
&\equiv
\gamma \cdot \Psi(t).
\end{align*}
Similarly we can write any cumulative hazard thresholds as scaled physical 
stimulus thresholds,
\begin{equation*}
\Lambda_{0} = \gamma \cdot \Psi_{0}.
\end{equation*}

With these scalings the event probability density function for the threshold survival 
model becomes
\begin{align*}
\pi(t) 
&=
\lambda(t) \, \mathrm{logistic} ( \Lambda(t) ; \Lambda_{0}, \alpha)
\\
&=
\lambda(t) \, \frac{1}{\alpha}  \,
\frac{\exp \left( \frac{ \Lambda(t) - \Lambda_{0} }{ \alpha } \right)}
{ \left(  1 + \exp \left( \frac{ \Lambda(t) - \Lambda_{0} }{ \alpha } \right) \right)^{2} }
\\
&=
\gamma \cdot \psi(t) \, \frac{1}{\alpha}  \,
\frac{\exp \left( \frac{ \gamma \cdot \Psi(t) - \gamma \cdot \Psi_{0} }{ \alpha } \right)}
{ \left(  1 + \exp \left( \frac{ \gamma \cdot \Psi(t) - \gamma \cdot \Psi_{0} }{ \alpha } \right) \right)^{2} }
\\
&=
\psi(t) \, \frac{1}{\gamma^{-1} \cdot \alpha}  \,
\frac{\exp \left( \frac{ \Psi(t) - \Psi_{0} }{ \gamma^{-1} \cdot \alpha } \right)}
{ \left(  1 + \exp \left( \frac{ \Psi(t) - \Psi_{0} }{ \gamma^{-1} \cdot \alpha } \right) \right)^{2} }
\\
&=
\psi(t) \, \mathrm{logistic} ( \Psi(t) ; \Psi_{0}, \gamma^{-1} \cdot \alpha).
\end{align*}

Because they always appear together the scaling of the hazard function, $\gamma$, and the 
scaling of the warping function, $\alpha$, are inherently non-identified and cannot be 
jointly inferred from event observations.  Inferring only the ratio 
$\sigma = \gamma^{-1} \cdot \alpha$ avoids this ambiguity but at the expense of 
preventing inference of the warping function configuration.  That said 
$\sigma$ does admit a direct interpretation in terms of the variance of the event 
times which makes it straightforward to incorporate into many modeling applications.

\subsection{Alternative Threshold Models}

Applications with events that occur only rarely before some threshold is reached 
are common across many fields, and they have stimulated a variety of modeling 
techniques to capture this behavior.

For example ballistic accumulator models \citep{BrownEtAl:2005} delay events until a
ballistic trajectory passes a fixed constant threshold.  Similarly first-passage 
models \citep{LeeEtAl:2006} delay events until the realization of a stochastic process 
is absorbed by a fixed boundary.  

As with the hazard function in survival models, the latent function in these two 
models can be considered as a heuristic engineered to give the desired threshold 
behavior or as a representation of a particular physical phenomenon.  Consequently
these models suffer from the same problem that we encountered with standard survival
models: the functional behavior needed to ensure the desired behavior is often too 
rigid to accommodate domain expertise about the stimulating phenomenon and vice 
versa.  The advantage of survival modeling is that the generalization to modified 
survival models is straightforward to implement whereas most changes to the latent 
function in ballistic accumulator and first-passage models frustrate the necessary 
analytic results.

\section{Ecological Demonstration}

To demonstrate the utility of modified survival models, in particular modified 
survival models with a soft thresholding warping function, I will apply them to 
plant \emph{phenology} in this section.  I will start by presenting the most
common ecological modeling approach before detailed how survival models can be 
applied.  Finally I will implement a Bayesian analysis with all of these models 
and compare their performance.

\subsection{Conventional Phenological Modeling} \label{sec:conventional_pheno}

Phenology is the study of the different stages of plant growth and the transitions 
between them \citep{LambersEtAl:2008}.  For example a plant might be initially dormant 
during the winter before the sequential appearance of buds, leaves, flowers, and then 
finally fruit as the weather heats up.  Given observations of these appearances we can 
model what environmental circumstances stimulate these phenological transitions.

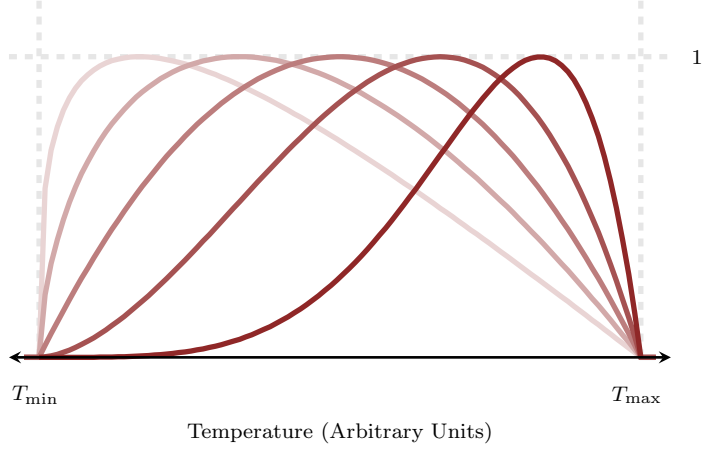
\begin{figure}
\centering
\begin{tikzpicture}[scale=1]

  \pgfmathsetmacro{\sx}{4}
  \pgfmathsetmacro{\sy}{4}

  \draw[gray90, dashed, line width=2] (-1.1 * \sx, \sy) -- (1.1 * \sx, \sy);
  \node[black, align=center] at (1.2 * \sx, \sy) { $1$ };
  
  \draw[gray90, dashed, line width=2] (-\sx, 0) -- +(0, 1.2 * \sy);
  \node[black, align=center] at (-\sx, -0.5) { $T_{\min}$ };
  
  \draw[gray90, dashed, line width=2] (\sx, 0) -- +(0, 1.2 * \sy);
  \node[black, align=center] at (\sx, -0.5) { $T_{\max}$ };

  \colorlet{custom}{dark!20!white};

  \draw[custom, line width=2] (-\sx * 1.050, \sy * 0.000) 
    \foreach \x/\y in {-1.000/0.000, -0.980/0.562, -0.960/0.689, -0.939/0.767, -0.919/0.822, 
                       -0.899/0.863, -0.879/0.895, -0.859/0.920, -0.838/0.940, -0.818/0.956, 
                       -0.798/0.969, -0.778/0.979, -0.758/0.987, -0.737/0.992, -0.717/0.996, 
                       -0.697/0.999, -0.677/1.000, -0.657/1.000, -0.636/0.999, -0.616/0.997, 
                       -0.596/0.994, -0.576/0.990, -0.556/0.986, -0.535/0.981, -0.515/0.976, 
                       -0.495/0.970, -0.475/0.963, -0.455/0.956, -0.434/0.948, -0.414/0.941, 
                       -0.394/0.932, -0.374/0.924, -0.354/0.915, -0.333/0.905, -0.313/0.896, 
                       -0.293/0.886, -0.273/0.876, -0.253/0.866, -0.232/0.855, -0.212/0.844, 
                       -0.192/0.833, -0.172/0.822, -0.152/0.810, -0.131/0.799, -0.111/0.787, 
                       -0.091/0.775, -0.071/0.763, -0.051/0.751, -0.030/0.738, -0.010/0.726, 
                       0.010/0.713, 0.030/0.700, 0.051/0.687, 0.071/0.674, 0.091/0.661, 
                       0.111/0.648, 0.131/0.635, 0.152/0.621, 0.172/0.608, 0.192/0.594, 
                       0.212/0.580, 0.232/0.567, 0.253/0.553, 0.273/0.539, 0.293/0.525, 
                       0.313/0.511, 0.333/0.496, 0.354/0.482, 0.374/0.468, 0.394/0.453, 
                       0.414/0.439, 0.434/0.424, 0.455/0.410, 0.475/0.395, 0.495/0.381, 
                       0.515/0.366, 0.535/0.351, 0.556/0.336, 0.576/0.321, 0.596/0.306, 
                       0.616/0.291, 0.636/0.276, 0.657/0.261, 0.677/0.246, 0.697/0.231,
                       0.717/0.216, 0.737/0.201, 0.758/0.186, 0.778/0.170, 0.798/0.155,
                       0.818/0.140, 0.838/0.124, 0.859/0.109, 0.879/0.093, 0.899/0.078, 
                       0.919/0.062, 0.939/0.047, 0.960/0.031, 0.980/0.016, 1.000/0.000} {
    -- (\sx * \x, \sy * \y)
  } -- (\sx * 1.050, \sy * 0.000);
  
  \colorlet{custom}{dark!40!white};

  \draw[custom, line width=2] (\sx * -1.050, \sy * 0.000) 
    \foreach \x/\y in  {-1.000/0.000, -0.980/0.208, -0.960/0.312, -0.939/0.392, -0.919/0.458, 
                        -0.899/0.516, -0.879/0.566, -0.859/0.611, -0.838/0.651, -0.818/0.687, 
                        -0.798/0.720, -0.778/0.750, -0.758/0.777, -0.737/0.802, -0.717/0.825, 
                        -0.697/0.846, -0.677/0.866, -0.657/0.883, -0.636/0.899, -0.616/0.913, 
                        -0.596/0.927, -0.576/0.938, -0.556/0.949, -0.535/0.959, -0.515/0.967, 
                        -0.495/0.974, -0.475/0.980, -0.455/0.986, -0.434/0.990, -0.414/0.994, 
                        -0.394/0.997, -0.374/0.999, -0.354/1.000, -0.333/1.000, -0.313/1.000, 
                        -0.293/0.999, -0.273/0.997, -0.253/0.994, -0.232/0.991, -0.212/0.988, 
                        -0.192/0.983, -0.172/0.978, -0.152/0.973, -0.131/0.967, -0.111/0.960, 
                        -0.091/0.953, -0.071/0.946, -0.051/0.938, -0.030/0.929, -0.010/0.920, 
                        0.010/0.910, 0.030/0.900, 0.051/0.890, 0.071/0.879, 0.091/0.867, 
                        0.111/0.855, 0.131/0.843, 0.152/0.830, 0.172/0.817, 0.192/0.804, 
                        0.212/0.790, 0.232/0.776, 0.253/0.761, 0.273/0.746, 0.293/0.731, 
                        0.313/0.715, 0.333/0.699, 0.354/0.683, 0.374/0.666, 0.394/0.649,
                        0.414/0.631, 0.434/0.614, 0.455/0.596, 0.475/0.577, 0.495/0.558,
                        0.515/0.539, 0.535/0.520, 0.556/0.501, 0.576/0.481, 0.596/0.460, 
                        0.616/0.440, 0.636/0.419, 0.657/0.398, 0.677/0.377, 0.697/0.355, 
                        0.717/0.333, 0.737/0.311, 0.758/0.289, 0.778/0.266, 0.798/0.243, 
                        0.818/0.220, 0.838/0.196, 0.859/0.173, 0.879/0.149, 0.899/0.125, 
                        0.919/0.100, 0.939/0.075, 0.960/0.051, 0.980/0.025, 1.000/0.000} {
    -- (\sx * \x, \sy * \y)
  } -- (\sx * 1.050, \sy * 0.000);

  \colorlet{custom}{dark!60!white};

  \draw[custom, line width=2] (\sx * -1.050, \sy * 0.000) 
    \foreach \x/\y in {-1.000/0.000, -0.980/0.040, -0.960/0.079, -0.939/0.118, -0.919/0.155, 
                       -0.899/0.192, -0.879/0.228, -0.859/0.263, -0.838/0.297, -0.818/0.331, 
                       -0.798/0.363, -0.778/0.395, -0.758/0.426, -0.737/0.456, -0.717/0.486, 
                       -0.697/0.514, -0.677/0.542, -0.657/0.569, -0.636/0.595, -0.616/0.620, 
                       -0.596/0.645, -0.576/0.669, -0.556/0.691, -0.535/0.713, -0.515/0.735, 
                       -0.495/0.755, -0.475/0.775, -0.455/0.793, -0.434/0.811, -0.414/0.828, 
                       -0.394/0.845, -0.374/0.860, -0.354/0.875, -0.333/0.889, -0.313/0.902,
                       -0.293/0.914, -0.273/0.926, -0.253/0.936, -0.232/0.946, -0.212/0.955, 
                       -0.192/0.963, -0.172/0.971, -0.152/0.977, -0.131/0.983, -0.111/0.988, 
                       -0.091/0.992, -0.071/0.995, -0.051/0.997, -0.030/0.999, -0.010/1.000, 
                       0.010/1.000, 0.030/0.999, 0.051/0.997, 0.071/0.995, 0.091/0.992, 
                       0.111/0.988, 0.131/0.983, 0.152/0.977, 0.172/0.971, 0.192/0.963, 
                       0.212/0.955, 0.232/0.946, 0.253/0.936, 0.273/0.926, 0.293/0.914, 
                       0.313/0.902, 0.333/0.889, 0.354/0.875, 0.374/0.860, 0.394/0.845, 
                       0.414/0.828, 0.434/0.811, 0.455/0.793, 0.475/0.775, 0.495/0.755, 
                       0.515/0.735, 0.535/0.713, 0.556/0.691, 0.576/0.669, 0.596/0.645, 
                       0.616/0.620, 0.636/0.595, 0.657/0.569, 0.677/0.542, 0.697/0.514, 
                       0.717/0.486, 0.737/0.456, 0.758/0.426, 0.778/0.395, 0.798/0.363, 
                       0.818/0.331, 0.838/0.297, 0.859/0.263, 0.879/0.228, 0.899/0.192, 
                       0.919/0.155, 0.939/0.118, 0.960/0.079, 0.980/0.040, 1.000/0.000} {
    -- (\sx * \x, \sy * \y)
  } -- (\sx * 1.050, \sy * 0.000);

  \colorlet{custom}{dark!80!white};

  \draw[custom, line width=2] (\sx * -1.050, \sy * 0.000) 
    \foreach \x/\y in {-1.000/0.000, -0.980/0.002, -0.960/0.005, -0.939/0.010, -0.919/0.017, 
                       -0.899/0.024, -0.879/0.033, -0.859/0.043, -0.838/0.054, -0.818/0.065, 
                       -0.798/0.078, -0.778/0.091, -0.758/0.106, -0.737/0.121, -0.717/0.136, 
                       -0.697/0.153, -0.677/0.170, -0.657/0.187, -0.636/0.205, -0.616/0.224, 
                       -0.596/0.243, -0.576/0.262, -0.556/0.282, -0.535/0.303, -0.515/0.323, 
                       -0.495/0.344, -0.475/0.365, -0.455/0.387, -0.434/0.408, -0.414/0.430, 
                       -0.394/0.452, -0.374/0.474, -0.354/0.496, -0.333/0.518, -0.313/0.540, 
                       -0.293/0.562, -0.273/0.584, -0.253/0.605, -0.232/0.627, -0.212/0.648, 
                       -0.192/0.669, -0.172/0.690, -0.152/0.710, -0.131/0.730, -0.111/0.750, 
                       -0.091/0.769, -0.071/0.788, -0.051/0.806, -0.030/0.824, -0.010/0.841, 
                       0.010/0.857, 0.030/0.873, 0.051/0.888, 0.071/0.902, 0.091/0.916, 
                       0.111/0.928, 0.131/0.940, 0.152/0.951, 0.172/0.961, 0.192/0.970, 
                       0.212/0.977, 0.232/0.984, 0.253/0.990, 0.273/0.994, 0.293/0.997, 
                       0.313/0.999, 0.333/1.000, 0.354/0.999, 0.374/0.997, 0.394/0.994, 
                       0.414/0.989, 0.434/0.982, 0.455/0.974, 0.475/0.965, 0.495/0.953, 
                       0.515/0.940, 0.535/0.926, 0.556/0.909, 0.576/0.891, 0.596/0.871, 
                       0.616/0.848, 0.636/0.824, 0.657/0.798, 0.677/0.770, 0.697/0.740, 
                       0.717/0.707, 0.737/0.673, 0.758/0.636, 0.778/0.596, 0.798/0.555, 
                       0.818/0.511, 0.838/0.465, 0.859/0.416, 0.879/0.364, 0.899/0.310, 
                       0.919/0.254, 0.939/0.194, 0.960/0.132, 0.980/0.068, 1.000/0.000} {
    -- (\sx * \x, \sy * \y)
  } -- (\sx * 1.050, \sy * 0.000);

  \colorlet{custom}{dark!100!white};

  \draw[custom, line width=2] (\sx * -1.000, \sy * 0.000) 
    \foreach \x/\y in {-1.000/0.000, -0.980/0.000, -0.960/0.000, -0.939/0.000, -0.919/0.000, 
                       -0.899/0.000, -0.879/0.000, -0.859/0.000, -0.838/0.000, -0.818/0.000, 
                       -0.798/0.001, -0.778/0.001, -0.758/0.001, -0.737/0.002, -0.717/0.002, 
                       -0.697/0.003, -0.677/0.004, -0.657/0.005, -0.636/0.006, -0.616/0.008, 
                       -0.596/0.009, -0.576/0.011, -0.556/0.013, -0.535/0.016, -0.515/0.018, 
                       -0.495/0.021, -0.475/0.025, -0.455/0.028, -0.434/0.033, -0.414/0.037, 
                       -0.394/0.042, -0.374/0.048, -0.354/0.054, -0.333/0.060, -0.313/0.068, 
                       -0.293/0.075, -0.273/0.084, -0.253/0.093, -0.232/0.102, -0.212/0.113, 
                       -0.192/0.124, -0.172/0.135, -0.152/0.148, -0.131/0.161, -0.111/0.175, 
                       -0.091/0.190, -0.071/0.205, -0.051/0.222, -0.030/0.239, -0.010/0.257, 
                       0.010/0.276, 0.030/0.295, 0.051/0.316, 0.071/0.337, 0.091/0.359, 
                       0.111/0.382, 0.131/0.406, 0.152/0.430, 0.172/0.455, 0.192/0.481, 
                       0.212/0.507, 0.232/0.534, 0.253/0.561, 0.273/0.589, 0.293/0.617, 
                       0.313/0.645, 0.333/0.673, 0.354/0.701, 0.374/0.729, 0.394/0.757,
                       0.414/0.784, 0.434/0.811, 0.455/0.837, 0.475/0.862, 0.495/0.885, 
                       0.515/0.908, 0.535/0.928, 0.556/0.947, 0.576/0.963, 0.596/0.977, 
                       0.616/0.988, 0.636/0.995, 0.657/0.999, 0.677/0.999, 0.697/0.995, 
                       0.717/0.986, 0.737/0.971, 0.758/0.950, 0.778/0.923, 0.798/0.888, 
                       0.818/0.846, 0.838/0.796, 0.859/0.736, 0.879/0.667, 0.899/0.587,
                       0.919/0.496, 0.939/0.393, 0.960/0.276, 0.980/0.146, 1.000/0.000} {
    -- (\sx * \x, \sy * \y)
  };
  
  \draw [<->, >=stealth, line width=1] (-1.1 * \sx, 0) -- (1.1 * \sx, 0);
  \node[] at (0, -1) { Temperature (Arbitrary Units) };

\end{tikzpicture}
\caption{The Wang-Engel forcing function is never differentiable at $T = T_{\max}$; the
derivative is negative for temperatures just below the boundary and zero for temperatures 
just above the boundary.  If $T_{\mathrm{opt}}$ is closer to $T_{\min}$ than $T_{\max}$
then the forcing function is also not differentiable at $T = T_{\min}$.  Here the curves 
become darker as $T_{\mathrm{opt}}$ moves from $T_{\min}$ towards $T_{\max}$.}
\label{fig:we_forcing} 
\end{figure}

Conventional phenological models assume that the transitions are fueled by the 
accumulation of daily \emph{forcings} that depend on the ambient temperature.  
One common model for these forcings is the three-parameter Wang-Engel forcing 
model \citep{WangEtAl:1998} given by
\begin{equation*}
f(T; T_{\min}, T_{\mathrm{opt}}, T_{\max}) =
\left\{
\begin{array}{rr}
0, & T < T_{\min} \\
\left( \frac{ T - T_{\min} }{ T_{\mathrm{opt}} - T_{\min} } \right)^{a} \,
\left[ 2 - \left( \frac{ T - T_{\min} }{ T_{\mathrm{opt}} - T_{\min} } \right)^{a} \right]
& T_{\min} \le T \le T_{\max} \\
0, & T > T_{\max}
\end{array}
\right. ,
\end{equation*}
where
\begin{equation*}
a = \frac{ \log 2 }{ \log \frac{T_{\max} - T_{\min}}{ T_{\mathrm{opt}} - T_{\min} } }
\end{equation*}
and
\begin{equation*}
T_{\min} < T_{\mathrm{opt}} < T_{\max}.
\end{equation*}
The Wang-Engel model captures the ecological assumption that forcings vanish below a 
minimum temperature $T_{\min}$ and above a maximum temperature $T_{\max}$.  The forcings 
are maximized at the temperature $T_{\mathrm{opt}}$ with the maximum forcing set to one,
$f(T_{\mathrm{opt}}) = 1$, by convention.  One drawback of the forcing model, 
however, is that it is not differentiable at $T_{\max}$ and, depending on the value 
of $T_{\mathrm{opt}}$, sometimes also at $T_{\min}$ (Figure \ref{fig:we_forcing}).  
These cusps can compromise gradient-based inferential computational algorithms.

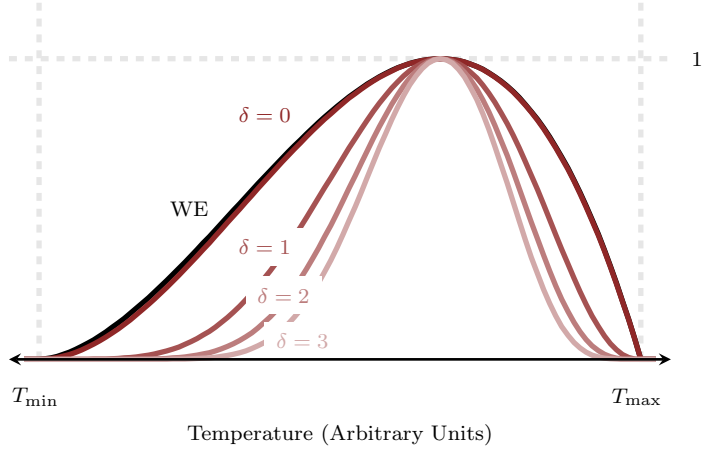
\begin{figure}
\centering
\begin{tikzpicture}[scale=1]

  \pgfmathsetmacro{\sx}{4}
  \pgfmathsetmacro{\sy}{4}

  \draw[gray90, dashed, line width=2] (-1.1 * \sx, \sy) -- (1.1 * \sx, \sy);
  \node[black, align=center] at (1.2 * \sx, \sy) { $1$ };
  
  \draw[gray90, dashed, line width=2] (-\sx, 0) -- +(0, 1.2 * \sy);
  \node[black, align=center] at (-\sx, -0.5) { $T_{\min}$ };
  
  \draw[gray90, dashed, line width=2] (\sx, 0) -- +(0, 1.2 * \sy);
  \node[black, align=center] at (\sx, -0.5) { $T_{\max}$ };

  \draw[black, line width=2] (\sx * -1.050, \sy * 0.000) 
    \foreach \x/\y in {-1.000/0.000, -0.980/0.002, -0.960/0.005, -0.939/0.010, -0.919/0.017, 
                       -0.899/0.024, -0.879/0.033, -0.859/0.043, -0.838/0.054, -0.818/0.065, 
                       -0.798/0.078, -0.778/0.091, -0.758/0.106, -0.737/0.121, -0.717/0.136, 
                       -0.697/0.153, -0.677/0.170, -0.657/0.187, -0.636/0.205, -0.616/0.224, 
                       -0.596/0.243, -0.576/0.262, -0.556/0.282, -0.535/0.303, -0.515/0.323, 
                       -0.495/0.344, -0.475/0.365, -0.455/0.387, -0.434/0.408, -0.414/0.430, 
                       -0.394/0.452, -0.374/0.474, -0.354/0.496, -0.333/0.518, -0.313/0.540, 
                       -0.293/0.562, -0.273/0.584, -0.253/0.605, -0.232/0.627, -0.212/0.648, 
                       -0.192/0.669, -0.172/0.690, -0.152/0.710, -0.131/0.730, -0.111/0.750, 
                       -0.091/0.769, -0.071/0.788, -0.051/0.806, -0.030/0.824, -0.010/0.841, 
                       0.010/0.857, 0.030/0.873, 0.051/0.888, 0.071/0.902, 0.091/0.916, 
                       0.111/0.928, 0.131/0.940, 0.152/0.951, 0.172/0.961, 0.192/0.970, 
                       0.212/0.977, 0.232/0.984, 0.253/0.990, 0.273/0.994, 0.293/0.997, 
                       0.313/0.999, 0.333/1.000, 0.354/0.999, 0.374/0.997, 0.394/0.994, 
                       0.414/0.989, 0.434/0.982, 0.455/0.974, 0.475/0.965, 0.495/0.953, 
                       0.515/0.940, 0.535/0.926, 0.556/0.909, 0.576/0.891, 0.596/0.871, 
                       0.616/0.848, 0.636/0.824, 0.657/0.798, 0.677/0.770, 0.697/0.740, 
                       0.717/0.707, 0.737/0.673, 0.758/0.636, 0.778/0.596, 0.798/0.555, 
                       0.818/0.511, 0.838/0.465, 0.859/0.416, 0.879/0.364, 0.899/0.310, 
                       0.919/0.254, 0.939/0.194, 0.960/0.132, 0.980/0.068, 1.000/0.000} {
    -- (\sx * \x, \sy * \y)
  } -- (\sx * 1.050, \sy * 0.000);

  \colorlet{custom}{dark!100!white};
  
  \draw[custom, line width=2] (\sx * -1.050, \sy * 0.000) 
    \foreach \x/\y in {-1.000/0.000, -0.980/0.001, -0.960/0.003, -0.939/0.006, -0.919/0.011, 
                       -0.899/0.016, -0.879/0.023, -0.859/0.031, -0.838/0.041, -0.818/0.051, 
                       -0.798/0.062, -0.778/0.074, -0.758/0.087, -0.737/0.101, -0.717/0.116, 
                       -0.697/0.131, -0.677/0.148, -0.657/0.165, -0.636/0.183, -0.616/0.201, 
                       -0.596/0.220, -0.576/0.239, -0.556/0.259, -0.535/0.280, -0.515/0.301, 
                       -0.495/0.322, -0.475/0.343, -0.455/0.365, -0.434/0.387, -0.414/0.410, 
                       -0.394/0.432, -0.374/0.455, -0.354/0.477, -0.333/0.500, -0.313/0.523, 
                       -0.293/0.545, -0.273/0.568, -0.253/0.590, -0.232/0.613, -0.212/0.635,
                       -0.192/0.657, -0.172/0.678, -0.152/0.699, -0.131/0.720, -0.111/0.741, 
                       -0.091/0.761, -0.071/0.780, -0.051/0.799, -0.030/0.817, -0.010/0.835, 
                       0.010/0.852, 0.030/0.869, 0.051/0.884, 0.071/0.899, 0.091/0.913, 
                       0.111/0.926, 0.131/0.938, 0.152/0.949, 0.172/0.959, 0.192/0.969, 
                       0.212/0.977, 0.232/0.984, 0.253/0.989, 0.273/0.994, 0.293/0.997, 
                       0.313/0.999, 0.333/1.000, 0.354/0.999, 0.374/0.997, 0.394/0.994, 
                       0.414/0.989, 0.434/0.982, 0.455/0.974, 0.475/0.964, 0.495/0.952, 
                       0.515/0.939, 0.535/0.924, 0.556/0.907, 0.576/0.889, 0.596/0.868, 
                       0.616/0.846, 0.636/0.822, 0.657/0.795, 0.677/0.767, 0.697/0.736, 
                       0.717/0.704, 0.737/0.669, 0.758/0.632, 0.778/0.593, 0.798/0.551, 
                       0.818/0.507, 0.838/0.461, 0.859/0.412, 0.879/0.361, 0.899/0.307, 
                       0.919/0.251, 0.939/0.192, 0.960/0.131, 0.980/0.067, 1.000/0.000} {
    -- (\sx * \x, \sy * \y)
  } -- (\sx * 1.050, \sy * 0.000);
  
  \colorlet{custom}{dark!80!white};
  
  \draw[custom, line width=2] (\sx * -1.050, \sy * 0.000) 
    \foreach \x/\y in {-1.000/0.000, -0.980/0.000, -0.960/0.000, -0.939/0.000, -0.919/0.000, 
                       -0.899/0.000, -0.879/0.000, -0.859/0.000, -0.838/0.000, -0.818/0.001, 
                       -0.798/0.001, -0.778/0.001, -0.758/0.002, -0.737/0.003, -0.717/0.005, 
                       -0.697/0.006, -0.677/0.008, -0.657/0.011, -0.636/0.014, -0.616/0.018, 
                       -0.596/0.023, -0.576/0.028, -0.556/0.034, -0.535/0.041, -0.515/0.050, 
                       -0.495/0.059, -0.475/0.069, -0.455/0.081, -0.434/0.093, -0.414/0.107, 
                       -0.394/0.123, -0.374/0.139, -0.354/0.157, -0.333/0.177, -0.313/0.198, 
                       -0.293/0.220, -0.273/0.243, -0.253/0.268, -0.232/0.294, -0.212/0.321, 
                       -0.192/0.349, -0.172/0.379, -0.152/0.409, -0.131/0.440, -0.111/0.472, 
                       -0.091/0.505, -0.071/0.538, -0.051/0.571, -0.030/0.604, -0.010/0.637, 
                       0.010/0.670, 0.030/0.703, 0.051/0.735, 0.071/0.766, 0.091/0.796, 
                       0.111/0.825, 0.131/0.852, 0.152/0.878, 0.172/0.902, 0.192/0.923, 
                       0.212/0.943, 0.232/0.960, 0.253/0.974, 0.273/0.985, 0.293/0.993, 
                       0.313/0.998, 0.333/1.000, 0.354/0.998, 0.374/0.993, 0.394/0.984, 
                       0.414/0.972, 0.434/0.955, 0.455/0.936, 0.475/0.912, 0.495/0.885, 
                       0.515/0.855, 0.535/0.821, 0.556/0.784, 0.576/0.745, 0.596/0.703, 
                       0.616/0.658, 0.636/0.612, 0.657/0.564, 0.677/0.515, 0.697/0.465, 
                       0.717/0.415, 0.737/0.366, 0.758/0.317, 0.778/0.270, 0.798/0.225, 
                       0.818/0.183, 0.838/0.144, 0.859/0.109, 0.879/0.078, 0.899/0.052, 
                       0.919/0.032, 0.939/0.016, 0.960/0.006, 0.980/0.001, 1.000/0.000} {
    -- (\sx * \x, \sy * \y)
  } -- (\sx * 1.050, \sy * 0.000);

  \colorlet{custom}{dark!60!white};

  \draw[custom, line width=2] (\sx * -1.050, \sy * 0.000) 
    \foreach \x/\y in {-1.000/0.000, -0.980/0.000, -0.960/0.000, -0.939/0.000, -0.919/0.000, 
                       -0.899/0.000, -0.879/0.000, -0.859/0.000, -0.838/0.000, -0.818/0.000, 
                       -0.798/0.000, -0.778/0.000, -0.758/0.000, -0.737/0.000, -0.717/0.000, 
                       -0.697/0.000, -0.677/0.000, -0.657/0.001, -0.636/0.001, -0.616/0.002, 
                       -0.596/0.002, -0.576/0.003, -0.556/0.005, -0.535/0.006, -0.515/0.008, 
                       -0.495/0.011, -0.475/0.014, -0.455/0.018, -0.434/0.022, -0.414/0.028, 
                       -0.394/0.035, -0.374/0.043, -0.354/0.052, -0.333/0.063, -0.313/0.075, 
                       -0.293/0.088, -0.273/0.104, -0.253/0.122, -0.232/0.141, -0.212/0.162, 
                       -0.192/0.186, -0.172/0.212, -0.152/0.239, -0.131/0.269, -0.111/0.301, 
                       -0.091/0.335, -0.071/0.370, -0.051/0.408, -0.030/0.446, -0.010/0.486, 
                       0.010/0.527, 0.030/0.569, 0.051/0.611, 0.071/0.653, 0.091/0.694, 
                       0.111/0.735, 0.131/0.774, 0.152/0.812, 0.172/0.848, 0.192/0.880, 
                       0.212/0.910, 0.232/0.936, 0.253/0.958, 0.273/0.976, 0.293/0.989, 
                       0.313/0.997, 0.333/1.000, 0.354/0.997, 0.374/0.989, 0.394/0.975, 
                       0.414/0.955, 0.434/0.930, 0.455/0.899, 0.475/0.863, 0.495/0.823, 
                       0.515/0.778, 0.535/0.729, 0.556/0.678, 0.576/0.624, 0.596/0.568, 
                       0.616/0.512, 0.636/0.456, 0.657/0.400, 0.677/0.346, 0.697/0.294, 
                       0.717/0.245, 0.737/0.200, 0.758/0.159, 0.778/0.123, 0.798/0.092, 
                       0.818/0.066, 0.838/0.045, 0.859/0.029, 0.879/0.017, 0.899/0.009, 
                       0.919/0.004, 0.939/0.001, 0.960/0.000, 0.980/0.000, 1.000/0.000} {
    -- (\sx * \x, \sy * \y)
  } -- (\sx * 1.050, \sy * 0.000);

  \colorlet{custom}{dark!40!white};

  \draw[custom, line width=2] (\sx * -1.050, \sy * 0.000) 
    \foreach \x/\y in {-1.000/0.000, -0.980/0.000, -0.960/0.000, -0.939/0.000, -0.919/0.000, 
                       -0.899/0.000, -0.879/0.000, -0.859/0.000, -0.838/0.000, -0.818/0.000, 
                       -0.798/0.000, -0.778/0.000, -0.758/0.000, -0.737/0.000, -0.717/0.000, 
                       -0.697/0.000, -0.677/0.000, -0.657/0.000, -0.636/0.000, -0.616/0.000, 
                       -0.596/0.000, -0.576/0.000, -0.556/0.001, -0.535/0.001, -0.515/0.001, 
                       -0.495/0.002, -0.475/0.003, -0.455/0.004, -0.434/0.005, -0.414/0.007, 
                       -0.394/0.010, -0.374/0.013, -0.354/0.017, -0.333/0.022, -0.313/0.028, 
                       -0.293/0.036, -0.273/0.045, -0.253/0.055, -0.232/0.068, -0.212/0.082, 
                       -0.192/0.099, -0.172/0.118, -0.152/0.140, -0.131/0.165, -0.111/0.192, 
                       -0.091/0.222, -0.071/0.255, -0.051/0.291, -0.030/0.330, -0.010/0.371, 
                       0.010/0.415, 0.030/0.461, 0.051/0.508, 0.071/0.556, 0.091/0.606, 
                       0.111/0.655, 0.131/0.704, 0.152/0.751, 0.172/0.797, 0.192/0.839, 
                       0.212/0.878, 0.232/0.913, 0.253/0.943, 0.273/0.967, 0.293/0.985, 
                       0.313/0.996, 0.333/1.000, 0.354/0.996, 0.374/0.985, 0.394/0.965, 
                       0.414/0.939, 0.434/0.904, 0.455/0.864, 0.475/0.817, 0.495/0.765, 
                       0.515/0.708, 0.535/0.648, 0.556/0.586, 0.576/0.523, 0.596/0.460, 
                       0.616/0.398, 0.636/0.339, 0.657/0.284, 0.677/0.232, 0.697/0.186, 
                       0.717/0.145, 0.737/0.109, 0.758/0.080, 0.778/0.056, 0.798/0.038, 
                       0.818/0.024, 0.838/0.014, 0.859/0.008, 0.879/0.004, 0.899/0.002, 
                       0.919/0.001, 0.939/0.000, 0.960/0.000, 0.980/0.000, 1.000/0.000} {
    -- (\sx * \x, \sy * \y)
  } -- (\sx * 1.050, \sy * 0.000);

  \node[black] at (-2, 2) { WE };

  \colorlet{custom}{dark!100!white};
  \node[custom] at (-1, 3.25) { $\delta = 0$ };

  \colorlet{custom}{dark!80!white};
  \fill[white] (-1.5, 1.25) rectangle +(1, 0.5);
  \node[custom] at (-1, 1.5) { $\delta = 1$ };

  \colorlet{custom}{dark!60!white}
  \fill[white] (-1.25, 0.6) rectangle +(1, 0.5);
  \node[custom] at (-0.75, 0.85) { $\delta = 2$ };

  \colorlet{custom}{dark!40!white}
  \fill[white] (-1, 0.0) rectangle +(1, 0.5);
  \node[custom] at (-0.5, 0.25) { $\delta = 3$ };

  \draw [<->, >=stealth, line width=1] (-1.1 * \sx, 0) -- (1.1 * \sx, 0);
  \node[] at (0, -1) { Temperature (Arbitrary Units) };

\end{tikzpicture}
\caption{The generalized forcing function derived in Appendix \ref{sec:diff_forcing}
introduces a new parameter $\delta$ that determines how strongly the forcing functions 
concentrate around the optimal temperature, as well as how differentiable the forcing 
functions are at $T_{\min}$ and $T_{\max}$.  As $\delta \rightarrow 0$ this family almost
exactly recovers the Wang-Engel family of forcing functions.}
\label{fig:diff_forcing} 
\end{figure}

The cusps in the Wang-Engel forcing model are often taken for granted when using 
gradient-based optimization methods to compute point estimates for the model configurations, 
but it can be harder to ignore when using Hamiltonian Monte Carlo to explore a corresponding 
posterior distribution.  Fortunately we can avoid any numerical problems the 
non-differentiable points might provoke with a slight modification to the forcing model.  
In Appendix \ref{sec:diff_forcing} I derive an alternative forcing function in the spirit 
of \cite{YinEtAl:1995} that smooths out the cusp without affecting the qualitative features 
(Figure \ref{fig:diff_forcing}),
\begin{equation*}
f(T; T_{\min}, T_{\mathrm{opt}}, T_{\max}, \delta) =
\left\{
\begin{array}{rr}
0, & T < T_{\min} \\
\left( 
\left( \frac{T - T_{\mathrm{min}}}{T_{\mathrm{opt}} - T_{\mathrm{min}}} \right)^{ \eta } \, 
\left( \frac{T_{\mathrm{max}} - T}{T_{\mathrm{max}} - T_{\mathrm{opt}}} \right)^{ \kappa }
\right)^{\gamma},
& T_{\min} \le T \le T_{\max} \\
0, & T > T_{\max}
\end{array}
\right.
\end{equation*}
with
\begin{align*}
\eta &= 1
\\
\kappa &= 
\frac{T_{\mathrm{max}} - T_{\mathrm{opt}}}{T_{\mathrm{opt}} - T_{\mathrm{min}}}
\\
\gamma &= 
\frac{\delta \, T_{\mathrm{max}} + T_{\mathrm{opt}} - (\delta + 1) \, T_{\mathrm{min}}}
{ T_{\mathrm{max}} - T_{\mathrm{opt}} }
\end{align*}
for $T_{\mathrm{opt}} > \frac{1}{2}( T_{\min} + T_{\max} )$ and 
\begin{align*}
\eta &= \frac{T_{\mathrm{opt}} - T_{\mathrm{min}}}{T_{\mathrm{max}} - T_{\mathrm{opt}}}
\\
\kappa &= 1
\\
\gamma &= 
\frac{ (\delta + 1) \, T_{\mathrm{max}} - T_{\mathrm{opt}} - \delta \, T_{\mathrm{min}} }
{ T_{\mathrm{opt}} - T_{\mathrm{min}} }
\end{align*}
for $ T_{\mathrm{opt}} \le \frac{1}{2} ( T_{\min} + T_{\max} )$.  The positive parameter
$\delta$ controls how strongly the cusp is smoothed; in the limit $\delta \rightarrow 0$
this model almost exactly recovers the Wang-Engel model.

Regardless  of the specific forcing function phenological events are typically assumed 
to occur immediately after the forcings accumulated over a sequence of days surpasses a 
given threshold $E$,
\begin{equation*}
\sum_{n} f(T_{n}; T_{\min}, T_{\mathrm{opt}}, T_{\max}) > E.
\end{equation*}
This deterministic condition results in a singular observational model for the event day 
$n$,
\begin{equation*}
\pi(n \mid T_{\min}, T_{\mathrm{opt}}, T_{\max}, E) 
= 
\delta \left( \sum_{n} f(T_{n}; T_{\min}, T_{\mathrm{opt}}, T_{\max}) - E \right),
\end{equation*}

Variation in the threshold from observation to observation,
\begin{equation*}
\pi(E \mid \Psi_{0}, \sigma) = \text{normal}(E \mid \Psi_{0}, \sigma),
\end{equation*}
smoothes out this singular behavior into the final observational model
\begin{align*}
\pi(n \mid T_{\min}, T_{\mathrm{opt}}, T_{\max}, \Psi_{0}, \sigma)
&=
\int \mathrm{d}E \, 
\pi(n \mid T_{\min}, T_{\mathrm{opt}}, T_{\max}, E) \,
\pi(E \mid \Psi_{0}, \sigma)
\\
&=
\int \mathrm{d}E \, 
\delta \left( \sum_{n} f(T_{n}; T_{\min}, T_{\mathrm{opt}}, T_{\max}) - E \right) \,
\text{normal}(E \mid \Psi_{0}, \sigma)
\\
&=
\text{normal} \left( \sum_{n} f(T_{n}; T_{\min}, T_{\mathrm{opt}}, T_{\max}) 
\mid \Psi_{0}, \sigma \right),
\end{align*}
which is then used to construct maximum likelihood estimates for the parameters
$T_{\min}$, $T_{\mathrm{opt}}$, $T_{\max}$, and $\Psi_{0}$.

\subsection{Survival Phenology}

From a survival modeling perspective a phenological forcing function defines a
natural hazard function, 
\begin{align*}
\lambda(t) 
&= \gamma \cdot \phi(t)
\\
&= \gamma \cdot f(T(t'); T_{\min}, T_{\mathrm{opt}}, T_{\max}),
\end{align*}
at least up to the daily discretization of the input temperatures.  If we consider 
the forcing function to be piecewise constant in between the daily temperature 
measurements and parameterize time in units of days then the cumulative hazard 
function becomes
\begin{align*}
\Lambda(t) 
&= \gamma \cdot \Psi(t)
\\
&= \gamma \cdot \int_{t_{i}}^{t} \mathrm{d} t' \, f(T(t'); T_{\min}, T_{\mathrm{opt}}, T_{\max})
\\
&= \gamma \cdot \sum_{n = \lfloor t_{i} \rfloor }^{ \lfloor t \rfloor} 
   f(T_{n}; T_{\min}, T_{\mathrm{opt}}, T_{\max}).
\end{align*}

Given this cumulative hazard function the standard survival model is given by 
the survival function
\begin{equation*}
S(t) = \exp \left( - \gamma \, \sum_{n = \lfloor t_{i} \rfloor}^{ \lfloor t \rfloor} 
                               f(T_{n}; T_{\min}, T_{\mathrm{opt}}, T_{\max}) \right),
\end{equation*}
where $\gamma$ controls the decay of the survival function.
Unfortunately this model is too rigid to accommodate the expected thresholding
of phenological transitions; the survival function will decay as soon as the
accumulated forcings rise above zero and long before the accumulation reaches
any fixed threshold.  This leads to an excess of premature events relative to 
what we would expect from thresholding behavior.

To ensure thresholding behavior we can introduce a modified survival model
\begin{equation*}
S(t) = \exp \left( -
g \left( \gamma \, \sum_{n = \lfloor t_{i} \rfloor}^{ \lfloor t \rfloor} 
         f(T_{n}; T_{\min}, T_{\mathrm{opt}}, T_{\max}); \Psi_{0}, \alpha \right)
\right),
\end{equation*}
where $g$ is the soft thresholding function introduced in Section \ref{sec:threshold_survival}.
With this modification the survival function will persist near unity until the 
accumulated forcings start to approach the threshold $\Psi_{0}$.  

In this case the scaling parameter $\gamma$ now influences how strongly the phenological 
events concentrate on the day where this minimal threshold is reached; the larger $\gamma$ 
the faster the survival function will decay once $\Psi(t) > \Psi_{0}$.  The scaling 
parameter $\alpha$ controls the rigidity of the warping function which also influences
the concentration of phenological events.  Indeed as I discussed in Section 
\ref{sec:scaling_degeneracies} these two parameters are non-identified and in practice 
best replaced by a single scaling parameter $\sigma = \gamma^{-1} \cdot \alpha$.

Using the generalized forcing model along with the soft thresholding function ensures 
a differentiable observational model which can then be used to inform maximum likelihood 
estimates of the parameters or, even better, Bayesian inference with Hamiltonian Monte 
Carlo.

\subsection{Application to Winegrape Phenology}

In this section I demonstrate the limitations of standard survival modeling, and the 
utility of modified survival modeling, in the ecological setting with an application 
to winegrape data.  Winegrape phenology is critical to the sustainability of 
winemaking in consideration of the rapidly evolving climate.

\subsubsection{Data}

Here I consider \emph{veraison} phenology which follows the development of ripe
berries on a single plant or collection of plants.  Veraison events begin with the 
appearance of flowers, formally 50\% capfall or equivalently 65 on the BBCH scale, and 
end with the ripening of berries, formally 50\% of berries showing color change or 
equivalently 85 of the BBCH scale \citep{CoombeEtAl:1992, LorenzEtAl:1994, CorneliusEtAl:2011}.

Phenology data were collected from Sauvignon Blanc varieties of Vitis vinifera grown
in Domaine de Vassal, an experimental research vineyard in Marseillan, France managed 
by INRAE \url{https://www6.montpellier.inrae.fr/vassal_eng/}.  Overall $112$ observations 
were made across the contiguous years 1987 to 2014.  These data are publicly accessible at 
\url{https://data.pheno.fr/}, albeit in it a relative time format that has to be converted 
to absolute day of year.  Daily temperatures are taken from recordings at the nearby 
Montpellier Airport weather station, 2207 MONTPELLIER-AEROPORT FR \citep{TankEtAl:2002}, 
and are publicly accessible at \url{https://climexp.knmi.nl/start.cgi}.

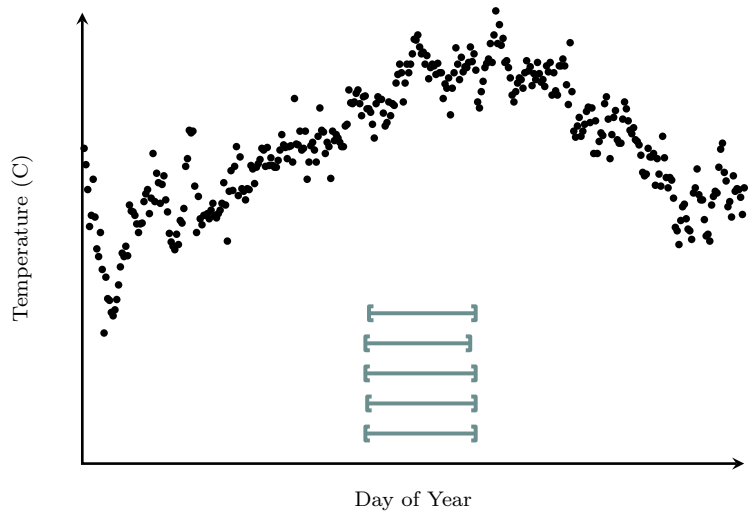
\begin{figure}
\centering
\begin{tikzpicture}[scale=1]

  \pgfmathsetmacro{\sx}{4}
  \pgfmathsetmacro{\sy}{4}
  
  \draw[white] (-1.5 * \sx, - 0.25 * \sy) rectangle (1.5 * \sx, 1.75 * \sy);

  \draw [<->, >=stealth, line width=1] (-1.1 * \sx, 1.5 * \sy) -- (-1.1 * \sx, 0) -- (1.1 * \sx, 0);
  \node at (0, -0.5) { Day of Year };
  \node[rotate=90] at (-1.3 * \sx, 0.75 * \sy) { Temperature (C) };
  
  \foreach \x/\y in {1.000/14.200, 2.000/12.300, 3.000/9.400, 4.000/5.100, 5.000/6.400, 6.000/10.600, 7.000/6.200, 8.000/2.500, 9.000/1.600, 10.000/4.400, 11.000/0.100, 12.000/-7.300, 13.000/-0.800, 14.000/-3.300, 15.000/-3.500, 16.000/-4.900, 17.000/-5.300, 18.000/-4.600, 19.000/-3.400, 20.000/-1.700, 21.000/0.400, 22.000/2.000, 23.000/1.600, 24.000/2.800, 25.000/1.700, 26.000/7.600, 27.000/6.900, 28.000/6.800, 29.000/6.400, 30.000/5.400, 31.000/4.400, 32.000/5.400, 33.000/5.500, 34.000/8.000, 35.000/8.900, 36.000/9.400, 37.000/6.600, 38.000/8.400, 39.000/13.600, 40.000/8.000, 41.000/11.300, 42.000/10.200, 43.000/7.800, 44.000/10.000, 45.000/11.000, 46.000/8.200, 47.000/6.800, 48.000/4.300, 49.000/3.300, 50.000/2.800, 51.000/2.400, 52.000/4.200, 53.000/3.000, 54.000/5.800, 55.000/5.500, 56.000/7.200, 57.000/12.100, 58.000/13.000, 59.000/16.300, 60.000/16.000, 61.000/16.200, 62.000/9.800, 63.000/4.400, 64.000/8.600, 65.000/6.000, 66.000/5.300, 67.000/6.400, 68.000/6.400, 69.000/6.600, 70.000/6.000, 71.000/7.000, 72.000/7.800, 73.000/6.400, 74.000/6.100, 75.000/6.600, 76.000/7.000, 77.000/7.800, 78.000/10.200, 79.000/7.600, 80.000/3.400, 81.000/9.200, 82.000/11.600, 83.000/9.300, 84.000/12.100, 85.000/14.400, 86.000/8.700, 87.000/13.200, 88.000/9.500, 89.000/9.200, 90.000/8.200, 91.000/9.500, 92.000/9.400, 93.000/12.200, 94.000/11.800, 95.000/10.100, 96.000/12.000, 97.000/12.400, 98.000/12.900, 99.000/14.200, 100.000/14.000, 101.000/12.100, 102.000/14.300, 103.000/12.200, 104.000/13.100, 105.000/12.200, 106.000/12.000, 107.000/12.000, 108.000/12.800, 109.000/14.000, 110.000/15.800, 111.000/14.600, 112.000/14.300, 113.000/13.000, 114.000/14.400, 115.000/14.500, 116.000/12.600, 117.000/20.000, 118.000/14.700, 119.000/14.600, 120.000/16.200, 121.000/14.000, 122.000/14.200, 123.000/16.200, 124.000/10.600, 125.000/12.600, 126.000/14.500, 127.000/14.900, 128.000/14.000, 129.000/12.600, 130.000/15.600, 131.000/18.900, 132.000/16.300, 133.000/16.100, 134.000/13.600, 135.000/14.900, 136.000/14.400, 137.000/10.700, 138.000/16.200, 139.000/15.100, 140.000/15.800, 141.000/15.400, 142.000/14.900, 143.000/14.400, 144.000/14.400, 145.000/16.800, 146.000/17.000, 147.000/21.000, 148.000/19.500, 149.000/19.400, 150.000/19.600, 151.000/20.600, 152.000/21.000, 153.000/19.400, 154.000/18.000, 155.000/20.200, 156.000/20.300, 157.000/18.800, 158.000/18.800, 159.000/18.400, 160.000/17.000, 161.000/15.400, 162.000/19.400, 163.000/20.200, 164.000/19.900, 165.000/19.900, 166.000/18.300, 167.000/16.800, 168.000/17.200, 169.000/19.600, 170.000/18.000, 171.000/19.400, 172.000/19.200, 173.000/21.800, 174.000/22.800, 175.000/24.000, 176.000/23.000, 177.000/19.600, 178.000/21.800, 179.000/24.000, 180.000/23.000, 181.000/24.000, 182.000/25.200, 183.000/26.900, 184.000/26.800, 185.000/27.400, 186.000/26.000, 187.000/25.200, 188.000/25.400, 189.000/24.000, 190.000/25.000, 191.000/22.300, 192.000/23.600, 193.000/24.000, 194.000/23.800, 195.000/22.800, 196.000/24.100, 197.000/23.000, 198.000/23.000, 199.000/20.000, 200.000/21.700, 201.000/21.200, 202.000/22.400, 203.000/18.100, 204.000/21.200, 205.000/23.000, 206.000/25.900, 207.000/24.000, 208.000/22.200, 209.000/24.800, 210.000/23.100, 211.000/23.700, 212.000/23.600, 213.000/24.200, 214.000/26.000, 215.000/24.300, 216.000/25.200, 217.000/23.300, 218.000/19.600, 219.000/18.900, 220.000/20.800, 221.000/22.000, 222.000/23.600, 223.000/24.100, 224.000/24.600, 225.000/26.400, 226.000/24.200, 227.000/27.400, 228.000/30.200, 229.000/26.400, 230.000/28.600, 231.000/26.200, 232.000/27.000, 233.000/24.200, 234.000/24.400, 235.000/23.600, 236.000/22.400, 237.000/20.400, 238.000/19.600, 239.000/20.800, 240.000/21.800, 241.000/22.400, 242.000/22.200, 243.000/21.200, 244.000/23.000, 245.000/24.400, 246.000/23.200, 247.000/23.700, 248.000/21.500, 249.000/22.900, 250.000/22.200, 251.000/24.200, 252.000/22.900, 253.000/21.900, 254.000/22.200, 255.000/21.400, 256.000/23.800, 257.000/23.100, 258.000/24.200, 259.000/20.400, 260.000/23.000, 261.000/20.900, 262.000/20.400, 263.000/23.800, 264.000/23.800, 265.000/24.600, 266.000/23.800, 267.000/21.600, 268.000/19.300, 269.000/26.500, 270.000/20.800, 271.000/16.200, 272.000/15.500, 273.000/16.800, 274.000/17.800, 275.000/15.600, 276.000/16.200, 277.000/17.000, 278.000/18.500, 279.000/19.000, 280.000/17.200, 281.000/18.900, 282.000/13.600, 283.000/17.400, 284.000/13.600, 285.000/15.000, 286.000/16.000, 287.000/15.700, 288.000/19.400, 289.000/20.200, 290.000/17.600, 291.000/13.000, 292.000/15.800, 293.000/17.800, 294.000/18.600, 295.000/15.200, 296.000/14.800, 297.000/14.800, 298.000/17.600, 299.000/19.600, 300.000/18.800, 301.000/17.400, 302.000/16.300, 303.000/15.900, 304.000/16.600, 305.000/15.200, 306.000/15.000, 307.000/14.600, 308.000/13.400, 309.000/11.400, 310.000/11.000, 311.000/11.200, 312.000/14.200, 313.000/13.200, 314.000/13.200, 315.000/13.600, 316.000/9.900, 317.000/13.000, 318.000/11.200, 319.000/9.800, 320.000/15.600, 321.000/13.100, 322.000/10.800, 323.000/9.600, 324.000/10.800, 325.000/10.500, 326.000/8.400, 327.000/5.000, 328.000/4.600, 329.000/3.000, 330.000/5.800, 331.000/7.400, 332.000/7.700, 333.000/6.000, 334.000/4.800, 335.000/4.600, 336.000/3.600, 337.000/9.000, 338.000/9.600, 339.000/11.800, 340.000/8.800, 341.000/8.200, 342.000/8.200, 343.000/5.900, 344.000/4.000, 345.000/4.200, 346.000/3.400, 347.000/7.600, 348.000/8.600, 349.000/9.200, 350.000/9.000, 351.000/12.000, 352.000/14.100, 353.000/14.800, 354.000/11.400, 355.000/10.400, 356.000/7.400, 357.000/8.000, 358.000/6.200, 359.000/11.000, 360.000/9.200, 361.000/9.500, 362.000/9.400, 363.000/8.500, 364.000/6.500, 365.000/9.600} {
    \fill[black] ({2.2 * \sx * \x / 365 - 1.1 * \sx}, {(\y + 5) * 1.0 * \sy / 35 + 0.5 * \sy}) circle (0.05);
  }
  
  \foreach \xi/\xf [count=\n] in {155.000/218.000, 156.000/218.000, 155.000/218.000,
                                  155.000/215.000, 157.000/218.000} {
    \draw[line width=1.5, light_teal,
          arrows={Bracket[line join=round, scale=0.75]-Bracket[line join=round, scale=0.75]}] 
      ({2.2 * \sx * \xi / 365 - 1.1 * \sx}, {0.4 * (\n)}) --
      ({2.2 * \sx * \xf / 365 - 1.1 * \sx}, {0.4 * (\n)});
  }

\end{tikzpicture}
\caption{Each phenology transition extends from the day of the previous phenology event to the next,
here from when grapevines flower to when those flowers produce ripe fruit or veraison.  In 1987
Domaine de Vassal collected five events from Sauvignon Blanc grape varietals, each beginning in
late spring and ending in the middle of summer.}
\label{fig:phenology_data} 
\end{figure}

Within a year the data can be visualized by overlaying the daily temperatures with the
time intervals spanning the start and end of each event (Figure \ref{fig:phenology_data}).
Typically flowering occurs in late spring and then veraison begins by the middle of 
summer.

\subsubsection{Bayesian Analysis}

To infer the configuration of the forcing function I ran a Bayesian analysis using both 
the standard survival and modified survival models.  I elicited prior models from domain 
experts \citep{WolkovichEtAl:2022} and estimated posterior expectation values with dynamic 
Hamiltonian Monte Carlo \citep{Betancourt:2018b} implemented in Stan 2.32.3 \citep{Stan:2019a} 
through the RStan interface \citep{RStan:2019}.

The standard survival model required an \texttt{adapt\_delta} of $0.9$ to avoid divergent
Hamiltonian Monte Carlo transitions while the modified survival model was run with
the default RStan configuration.  Otherwise no diagnostics indicated biased posterior 
quantification.  Processed data, Stan programs, Markov chain chain Monte Carlo diagnostic
code, and analysis code are included in the supplementary material.

\subsubsection{Posterior Retrodictive Checks}

\begin{figure}
\centering
\subfigure[]{ 
\begin{tikzpicture}[scale=1.0]

  \pgfmathsetmacro{\sx}{2.4}
  \pgfmathsetmacro{\sy}{2.4}
  
  \draw[white] (-1.5 * \sx, - 0.35 * \sy) rectangle (1.5 * \sx, 1.85 * \sy);

  \pgfmathsetmacro{\prop}{20 + 15 * 1};
  \colorlet{custom}{dark!\prop!white};
  \foreach \a/\b/\c/\d in {-0.500/14.500/0.000/0.000, 14.500/29.500/0.000/0.000, 29.500/44.500/0.000/0.000, 44.500/59.500/0.000/0.000, 59.500/74.500/0.000/0.000, 74.500/89.500/0.000/0.000, 89.500/104.500/0.000/0.000, 104.500/119.500/0.000/0.000, 119.500/134.500/0.000/0.000, 134.500/149.500/1.000/6.000, 149.500/164.500/11.000/21.000, 164.500/179.500/15.000/26.000, 179.500/194.500/13.000/23.000, 194.500/209.500/9.000/18.000, 209.500/224.500/6.000/14.000, 224.500/239.500/4.000/11.000, 239.500/254.500/2.000/8.000, 254.500/269.500/1.000/5.000, 269.500/284.500/0.000/3.000, 284.500/299.500/0.000/2.000, 299.500/314.500/0.000/1.000, 314.500/329.500/0.000/0.000, 329.500/344.500/0.000/0.000, 344.500/359.500/0.000/0.000, 359.500/374.500/7.000/18.000} {
    \fill[custom] (2.2 * \sx * \a / 380 - 1.1 * \sx, 1.4 * \sy * \c / 75) rectangle 
                  (2.2 * \sx * \b / 380 - 1.1 * \sx, 1.4 * \sy * \d / 75);
  }
  
  \pgfmathsetmacro{\prop}{20 + 15 * 2};
  \colorlet{custom}{dark!\prop!white};
  \foreach \a/\b/\c/\d in {-0.500/14.500/0.000/0.000, 14.500/29.500/0.000/0.000, 29.500/44.500/0.000/0.000, 44.500/59.500/0.000/0.000, 59.500/74.500/0.000/0.000, 74.500/89.500/0.000/0.000, 89.500/104.500/0.000/0.000, 104.500/119.500/0.000/0.000, 119.500/134.500/0.000/0.000, 134.500/149.500/2.000/5.000, 149.500/164.500/12.000/19.000, 164.500/179.500/17.000/24.000, 179.500/194.500/14.000/21.000, 194.500/209.500/10.000/16.000, 209.500/224.500/8.000/13.000, 224.500/239.500/5.000/10.000, 239.500/254.500/3.000/7.000, 254.500/269.500/1.000/4.000, 269.500/284.500/0.000/2.000, 284.500/299.500/0.000/1.000, 299.500/314.500/0.000/1.000, 314.500/329.500/0.000/0.000, 329.500/344.500/0.000/0.000, 344.500/359.500/0.000/0.000, 359.500/374.500/9.000/16.000} {
    \fill[custom] (2.2 * \sx * \a / 380 - 1.1 * \sx, 1.4 * \sy * \c / 75) rectangle 
                  (2.2 * \sx * \b / 380 - 1.1 * \sx, 1.4 * \sy * \d / 75);
  }
    
  \pgfmathsetmacro{\prop}{20 + 15 * 3};
  \colorlet{custom}{dark!\prop!white};
  \foreach \a/\b/\c/\d in {-0.500/14.500/0.000/0.000, 14.500/29.500/0.000/0.000, 29.500/44.500/0.000/0.000, 44.500/59.500/0.000/0.000, 59.500/74.500/0.000/0.000, 74.500/89.500/0.000/0.000, 89.500/104.500/0.000/0.000, 104.500/119.500/0.000/0.000, 119.500/134.500/0.000/0.000, 134.500/149.500/2.000/4.000, 149.500/164.500/14.000/18.000, 164.500/179.500/18.000/23.000, 179.500/194.500/15.000/20.000, 194.500/209.500/12.000/15.000, 209.500/224.500/9.000/12.000, 224.500/239.500/6.000/9.000, 239.500/254.500/4.000/6.000, 254.500/269.500/2.000/4.000, 269.500/284.500/1.000/2.000, 284.500/299.500/0.000/1.000, 299.500/314.500/0.000/0.000, 314.500/329.500/0.000/0.000, 329.500/344.500/0.000/0.000, 344.500/359.500/0.000/0.000, 359.500/374.500/10.000/15.000} {
    \fill[custom] (2.2 * \sx * \a / 380 - 1.1 * \sx, 1.4 * \sy * \c / 75) rectangle 
                  (2.2 * \sx * \b / 380 - 1.1 * \sx, 1.4 * \sy * \d / 75);
  }
    
  \pgfmathsetmacro{\prop}{20 + 15 * 4};
  \colorlet{custom}{dark!\prop!white};
  \foreach \a/\b/\c/\d in {-0.500/14.500/0.000/0.000, 14.500/29.500/0.000/0.000, 29.500/44.500/0.000/0.000, 44.500/59.500/0.000/0.000, 59.500/74.500/0.000/0.000, 74.500/89.500/0.000/0.000, 89.500/104.500/0.000/0.000, 104.500/119.500/0.000/0.000, 119.500/134.500/0.000/0.000, 134.500/149.500/3.000/4.000, 149.500/164.500/15.000/17.000, 164.500/179.500/19.000/22.000, 179.500/194.500/17.000/19.000, 194.500/209.500/12.000/14.000, 209.500/224.500/9.000/11.000, 224.500/239.500/7.000/8.000, 239.500/254.500/4.000/6.000, 254.500/269.500/2.000/3.000, 269.500/284.500/1.000/1.000, 284.500/299.500/0.000/1.000, 299.500/314.500/0.000/0.000, 314.500/329.500/0.000/0.000, 329.500/344.500/0.000/0.000, 344.500/359.500/0.000/0.000, 359.500/374.500/11.000/13.000} {
    \fill[custom] (2.2 * \sx * \a / 380 - 1.1 * \sx, 1.4 * \sy * \c / 75) rectangle 
                  (2.2 * \sx * \b / 380 - 1.1 * \sx, 1.4 * \sy * \d / 75);
  }
    
  \foreach \a/\b/\c in {-0.500/14.500/0.000, 14.500/29.500/0.000, 29.500/44.500/0.000, 44.500/59.500/0.000, 59.500/74.500/0.000, 74.500/89.500/0.000, 89.500/104.500/0.000, 104.500/119.500/0.000, 119.500/134.500/0.000, 134.500/149.500/3.000, 149.500/164.500/16.000, 164.500/179.500/20.000, 179.500/194.500/18.000, 194.500/209.500/13.000, 209.500/224.500/10.000, 224.500/239.500/7.000, 239.500/254.500/5.000, 254.500/269.500/3.000, 269.500/284.500/1.000, 284.500/299.500/1.000, 299.500/314.500/0.000, 314.500/329.500/0.000, 329.500/344.500/0.000, 344.500/359.500/0.000, 359.500/374.500/12.000} {
  \draw[dark, line width=1] (2.2 * \sx * \a / 380 - 1.1 * \sx, 1.4 * \sy * \c / 75) --
                            (2.2 * \sx * \b / 380 - 1.1 * \sx, 1.4 * \sy * \c / 75);
  }

  \draw[white, line width=2] (-1.1 * \sx, 1.4 * \sy) 
  \foreach \x/\y in {-0.500/0.000, 14.500/0.000, 14.500/0.000, 29.500/0.000, 29.500/0.000, 44.500/0.000, 44.500/0.000, 59.500/0.000, 59.500/0.000, 74.500/0.000, 74.500/0.000, 89.500/0.000, 89.500/0.000, 104.500/0.000, 104.500/0.000, 119.500/0.000, 119.500/0.000, 134.500/0.000, 134.500/0.000, 149.500/0.000, 149.500/0.000, 164.500/0.000, 164.500/0.000, 179.500/0.000, 179.500/5.000, 194.500/5.000, 194.500/62.000, 209.500/62.000, 209.500/45.000, 224.500/45.000, 224.500/0.000, 239.500/0.000, 239.500/0.000, 254.500/0.000, 254.500/0.000, 269.500/0.000, 269.500/0.000, 284.500/0.000, 284.500/0.000, 299.500/0.000, 299.500/0.000, 314.500/0.000, 314.500/0.000, 329.500/0.000, 329.500/0.000, 344.500/0.000, 344.500/0.000, 359.500/0.000, 359.500/0.000, 374.500/0.000} {
    -- ({2.2 * \sx * \x / 380 - 1.1 * \sx}, {1.4 * \sy * \y / 75})
  };

  \draw[black, line width=1] (-1.1 * \sx, 1.4 * \sy) 
  \foreach \x/\y in {-0.500/0.000, 14.500/0.000, 14.500/0.000, 29.500/0.000, 29.500/0.000, 44.500/0.000, 44.500/0.000, 59.500/0.000, 59.500/0.000, 74.500/0.000, 74.500/0.000, 89.500/0.000, 89.500/0.000, 104.500/0.000, 104.500/0.000, 119.500/0.000, 119.500/0.000, 134.500/0.000, 134.500/0.000, 149.500/0.000, 149.500/0.000, 164.500/0.000, 164.500/0.000, 179.500/0.000, 179.500/5.000, 194.500/5.000, 194.500/62.000, 209.500/62.000, 209.500/45.000, 224.500/45.000, 224.500/0.000, 239.500/0.000, 239.500/0.000, 254.500/0.000, 254.500/0.000, 269.500/0.000, 269.500/0.000, 284.500/0.000, 284.500/0.000, 299.500/0.000, 299.500/0.000, 314.500/0.000, 314.500/0.000, 329.500/0.000, 329.500/0.000, 344.500/0.000, 344.500/0.000, 359.500/0.000, 359.500/0.000, 374.500/0.000} {
    -- ({2.2 * \sx * \x / 380 - 1.1 * \sx}, {1.4 * \sy * \y / 75})
  };

  \draw [<->, >=stealth, line width=1] (-1.1 * \sx, 1.5 * \sy) -- (-1.1 * \sx, 0) -- (1.1 * \sx, 0);
  \node at (0, -0.5) { Day of Year };
  \node[rotate=90] at (-1.4 * \sx, 0.75 * \sy) { Number of Phenology Events };
  \node at (-1.2 * \sx, 0) { $0$ };
  \node at (-1.2 * \sx, 1.4 * \sy) { $1$ };
  
  \node at (0, 1.7 * \sy) { Standard Survival Model };

\end{tikzpicture} 
}
\subfigure[]{
\begin{tikzpicture}[scale=1.0]
  \pgfmathsetmacro{\sx}{2.4}
  \pgfmathsetmacro{\sy}{2.4}
  
  \draw[white] (-1.5 * \sx, - 0.35 * \sy) rectangle (1.5 * \sx, 1.85 * \sy);
  
  \pgfmathsetmacro{\prop}{20 + 15 * 1};
  \colorlet{custom}{dark!\prop!white};
  \foreach \a/\b/\c/\d in {-0.500/14.500/0.000/0.000, 14.500/29.500/0.000/0.000, 29.500/44.500/0.000/0.000, 44.500/59.500/0.000/0.000, 59.500/74.500/0.000/0.000, 74.500/89.500/0.000/0.000, 89.500/104.500/0.000/0.000, 104.500/119.500/0.000/0.000, 119.500/134.500/0.000/0.000, 134.500/149.500/0.000/0.000, 149.500/164.500/0.000/0.000, 164.500/179.500/0.000/0.000, 179.500/194.500/0.000/3.000, 194.500/209.500/47.000/59.000, 209.500/224.500/50.000/63.000, 224.500/239.500/0.000/2.000, 239.500/254.500/0.000/0.000, 254.500/269.500/0.000/0.000, 269.500/284.500/0.000/0.000, 284.500/299.500/0.000/0.000, 299.500/314.500/0.000/0.000, 314.500/329.500/0.000/0.000, 329.500/344.500/0.000/0.000, 344.500/359.500/0.000/0.000, 359.500/374.500/0.000/0.000} {
    \fill[custom] (2.2 * \sx * \a / 380 - 1.1 * \sx, 1.4 * \sy * \c / 75) rectangle 
                  (2.2 * \sx * \b / 380 - 1.1 * \sx, 1.4 * \sy * \d / 75);
  }
  
  \pgfmathsetmacro{\prop}{20 + 15 * 2};
  \colorlet{custom}{dark!\prop!white};
  \foreach \a/\b/\c/\d in {-0.500/14.500/0.000/0.000, 14.500/29.500/0.000/0.000, 29.500/44.500/0.000/0.000, 44.500/59.500/0.000/0.000, 59.500/74.500/0.000/0.000, 74.500/89.500/0.000/0.000, 89.500/104.500/0.000/0.000, 104.500/119.500/0.000/0.000, 119.500/134.500/0.000/0.000, 134.500/149.500/0.000/0.000, 149.500/164.500/0.000/0.000, 164.500/179.500/0.000/0.000, 179.500/194.500/1.000/3.000, 194.500/209.500/49.000/57.000, 209.500/224.500/52.000/60.000, 224.500/239.500/0.000/2.000, 239.500/254.500/0.000/0.000, 254.500/269.500/0.000/0.000, 269.500/284.500/0.000/0.000, 284.500/299.500/0.000/0.000, 299.500/314.500/0.000/0.000, 314.500/329.500/0.000/0.000, 329.500/344.500/0.000/0.000, 344.500/359.500/0.000/0.000, 359.500/374.500/0.000/0.000} {
    \fill[custom] (2.2 * \sx * \a / 380 - 1.1 * \sx, 1.4 * \sy * \c / 75) rectangle 
                  (2.2 * \sx * \b / 380 - 1.1 * \sx, 1.4 * \sy * \d / 75);
  }
    
  \pgfmathsetmacro{\prop}{20 + 15 * 3};
  \colorlet{custom}{dark!\prop!white};
  \foreach \a/\b/\c/\d in {-0.500/14.500/0.000/0.000, 14.500/29.500/0.000/0.000, 29.500/44.500/0.000/0.000, 44.500/59.500/0.000/0.000, 59.500/74.500/0.000/0.000, 74.500/89.500/0.000/0.000, 89.500/104.500/0.000/0.000, 104.500/119.500/0.000/0.000, 119.500/134.500/0.000/0.000, 134.500/149.500/0.000/0.000, 149.500/164.500/0.000/0.000, 164.500/179.500/0.000/0.000, 179.500/194.500/1.000/2.000, 194.500/209.500/50.000/55.000, 209.500/224.500/54.000/59.000, 224.500/239.500/0.000/1.000, 239.500/254.500/0.000/0.000, 254.500/269.500/0.000/0.000, 269.500/284.500/0.000/0.000, 284.500/299.500/0.000/0.000, 299.500/314.500/0.000/0.000, 314.500/329.500/0.000/0.000, 329.500/344.500/0.000/0.000, 344.500/359.500/0.000/0.000, 359.500/374.500/0.000/0.000} {
    \fill[custom] (2.2 * \sx * \a / 380 - 1.1 * \sx, 1.4 * \sy * \c / 75) rectangle 
                  (2.2 * \sx * \b / 380 - 1.1 * \sx, 1.4 * \sy * \d / 75);
  }
    
  \pgfmathsetmacro{\prop}{20 + 15 * 4};
  \colorlet{custom}{dark!\prop!white};
  \foreach \a/\b/\c/\d in {-0.500/14.500/0.000/0.000, 14.500/29.500/0.000/0.000, 29.500/44.500/0.000/0.000, 44.500/59.500/0.000/0.000, 59.500/74.500/0.000/0.000, 74.500/89.500/0.000/0.000, 89.500/104.500/0.000/0.000, 104.500/119.500/0.000/0.000, 119.500/134.500/0.000/0.000, 134.500/149.500/0.000/0.000, 149.500/164.500/0.000/0.000, 164.500/179.500/0.000/0.000, 179.500/194.500/1.000/2.000, 194.500/209.500/52.000/54.000, 209.500/224.500/55.000/58.000, 224.500/239.500/1.000/1.000, 239.500/254.500/0.000/0.000, 254.500/269.500/0.000/0.000, 269.500/284.500/0.000/0.000, 284.500/299.500/0.000/0.000, 299.500/314.500/0.000/0.000, 314.500/329.500/0.000/0.000, 329.500/344.500/0.000/0.000, 344.500/359.500/0.000/0.000, 359.500/374.500/0.000/0.000} {
    \fill[custom] (2.2 * \sx * \a / 380 - 1.1 * \sx, 1.4 * \sy * \c / 75) rectangle 
                  (2.2 * \sx * \b / 380 - 1.1 * \sx, 1.4 * \sy * \d / 75);
  }
    
  \foreach \a/\b/\c in {-0.500/14.500/0.000, 14.500/29.500/0.000, 29.500/44.500/0.000, 44.500/59.500/0.000, 59.500/74.500/0.000, 74.500/89.500/0.000, 89.500/104.500/0.000, 104.500/119.500/0.000, 119.500/134.500/0.000, 134.500/149.500/0.000, 149.500/164.500/0.000, 164.500/179.500/0.000, 179.500/194.500/1.000, 194.500/209.500/53.000, 209.500/224.500/56.000, 224.500/239.500/1.000, 239.500/254.500/0.000, 254.500/269.500/0.000, 269.500/284.500/0.000, 284.500/299.500/0.000, 299.500/314.500/0.000, 314.500/329.500/0.000, 329.500/344.500/0.000, 344.500/359.500/0.000, 359.500/374.500/0.000} {
  \draw[dark, line width=1] (2.2 * \sx * \a / 380 - 1.1 * \sx, 1.4 * \sy * \c / 75) --
                            (2.2 * \sx * \b / 380 - 1.1 * \sx, 1.4 * \sy * \c / 75);
  }
  
  \draw[white, line width=2] (-1.1 * \sx, 1.4 * \sy) 
  \foreach \x/\y in {-0.500/0.000, 14.500/0.000, 14.500/0.000, 29.500/0.000, 29.500/0.000, 44.500/0.000, 44.500/0.000, 59.500/0.000, 59.500/0.000, 74.500/0.000, 74.500/0.000, 89.500/0.000, 89.500/0.000, 104.500/0.000, 104.500/0.000, 119.500/0.000, 119.500/0.000, 134.500/0.000, 134.500/0.000, 149.500/0.000, 149.500/0.000, 164.500/0.000, 164.500/0.000, 179.500/0.000, 179.500/5.000, 194.500/5.000, 194.500/62.000, 209.500/62.000, 209.500/45.000, 224.500/45.000, 224.500/0.000, 239.500/0.000, 239.500/0.000, 254.500/0.000, 254.500/0.000, 269.500/0.000, 269.500/0.000, 284.500/0.000, 284.500/0.000, 299.500/0.000, 299.500/0.000, 314.500/0.000, 314.500/0.000, 329.500/0.000, 329.500/0.000, 344.500/0.000, 344.500/0.000, 359.500/0.000, 359.500/0.000, 374.500/0.000} {
    -- ({2.2 * \sx * \x / 380 - 1.1 * \sx}, {1.4 * \sy * \y / 75})
  };

  \draw[black, line width=1] (-1.1 * \sx, 1.4 * \sy) 
  \foreach \x/\y in {-0.500/0.000, 14.500/0.000, 14.500/0.000, 29.500/0.000, 29.500/0.000, 44.500/0.000, 44.500/0.000, 59.500/0.000, 59.500/0.000, 74.500/0.000, 74.500/0.000, 89.500/0.000, 89.500/0.000, 104.500/0.000, 104.500/0.000, 119.500/0.000, 119.500/0.000, 134.500/0.000, 134.500/0.000, 149.500/0.000, 149.500/0.000, 164.500/0.000, 164.500/0.000, 179.500/0.000, 179.500/5.000, 194.500/5.000, 194.500/62.000, 209.500/62.000, 209.500/45.000, 224.500/45.000, 224.500/0.000, 239.500/0.000, 239.500/0.000, 254.500/0.000, 254.500/0.000, 269.500/0.000, 269.500/0.000, 284.500/0.000, 284.500/0.000, 299.500/0.000, 299.500/0.000, 314.500/0.000, 314.500/0.000, 329.500/0.000, 329.500/0.000, 344.500/0.000, 344.500/0.000, 359.500/0.000, 359.500/0.000, 374.500/0.000} {
    -- ({2.2 * \sx * \x / 380 - 1.1 * \sx}, {1.4 * \sy * \y / 75})
  };
  
  \draw [<->, >=stealth, line width=1] (-1.1 * \sx, 1.5 * \sy) -- (-1.1 * \sx, 0) -- (1.1 * \sx, 0);
  \node at (0, -0.5) { Day of Year };
  \node[rotate=90] at (-1.4 * \sx, 0.75 * \sy) { Number of Phenology Events };
  \node at (-1.2 * \sx, 0) { $0$ };
  \node at (-1.2 * \sx, 1.4 * \sy) { $1$ };
  
  \node at (0, 1.7 * \sy) { Modified Survival Model };

\end{tikzpicture}
}
\caption{Posterior retrodictive checks demonstrate the superior fit of the (b)
modified survival model relative to the (a) standard survival model. Here the check 
is based on a histogram of the veraison events.  The standard survival model exhibits 
strong tension between the narrow observed histogram in black and the posterior 
predictive distribution of much wider histograms in the red ribbons, indicating model 
inadequacies.  On the other hand the posterior predictive distribution of the modified 
survival model concentrates on appropriately narrow histograms.}
\label{fig:retrodictive_checks} 
\end{figure}
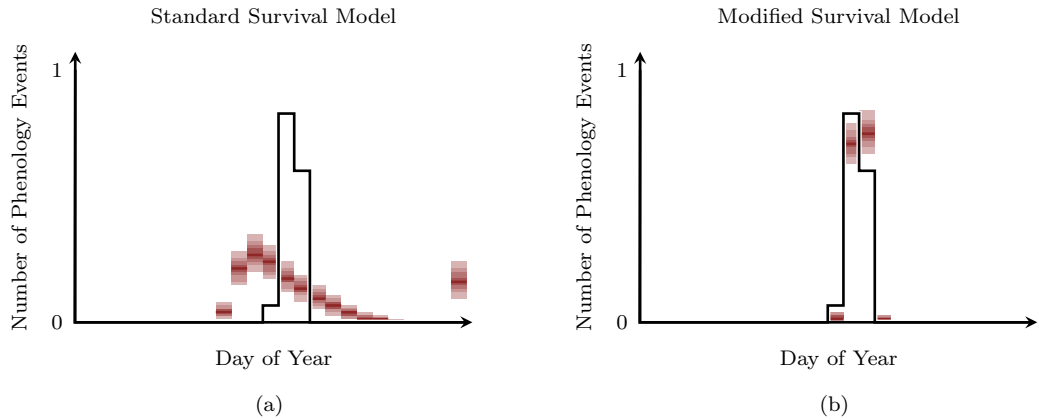

Comparison between the two models begins with posterior retrodictive checks \citep{Betancourt:2020a}
that contrast aspects of posterior predictive distribution to the same aspects in the 
observed data to qualify how well each model captures the relevant structure of the
observed data.  In particular I consider a histogram of all $112$ veraison events with 
bins fifteen days wide; the posterior predictive distribution of histograms is visualized
with ribbons showing the 10\%-90\%, 20\%-80\%, 30\%-70\%, 40\%-60\%, and 50\% marginal
quantiles of the bin counts (Figure \ref{fig:retrodictive_checks}).

Because the survival function decays as soon as the accumulated forcing becomes 
non-zero the standard survival model cannot cluster the versaison events as narrowly
as we see in the observed data.  The consistent model configurations spread out the
posterior predictive veraison events much earlier, and extend far past, what is seen 
in the observed data.  Indeed the bump in the last bin demonstrates a substantial 
posterior predictive probability for versaison events to not occur at all within 
the year.  On the other hand the modified survival model has no problem capturing
the narrow range of the observed veraison events days.  The fit is not perfect, but 
that isn't surprising given that the model does not take into account for example 
any variation between airport temperatures and local plant temperatures and 
heterogeneities in the ecological circumstances of each observation.

\subsubsection{Posterior Inferences}

\begin{figure}
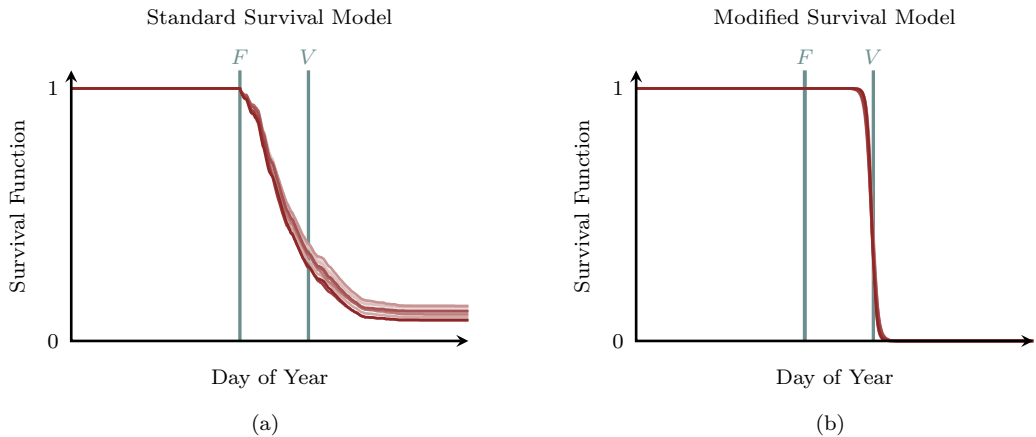

\centering
\subfigure[]{ 

}
\caption{Posterior inferences for the survival function of the first veraison event
in 1897 demonstrate the limitation of the standard survival model and the benefits
of the modified survival model.  (a) The inferred survival function for the standard 
survival model decays almost immediately after flowering (F) and persists until long
after the observed veraison (V).  (b) With the cumulative hazard suppressed by the
soft thresholding warping function the inferred survival function for the modified 
survival model does not start to decay until right before the observed veraison and 
then finishes decaying almost immediately afterwards.}
\label{fig:inferred_survival} 
\end{figure}

We can trace the source of the standard survival model's poor performance by
examining the corresponding posterior inferences.  Figure \ref{fig:inferred_survival} 
compares the inferred survival function behavior for first veraison event in 1987 
from both models.  The thresholding in the modified survival model allows the
survival function to stay at zero until right before the observed versaison and 
then decay completely almost immediately afterwards.  On the other hand the survival 
function in the standard survival model starts decaying almost immediately after 
flowering and then persists all the way to the end of year.

\begin{figure}
\centering
\subfigure[]{ 
\begin{tikzpicture}[scale=1.0]

  \pgfmathsetmacro{\sx}{2.4}
  \pgfmathsetmacro{\sy}{2.4}
  
  \draw[white] (-1.5 * \sx, - 0.35 * \sy) rectangle (1.5 * \sx, 1.85 * \sy);
  
  \colorlet{custom}{dark!10!white};
  \draw[custom, line width=1] (-1.1 * \sx, 0) 
  \foreach \y [count=\n] in {0.000, 0.000, 0.000, 0.000, 0.000, 0.000, 0.000, 0.000, 0.000, 0.000, 0.000, 0.000, 0.000, 0.000, 0.000, 0.000, 0.000, 0.000, 0.000, 0.000, 0.000, 0.000, 0.000, 0.000, 0.000, 0.000, 0.000, 0.000, 0.000, 0.000, 0.000, 0.000, 0.000, 0.000, 0.000, 0.000, 0.000, 0.000, 0.000, 0.000, 0.000, 0.000, 0.000, 0.000, 0.000, 0.000, 0.000, 0.000, 0.000, 0.000, 0.000, 0.000, 0.001, 0.001, 0.001, 0.001, 0.002, 0.003, 0.004, 0.005, 0.006, 0.007, 0.009, 0.011, 0.013, 0.016, 0.019, 0.023, 0.027, 0.032, 0.037, 0.042, 0.048, 0.055, 0.063, 0.071, 0.080, 0.090, 0.100, 0.111, 0.123, 0.136, 0.149, 0.164, 0.179, 0.195, 0.211, 0.229, 0.247, 0.266, 0.286, 0.307, 0.328, 0.349, 0.372, 0.395, 0.418, 0.442, 0.466, 0.490, 0.515, 0.540, 0.565, 0.590, 0.615, 0.640, 0.665, 0.690, 0.714, 0.738, 0.761, 0.783, 0.805, 0.826, 0.847, 0.866, 0.884, 0.902, 0.918, 0.932, 0.946, 0.958, 0.969, 0.978, 0.986, 0.992, 0.996, 0.999, 1.000, 0.999, 0.997, 0.993, 0.987, 0.979, 0.970, 0.959, 0.946, 0.932, 0.916, 0.898, 0.879, 0.859, 0.837, 0.814, 0.790, 0.764, 0.738, 0.711, 0.683, 0.654, 0.625, 0.595, 0.565, 0.535, 0.504, 0.474, 0.444, 0.414, 0.384, 0.355, 0.327, 0.300, 0.273, 0.247, 0.223, 0.199, 0.177, 0.156, 0.137, 0.118, 0.101, 0.086, 0.072, 0.060, 0.049, 0.039, 0.030, 0.023, 0.017, 0.012, 0.009, 0.006, 0.004, 0.002, 0.001, 0.000, 0.000, 0.000, 0.000, 0.000, 0.000, 0.000, 0.000, 0.000, 0.000, 0.000, 0.000, 0.000, 0.000, 0.000, 0.000, 0.000, 0.000, 0.000, 0.000, 0.000, 0.000, 0.000, 0.000, 0.000, 0.000, 0.000, 0.000, 0.000, 0.000, 0.000} {
    -- ({2.2 * \sx * \n / 216 - 1.1 * \sx}, {1.4 * \sy * \y})
  };
  
  \colorlet{custom}{dark!20!white};
  \draw[custom, line width=1] (-1.1 * \sx, 0) 
  \foreach \y [count=\n] in {0.000, 0.000, 0.000, 0.000, 0.000, 0.000, 0.000, 0.000, 0.000, 0.000, 0.000, 0.000, 0.000, 0.000, 0.000, 0.000, 0.000, 0.000, 0.000, 0.000, 0.000, 0.000, 0.000, 0.000, 0.000, 0.000, 0.000, 0.000, 0.000, 0.000, 0.000, 0.000, 0.000, 0.000, 0.000, 0.000, 0.000, 0.000, 0.000, 0.000, 0.000, 0.000, 0.000, 0.000, 0.000, 0.000, 0.000, 0.000, 0.000, 0.000, 0.000, 0.000, 0.000, 0.000, 0.000, 0.000, 0.000, 0.000, 0.000, 0.000, 0.000, 0.001, 0.001, 0.001, 0.001, 0.002, 0.002, 0.002, 0.003, 0.004, 0.004, 0.005, 0.007, 0.008, 0.009, 0.011, 0.013, 0.015, 0.018, 0.021, 0.025, 0.029, 0.033, 0.038, 0.044, 0.050, 0.057, 0.065, 0.074, 0.083, 0.094, 0.105, 0.118, 0.132, 0.146, 0.163, 0.180, 0.199, 0.219, 0.240, 0.263, 0.287, 0.312, 0.339, 0.367, 0.397, 0.427, 0.459, 0.492, 0.525, 0.560, 0.595, 0.630, 0.665, 0.701, 0.735, 0.770, 0.803, 0.835, 0.865, 0.893, 0.919, 0.942, 0.962, 0.978, 0.990, 0.997, 1.000, 0.997, 0.989, 0.976, 0.956, 0.930, 0.898, 0.860, 0.815, 0.765, 0.710, 0.650, 0.586, 0.519, 0.450, 0.381, 0.312, 0.246, 0.184, 0.128, 0.081, 0.043, 0.017, 0.003, 0.000, 0.000, 0.000, 0.000, 0.000, 0.000, 0.000, 0.000, 0.000, 0.000, 0.000, 0.000, 0.000, 0.000, 0.000, 0.000, 0.000, 0.000, 0.000, 0.000, 0.000, 0.000, 0.000, 0.000, 0.000, 0.000, 0.000, 0.000, 0.000, 0.000, 0.000, 0.000, 0.000, 0.000, 0.000, 0.000, 0.000, 0.000, 0.000, 0.000, 0.000, 0.000, 0.000, 0.000, 0.000, 0.000, 0.000, 0.000, 0.000, 0.000, 0.000, 0.000, 0.000, 0.000, 0.000, 0.000, 0.000, 0.000, 0.000, 0.000, 0.000, 0.000, 0.000, 0.000, 0.000} {
    -- ({2.2 * \sx * \n / 216 - 1.1 * \sx}, {1.4 * \sy * \y})
  };
  
  \colorlet{custom}{dark!30!white};
  \draw[custom, line width=1] (-1.1 * \sx, 0) 
  \foreach \y [count=\n] in {0.000, 0.000, 0.000, 0.000, 0.000, 0.000, 0.000, 0.000, 0.000, 0.000, 0.000, 0.000, 0.000, 0.000, 0.000, 0.000, 0.000, 0.000, 0.000, 0.000, 0.000, 0.000, 0.000, 0.000, 0.000, 0.000, 0.000, 0.000, 0.000, 0.000, 0.000, 0.000, 0.000, 0.000, 0.000, 0.000, 0.000, 0.000, 0.000, 0.000, 0.000, 0.000, 0.000, 0.000, 0.000, 0.000, 0.000, 0.000, 0.000, 0.000, 0.000, 0.000, 0.000, 0.000, 0.000, 0.000, 0.000, 0.000, 0.000, 0.000, 0.000, 0.000, 0.000, 0.001, 0.001, 0.001, 0.001, 0.002, 0.002, 0.002, 0.003, 0.004, 0.004, 0.005, 0.007, 0.008, 0.009, 0.011, 0.013, 0.016, 0.019, 0.022, 0.025, 0.030, 0.034, 0.040, 0.046, 0.053, 0.060, 0.069, 0.078, 0.088, 0.100, 0.112, 0.126, 0.141, 0.158, 0.175, 0.194, 0.215, 0.237, 0.261, 0.286, 0.313, 0.341, 0.370, 0.401, 0.434, 0.467, 0.502, 0.538, 0.574, 0.611, 0.648, 0.685, 0.722, 0.759, 0.794, 0.828, 0.860, 0.890, 0.918, 0.942, 0.963, 0.979, 0.991, 0.998, 1.000, 0.995, 0.985, 0.968, 0.944, 0.913, 0.875, 0.831, 0.780, 0.724, 0.662, 0.595, 0.525, 0.453, 0.381, 0.309, 0.241, 0.177, 0.121, 0.073, 0.037, 0.013, 0.002, 0.000, 0.000, 0.000, 0.000, 0.000, 0.000, 0.000, 0.000, 0.000, 0.000, 0.000, 0.000, 0.000, 0.000, 0.000, 0.000, 0.000, 0.000, 0.000, 0.000, 0.000, 0.000, 0.000, 0.000, 0.000, 0.000, 0.000, 0.000, 0.000, 0.000, 0.000, 0.000, 0.000, 0.000, 0.000, 0.000, 0.000, 0.000, 0.000, 0.000, 0.000, 0.000, 0.000, 0.000, 0.000, 0.000, 0.000, 0.000, 0.000, 0.000, 0.000, 0.000, 0.000, 0.000, 0.000, 0.000, 0.000, 0.000, 0.000, 0.000, 0.000, 0.000, 0.000, 0.000, 0.000, 0.000} {
    -- ({2.2 * \sx * \n / 216 - 1.1 * \sx}, {1.4 * \sy * \y})
  };
  
  \colorlet{custom}{dark!40!white};
  \draw[custom, line width=1] (-1.1 * \sx, 0) 
  \foreach \y [count=\n] in {0.000, 0.000, 0.000, 0.000, 0.000, 0.000, 0.000, 0.000, 0.000, 0.000, 0.000, 0.000, 0.000, 0.000, 0.000, 0.000, 0.000, 0.000, 0.000, 0.000, 0.000, 0.000, 0.000, 0.000, 0.000, 0.000, 0.000, 0.000, 0.000, 0.000, 0.000, 0.000, 0.000, 0.000, 0.000, 0.000, 0.000, 0.000, 0.000, 0.000, 0.000, 0.000, 0.000, 0.000, 0.000, 0.000, 0.000, 0.000, 0.000, 0.000, 0.000, 0.000, 0.000, 0.000, 0.000, 0.000, 0.000, 0.000, 0.000, 0.000, 0.001, 0.001, 0.001, 0.001, 0.001, 0.002, 0.002, 0.003, 0.004, 0.004, 0.005, 0.007, 0.008, 0.010, 0.012, 0.014, 0.017, 0.021, 0.024, 0.029, 0.034, 0.039, 0.046, 0.053, 0.061, 0.070, 0.081, 0.092, 0.104, 0.118, 0.133, 0.150, 0.167, 0.187, 0.208, 0.230, 0.254, 0.279, 0.306, 0.334, 0.364, 0.395, 0.427, 0.460, 0.494, 0.529, 0.565, 0.601, 0.637, 0.672, 0.708, 0.743, 0.777, 0.809, 0.840, 0.869, 0.896, 0.921, 0.942, 0.961, 0.976, 0.988, 0.996, 1.000, 0.999, 0.995, 0.987, 0.974, 0.957, 0.936, 0.911, 0.882, 0.850, 0.815, 0.777, 0.736, 0.694, 0.650, 0.604, 0.558, 0.512, 0.466, 0.421, 0.377, 0.335, 0.294, 0.256, 0.220, 0.187, 0.157, 0.130, 0.106, 0.085, 0.067, 0.052, 0.039, 0.029, 0.021, 0.015, 0.010, 0.006, 0.004, 0.002, 0.001, 0.001, 0.000, 0.000, 0.000, 0.000, 0.000, 0.000, 0.000, 0.000, 0.000, 0.000, 0.000, 0.000, 0.000, 0.000, 0.000, 0.000, 0.000, 0.000, 0.000, 0.000, 0.000, 0.000, 0.000, 0.000, 0.000, 0.000, 0.000, 0.000, 0.000, 0.000, 0.000, 0.000, 0.000, 0.000, 0.000, 0.000, 0.000, 0.000, 0.000, 0.000, 0.000, 0.000, 0.000, 0.000, 0.000, 0.000, 0.000, 0.000, 0.000, 0.000, 0.000} {
    -- ({2.2 * \sx * \n / 216 - 1.1 * \sx}, {1.4 * \sy * \y})
  };
  
  \colorlet{custom}{dark!50!white};
  \draw[custom, line width=1] (-1.1 * \sx, 0) 
  \foreach \y [count=\n] in {0.000, 0.000, 0.000, 0.000, 0.000, 0.000, 0.000, 0.000, 0.000, 0.000, 0.000, 0.000, 0.000, 0.000, 0.000, 0.000, 0.000, 0.000, 0.000, 0.000, 0.000, 0.000, 0.000, 0.000, 0.000, 0.000, 0.000, 0.000, 0.000, 0.000, 0.000, 0.000, 0.000, 0.000, 0.000, 0.000, 0.000, 0.000, 0.000, 0.000, 0.000, 0.000, 0.000, 0.000, 0.000, 0.000, 0.000, 0.000, 0.000, 0.000, 0.000, 0.000, 0.000, 0.000, 0.000, 0.001, 0.001, 0.001, 0.001, 0.001, 0.002, 0.002, 0.003, 0.003, 0.004, 0.005, 0.006, 0.007, 0.009, 0.010, 0.012, 0.014, 0.017, 0.019, 0.023, 0.026, 0.030, 0.035, 0.040, 0.046, 0.052, 0.059, 0.067, 0.076, 0.086, 0.096, 0.108, 0.120, 0.134, 0.148, 0.164, 0.181, 0.200, 0.220, 0.241, 0.263, 0.286, 0.311, 0.338, 0.365, 0.394, 0.424, 0.455, 0.486, 0.519, 0.553, 0.587, 0.621, 0.656, 0.690, 0.725, 0.758, 0.791, 0.823, 0.853, 0.882, 0.908, 0.932, 0.953, 0.970, 0.984, 0.994, 0.999, 1.000, 0.995, 0.985, 0.970, 0.949, 0.922, 0.889, 0.850, 0.806, 0.757, 0.702, 0.643, 0.581, 0.516, 0.449, 0.381, 0.315, 0.251, 0.190, 0.136, 0.089, 0.051, 0.023, 0.006, 0.000, 0.000, 0.000, 0.000, 0.000, 0.000, 0.000, 0.000, 0.000, 0.000, 0.000, 0.000, 0.000, 0.000, 0.000, 0.000, 0.000, 0.000, 0.000, 0.000, 0.000, 0.000, 0.000, 0.000, 0.000, 0.000, 0.000, 0.000, 0.000, 0.000, 0.000, 0.000, 0.000, 0.000, 0.000, 0.000, 0.000, 0.000, 0.000, 0.000, 0.000, 0.000, 0.000, 0.000, 0.000, 0.000, 0.000, 0.000, 0.000, 0.000, 0.000, 0.000, 0.000, 0.000, 0.000, 0.000, 0.000, 0.000, 0.000, 0.000, 0.000, 0.000, 0.000, 0.000, 0.000, 0.000, 0.000, 0.000, 0.000} {
    -- ({2.2 * \sx * \n / 216 - 1.1 * \sx}, {1.4 * \sy * \y})
  };
  
  \colorlet{custom}{dark!60!white};
  \draw[custom, line width=1] (-1.1 * \sx, 0) 
  \foreach \y [count=\n] in {0.000, 0.000, 0.000, 0.000, 0.000, 0.000, 0.000, 0.000, 0.000, 0.000, 0.000, 0.000, 0.000, 0.000, 0.000, 0.000, 0.000, 0.000, 0.000, 0.000, 0.000, 0.000, 0.000, 0.000, 0.000, 0.000, 0.000, 0.000, 0.000, 0.000, 0.000, 0.000, 0.000, 0.000, 0.000, 0.000, 0.000, 0.000, 0.000, 0.000, 0.000, 0.000, 0.000, 0.000, 0.000, 0.000, 0.000, 0.000, 0.000, 0.000, 0.000, 0.001, 0.001, 0.001, 0.001, 0.001, 0.002, 0.002, 0.003, 0.003, 0.004, 0.005, 0.006, 0.007, 0.008, 0.009, 0.011, 0.013, 0.015, 0.017, 0.020, 0.023, 0.026, 0.029, 0.033, 0.038, 0.043, 0.048, 0.054, 0.060, 0.067, 0.075, 0.083, 0.092, 0.101, 0.111, 0.122, 0.134, 0.146, 0.159, 0.173, 0.188, 0.204, 0.220, 0.238, 0.256, 0.275, 0.295, 0.315, 0.336, 0.359, 0.381, 0.405, 0.429, 0.454, 0.479, 0.505, 0.532, 0.558, 0.585, 0.612, 0.639, 0.667, 0.694, 0.720, 0.747, 0.773, 0.798, 0.822, 0.846, 0.868, 0.889, 0.909, 0.928, 0.944, 0.959, 0.972, 0.982, 0.990, 0.996, 0.999, 1.000, 0.997, 0.992, 0.984, 0.972, 0.957, 0.939, 0.918, 0.893, 0.866, 0.835, 0.801, 0.764, 0.725, 0.682, 0.638, 0.592, 0.544, 0.495, 0.445, 0.394, 0.344, 0.295, 0.247, 0.202, 0.159, 0.120, 0.086, 0.056, 0.032, 0.015, 0.004, 0.000, 0.000, 0.000, 0.000, 0.000, 0.000, 0.000, 0.000, 0.000, 0.000, 0.000, 0.000, 0.000, 0.000, 0.000, 0.000, 0.000, 0.000, 0.000, 0.000, 0.000, 0.000, 0.000, 0.000, 0.000, 0.000, 0.000, 0.000, 0.000, 0.000, 0.000, 0.000, 0.000, 0.000, 0.000, 0.000, 0.000, 0.000, 0.000, 0.000, 0.000, 0.000, 0.000, 0.000, 0.000, 0.000, 0.000, 0.000, 0.000, 0.000, 0.000, 0.000, 0.000} {
    -- ({2.2 * \sx * \n / 216 - 1.1 * \sx}, {1.4 * \sy * \y})
  };
  
  \colorlet{custom}{dark!70!white};
  \draw[custom, line width=1] (-1.1 * \sx, 0) 
  \foreach \y [count=\n] in {0.000, 0.000, 0.000, 0.000, 0.000, 0.000, 0.000, 0.000, 0.000, 0.000, 0.000, 0.000, 0.000, 0.000, 0.000, 0.000, 0.000, 0.000, 0.000, 0.000, 0.000, 0.000, 0.000, 0.000, 0.000, 0.000, 0.000, 0.000, 0.000, 0.000, 0.000, 0.000, 0.000, 0.000, 0.000, 0.000, 0.000, 0.000, 0.000, 0.000, 0.000, 0.000, 0.000, 0.000, 0.000, 0.000, 0.000, 0.000, 0.000, 0.000, 0.000, 0.000, 0.000, 0.000, 0.000, 0.000, 0.000, 0.000, 0.000, 0.000, 0.000, 0.000, 0.000, 0.000, 0.000, 0.000, 0.000, 0.000, 0.000, 0.000, 0.001, 0.001, 0.001, 0.001, 0.002, 0.002, 0.003, 0.004, 0.005, 0.006, 0.008, 0.009, 0.012, 0.014, 0.017, 0.021, 0.025, 0.031, 0.036, 0.043, 0.051, 0.060, 0.071, 0.083, 0.096, 0.111, 0.128, 0.146, 0.167, 0.189, 0.214, 0.240, 0.269, 0.300, 0.333, 0.367, 0.404, 0.442, 0.482, 0.523, 0.565, 0.608, 0.650, 0.693, 0.735, 0.775, 0.814, 0.850, 0.884, 0.915, 0.941, 0.963, 0.981, 0.993, 0.999, 0.999, 0.994, 0.982, 0.964, 0.940, 0.911, 0.875, 0.835, 0.790, 0.742, 0.690, 0.636, 0.580, 0.524, 0.468, 0.413, 0.360, 0.310, 0.263, 0.219, 0.180, 0.145, 0.114, 0.088, 0.066, 0.048, 0.034, 0.023, 0.015, 0.009, 0.005, 0.003, 0.001, 0.001, 0.000, 0.000, 0.000, 0.000, 0.000, 0.000, 0.000, 0.000, 0.000, 0.000, 0.000, 0.000, 0.000, 0.000, 0.000, 0.000, 0.000, 0.000, 0.000, 0.000, 0.000, 0.000, 0.000, 0.000, 0.000, 0.000, 0.000, 0.000, 0.000, 0.000, 0.000, 0.000, 0.000, 0.000, 0.000, 0.000, 0.000, 0.000, 0.000, 0.000, 0.000, 0.000, 0.000, 0.000, 0.000, 0.000, 0.000, 0.000, 0.000, 0.000, 0.000, 0.000, 0.000, 0.000, 0.000, 0.000, 0.000} {
    -- ({2.2 * \sx * \n / 216 - 1.1 * \sx}, {1.4 * \sy * \y})
  };
  
  \colorlet{custom}{dark!80!white};
  \draw[custom, line width=1] (-1.1 * \sx, 0) 
  \foreach \y [count=\n] in {0.000, 0.000, 0.000, 0.000, 0.000, 0.000, 0.000, 0.000, 0.000, 0.000, 0.000, 0.000, 0.000, 0.000, 0.000, 0.000, 0.000, 0.000, 0.000, 0.000, 0.000, 0.000, 0.000, 0.000, 0.000, 0.000, 0.000, 0.000, 0.000, 0.000, 0.000, 0.000, 0.000, 0.000, 0.000, 0.000, 0.000, 0.000, 0.000, 0.000, 0.000, 0.000, 0.000, 0.000, 0.000, 0.000, 0.000, 0.000, 0.000, 0.000, 0.000, 0.000, 0.000, 0.000, 0.000, 0.000, 0.000, 0.000, 0.000, 0.000, 0.000, 0.000, 0.000, 0.000, 0.000, 0.000, 0.001, 0.001, 0.001, 0.001, 0.002, 0.002, 0.003, 0.003, 0.004, 0.005, 0.006, 0.008, 0.010, 0.012, 0.014, 0.017, 0.021, 0.025, 0.029, 0.035, 0.041, 0.048, 0.056, 0.065, 0.076, 0.087, 0.100, 0.114, 0.130, 0.147, 0.166, 0.187, 0.210, 0.234, 0.260, 0.288, 0.317, 0.349, 0.382, 0.416, 0.452, 0.489, 0.527, 0.566, 0.606, 0.645, 0.685, 0.724, 0.762, 0.799, 0.834, 0.867, 0.897, 0.924, 0.948, 0.968, 0.983, 0.994, 0.999, 0.999, 0.994, 0.982, 0.965, 0.942, 0.913, 0.879, 0.840, 0.796, 0.747, 0.695, 0.640, 0.583, 0.525, 0.467, 0.409, 0.352, 0.298, 0.247, 0.201, 0.158, 0.121, 0.090, 0.063, 0.042, 0.027, 0.015, 0.008, 0.003, 0.001, 0.000, 0.000, 0.000, 0.000, 0.000, 0.000, 0.000, 0.000, 0.000, 0.000, 0.000, 0.000, 0.000, 0.000, 0.000, 0.000, 0.000, 0.000, 0.000, 0.000, 0.000, 0.000, 0.000, 0.000, 0.000, 0.000, 0.000, 0.000, 0.000, 0.000, 0.000, 0.000, 0.000, 0.000, 0.000, 0.000, 0.000, 0.000, 0.000, 0.000, 0.000, 0.000, 0.000, 0.000, 0.000, 0.000, 0.000, 0.000, 0.000, 0.000, 0.000, 0.000, 0.000, 0.000, 0.000, 0.000, 0.000, 0.000, 0.000, 0.000, 0.000} {
    -- ({2.2 * \sx * \n / 216 - 1.1 * \sx}, {1.4 * \sy * \y})
  };
  
  \colorlet{custom}{dark!90!white};
  \draw[custom, line width=1] (-1.1 * \sx, 0) 
  \foreach \y [count=\n] in {0.000, 0.000, 0.000, 0.000, 0.000, 0.000, 0.000, 0.000, 0.000, 0.000, 0.000, 0.000, 0.000, 0.000, 0.000, 0.000, 0.000, 0.000, 0.000, 0.000, 0.000, 0.000, 0.000, 0.000, 0.000, 0.000, 0.000, 0.000, 0.000, 0.000, 0.000, 0.000, 0.000, 0.000, 0.000, 0.000, 0.000, 0.000, 0.000, 0.000, 0.000, 0.000, 0.000, 0.000, 0.000, 0.000, 0.000, 0.000, 0.000, 0.000, 0.000, 0.000, 0.000, 0.000, 0.000, 0.000, 0.000, 0.000, 0.000, 0.000, 0.001, 0.001, 0.001, 0.001, 0.002, 0.002, 0.003, 0.003, 0.004, 0.005, 0.006, 0.007, 0.009, 0.011, 0.013, 0.015, 0.017, 0.020, 0.024, 0.027, 0.032, 0.036, 0.042, 0.048, 0.054, 0.061, 0.069, 0.078, 0.088, 0.098, 0.109, 0.122, 0.135, 0.149, 0.164, 0.181, 0.198, 0.217, 0.236, 0.257, 0.279, 0.302, 0.327, 0.352, 0.378, 0.406, 0.434, 0.463, 0.493, 0.524, 0.555, 0.587, 0.619, 0.651, 0.683, 0.714, 0.746, 0.776, 0.806, 0.835, 0.862, 0.888, 0.912, 0.933, 0.952, 0.968, 0.982, 0.992, 0.998, 1.000, 0.998, 0.992, 0.981, 0.965, 0.945, 0.919, 0.889, 0.853, 0.813, 0.767, 0.718, 0.664, 0.607, 0.546, 0.483, 0.419, 0.355, 0.291, 0.230, 0.172, 0.120, 0.074, 0.038, 0.013, 0.001, 0.000, 0.000, 0.000, 0.000, 0.000, 0.000, 0.000, 0.000, 0.000, 0.000, 0.000, 0.000, 0.000, 0.000, 0.000, 0.000, 0.000, 0.000, 0.000, 0.000, 0.000, 0.000, 0.000, 0.000, 0.000, 0.000, 0.000, 0.000, 0.000, 0.000, 0.000, 0.000, 0.000, 0.000, 0.000, 0.000, 0.000, 0.000, 0.000, 0.000, 0.000, 0.000, 0.000, 0.000, 0.000, 0.000, 0.000, 0.000, 0.000, 0.000, 0.000, 0.000, 0.000, 0.000, 0.000, 0.000, 0.000, 0.000, 0.000, 0.000, 0.000} {
    -- ({2.2 * \sx * \n / 216 - 1.1 * \sx}, {1.4 * \sy * \y})
  };
  
  \colorlet{custom}{dark!100!white};
  \draw[custom, line width=1] (-1.1 * \sx, 0) 
  \foreach \y [count=\n] in {0.000, 0.000, 0.000, 0.000, 0.000, 0.000, 0.000, 0.000, 0.000, 0.000, 0.000, 0.000, 0.000, 0.000, 0.000, 0.000, 0.000, 0.000, 0.000, 0.000, 0.000, 0.000, 0.000, 0.000, 0.000, 0.000, 0.000, 0.000, 0.000, 0.000, 0.000, 0.000, 0.000, 0.000, 0.000, 0.000, 0.000, 0.000, 0.000, 0.000, 0.000, 0.000, 0.000, 0.000, 0.000, 0.000, 0.000, 0.000, 0.000, 0.000, 0.000, 0.000, 0.000, 0.000, 0.000, 0.000, 0.000, 0.000, 0.000, 0.000, 0.000, 0.000, 0.000, 0.000, 0.000, 0.000, 0.000, 0.000, 0.000, 0.001, 0.001, 0.001, 0.001, 0.002, 0.003, 0.003, 0.004, 0.006, 0.007, 0.009, 0.012, 0.015, 0.019, 0.023, 0.028, 0.035, 0.042, 0.051, 0.060, 0.072, 0.085, 0.100, 0.116, 0.135, 0.155, 0.178, 0.203, 0.230, 0.259, 0.291, 0.325, 0.360, 0.398, 0.437, 0.478, 0.519, 0.562, 0.604, 0.647, 0.690, 0.731, 0.772, 0.810, 0.846, 0.879, 0.909, 0.936, 0.958, 0.976, 0.989, 0.997, 1.000, 0.998, 0.990, 0.977, 0.960, 0.937, 0.910, 0.878, 0.843, 0.804, 0.763, 0.719, 0.674, 0.627, 0.580, 0.533, 0.486, 0.440, 0.395, 0.353, 0.312, 0.274, 0.239, 0.206, 0.176, 0.149, 0.125, 0.104, 0.085, 0.069, 0.056, 0.044, 0.035, 0.027, 0.020, 0.015, 0.011, 0.008, 0.006, 0.004, 0.003, 0.002, 0.001, 0.001, 0.000, 0.000, 0.000, 0.000, 0.000, 0.000, 0.000, 0.000, 0.000, 0.000, 0.000, 0.000, 0.000, 0.000, 0.000, 0.000, 0.000, 0.000, 0.000, 0.000, 0.000, 0.000, 0.000, 0.000, 0.000, 0.000, 0.000, 0.000, 0.000, 0.000, 0.000, 0.000, 0.000, 0.000, 0.000, 0.000, 0.000, 0.000, 0.000, 0.000, 0.000, 0.000, 0.000, 0.000, 0.000, 0.000, 0.000, 0.000, 0.000, 0.000, 0.000} {
    -- ({2.2 * \sx * \n / 216 - 1.1 * \sx}, {1.4 * \sy * \y})
  };

  \draw [<->, >=stealth, line width=1] (-1.1 * \sx, 1.5 * \sy) -- (-1.1 * \sx, 0) -- (1.1 * \sx, 0);
  \node at (0, -0.5) { Temperature (Arbitrary Units) };
  \node[rotate=90] at (-1.4 * \sx, 0.75 * \sy) { Forcing Function };
  \node at (-1.2 * \sx, 0) { $0$ };
  \node at (-1.2 * \sx, 1.4 * \sy) { $1$ };
  
  \node at (0, 1.7 * \sy) { Standard Survival Model };

\end{tikzpicture} 
}
\subfigure[]{
\begin{tikzpicture}[scale=1.0]
  \pgfmathsetmacro{\sx}{2.4}
  \pgfmathsetmacro{\sy}{2.4}
  
  \draw[white] (-1.5 * \sx, - 0.35 * \sy) rectangle (1.5 * \sx, 1.85 * \sy);
  
  \colorlet{custom}{dark!10!white};
  \draw[custom, line width=1] (-1.1 * \sx, 0) 
  \foreach \y [count=\n] in {0.000, 0.000, 0.000, 0.000, 0.000, 0.000, 0.000, 0.000, 0.000, 0.000, 0.000, 0.000, 0.000, 0.000, 0.000, 0.000, 0.000, 0.000, 0.000, 0.001, 0.001, 0.002, 0.003, 0.004, 0.006, 0.007, 0.009, 0.012, 0.015, 0.018, 0.021, 0.025, 0.029, 0.034, 0.039, 0.044, 0.050, 0.056, 0.062, 0.069, 0.076, 0.084, 0.092, 0.100, 0.109, 0.118, 0.127, 0.137, 0.146, 0.157, 0.167, 0.178, 0.190, 0.201, 0.213, 0.225, 0.237, 0.250, 0.263, 0.276, 0.289, 0.303, 0.316, 0.330, 0.344, 0.358, 0.373, 0.387, 0.402, 0.416, 0.431, 0.446, 0.461, 0.476, 0.491, 0.506, 0.521, 0.536, 0.551, 0.566, 0.581, 0.596, 0.611, 0.626, 0.640, 0.655, 0.669, 0.683, 0.698, 0.712, 0.725, 0.739, 0.752, 0.765, 0.778, 0.791, 0.803, 0.815, 0.827, 0.839, 0.850, 0.861, 0.871, 0.882, 0.891, 0.901, 0.910, 0.919, 0.927, 0.935, 0.943, 0.950, 0.956, 0.963, 0.968, 0.974, 0.979, 0.983, 0.987, 0.990, 0.993, 0.996, 0.997, 0.999, 1.000, 1.000, 1.000, 0.999, 0.998, 0.996, 0.994, 0.991, 0.988, 0.984, 0.979, 0.974, 0.969, 0.963, 0.956, 0.949, 0.941, 0.933, 0.924, 0.915, 0.905, 0.895, 0.884, 0.873, 0.861, 0.849, 0.836, 0.823, 0.809, 0.795, 0.781, 0.766, 0.751, 0.735, 0.719, 0.702, 0.686, 0.669, 0.651, 0.633, 0.615, 0.597, 0.579, 0.560, 0.541, 0.522, 0.503, 0.483, 0.464, 0.445, 0.425, 0.405, 0.386, 0.366, 0.347, 0.328, 0.308, 0.289, 0.271, 0.252, 0.234, 0.216, 0.198, 0.181, 0.164, 0.148, 0.133, 0.117, 0.103, 0.089, 0.076, 0.064, 0.053, 0.042, 0.033, 0.024, 0.017, 0.011, 0.006, 0.002, 0.000, 0.000, 0.000, 0.000, 0.000, 0.000, 0.000, 0.000, 0.000, 0.000, 0.000, 0.000} {
    -- ({2.2 * \sx * \n / 216 - 1.1 * \sx}, {1.4 * \sy * \y})
  };

  \colorlet{custom}{dark!20!white};
  \draw[custom, line width=1] (-1.1 * \sx, 0) 
  \foreach \y [count=\n] in {0.000, 0.000, 0.000, 0.000, 0.000, 0.000, 0.000, 0.000, 0.000, 0.000, 0.000, 0.000, 0.000, 0.000, 0.000, 0.000, 0.000, 0.000, 0.000, 0.000, 0.000, 0.000, 0.000, 0.000, 0.000, 0.000, 0.000, 0.000, 0.000, 0.000, 0.000, 0.000, 0.000, 0.001, 0.001, 0.002, 0.004, 0.006, 0.008, 0.011, 0.014, 0.017, 0.021, 0.026, 0.031, 0.036, 0.042, 0.048, 0.055, 0.063, 0.070, 0.079, 0.088, 0.097, 0.107, 0.117, 0.127, 0.139, 0.150, 0.162, 0.174, 0.187, 0.200, 0.213, 0.227, 0.241, 0.255, 0.270, 0.285, 0.300, 0.316, 0.331, 0.347, 0.363, 0.379, 0.396, 0.412, 0.429, 0.446, 0.463, 0.479, 0.496, 0.513, 0.530, 0.547, 0.564, 0.581, 0.598, 0.614, 0.631, 0.647, 0.663, 0.680, 0.695, 0.711, 0.727, 0.742, 0.757, 0.771, 0.786, 0.800, 0.813, 0.827, 0.840, 0.852, 0.864, 0.876, 0.887, 0.898, 0.909, 0.918, 0.928, 0.937, 0.945, 0.953, 0.960, 0.967, 0.973, 0.978, 0.983, 0.988, 0.991, 0.994, 0.997, 0.998, 1.000, 1.000, 1.000, 0.999, 0.997, 0.995, 0.992, 0.988, 0.984, 0.979, 0.973, 0.967, 0.960, 0.952, 0.944, 0.934, 0.925, 0.914, 0.903, 0.891, 0.879, 0.865, 0.852, 0.837, 0.822, 0.807, 0.790, 0.774, 0.756, 0.738, 0.720, 0.701, 0.682, 0.662, 0.642, 0.621, 0.600, 0.579, 0.557, 0.535, 0.513, 0.490, 0.468, 0.445, 0.422, 0.399, 0.376, 0.354, 0.331, 0.308, 0.286, 0.264, 0.242, 0.220, 0.199, 0.179, 0.159, 0.139, 0.121, 0.103, 0.086, 0.070, 0.056, 0.042, 0.031, 0.020, 0.012, 0.005, 0.001, 0.000, 0.000, 0.000, 0.000, 0.000, 0.000, 0.000, 0.000, 0.000, 0.000, 0.000, 0.000, 0.000, 0.000, 0.000, 0.000, 0.000, 0.000, 0.000, 0.000, 0.000, 0.000} {
    -- ({2.2 * \sx * \n / 216 - 1.1 * \sx}, {1.4 * \sy * \y})
  };  
  
  \colorlet{custom}{dark!30!white};
  \draw[custom, line width=1] (-1.1 * \sx, 0) 
  \foreach \y [count=\n] in {0.000, 0.000, 0.000, 0.000, 0.000, 0.000, 0.000, 0.000, 0.000, 0.000, 0.000, 0.000, 0.000, 0.000, 0.000, 0.000, 0.000, 0.000, 0.000, 0.000, 0.000, 0.000, 0.000, 0.000, 0.000, 0.000, 0.000, 0.000, 0.000, 0.000, 0.000, 0.001, 0.002, 0.002, 0.004, 0.005, 0.007, 0.009, 0.012, 0.015, 0.018, 0.022, 0.026, 0.031, 0.036, 0.042, 0.048, 0.055, 0.062, 0.069, 0.077, 0.085, 0.094, 0.104, 0.113, 0.124, 0.134, 0.146, 0.157, 0.169, 0.181, 0.194, 0.207, 0.221, 0.235, 0.249, 0.264, 0.279, 0.294, 0.309, 0.325, 0.341, 0.357, 0.374, 0.390, 0.407, 0.424, 0.441, 0.458, 0.476, 0.493, 0.511, 0.528, 0.545, 0.563, 0.580, 0.598, 0.615, 0.632, 0.649, 0.666, 0.683, 0.699, 0.715, 0.731, 0.747, 0.763, 0.778, 0.793, 0.807, 0.822, 0.835, 0.849, 0.862, 0.874, 0.886, 0.898, 0.908, 0.919, 0.929, 0.938, 0.947, 0.955, 0.962, 0.969, 0.976, 0.981, 0.986, 0.990, 0.994, 0.996, 0.998, 1.000, 1.000, 1.000, 0.999, 0.997, 0.994, 0.991, 0.987, 0.982, 0.976, 0.969, 0.962, 0.953, 0.944, 0.934, 0.924, 0.912, 0.900, 0.887, 0.873, 0.858, 0.843, 0.827, 0.810, 0.793, 0.775, 0.756, 0.736, 0.716, 0.695, 0.674, 0.652, 0.630, 0.607, 0.584, 0.560, 0.536, 0.512, 0.487, 0.463, 0.438, 0.413, 0.387, 0.362, 0.337, 0.312, 0.288, 0.263, 0.239, 0.215, 0.192, 0.170, 0.148, 0.127, 0.107, 0.089, 0.071, 0.055, 0.040, 0.028, 0.017, 0.008, 0.002, 0.000, 0.000, 0.000, 0.000, 0.000, 0.000, 0.000, 0.000, 0.000, 0.000, 0.000, 0.000, 0.000, 0.000, 0.000, 0.000, 0.000, 0.000, 0.000, 0.000, 0.000, 0.000, 0.000, 0.000, 0.000, 0.000, 0.000, 0.000, 0.000, 0.000, 0.000} {
    -- ({2.2 * \sx * \n / 216 - 1.1 * \sx}, {1.4 * \sy * \y})
  };  
  
  \colorlet{custom}{dark!40!white};
  \draw[custom, line width=1] (-1.1 * \sx, 0) 
  \foreach \y [count=\n] in {0.000, 0.000, 0.000, 0.000, 0.000, 0.000, 0.000, 0.000, 0.000, 0.000, 0.000, 0.000, 0.000, 0.000, 0.000, 0.000, 0.000, 0.000, 0.000, 0.000, 0.000, 0.000, 0.000, 0.000, 0.000, 0.000, 0.000, 0.000, 0.000, 0.000, 0.000, 0.000, 0.000, 0.000, 0.000, 0.000, 0.000, 0.001, 0.001, 0.001, 0.002, 0.003, 0.004, 0.005, 0.007, 0.009, 0.011, 0.014, 0.017, 0.020, 0.024, 0.028, 0.032, 0.037, 0.043, 0.049, 0.055, 0.062, 0.069, 0.077, 0.086, 0.095, 0.104, 0.114, 0.125, 0.136, 0.147, 0.160, 0.172, 0.185, 0.199, 0.213, 0.227, 0.242, 0.258, 0.273, 0.290, 0.306, 0.323, 0.341, 0.358, 0.376, 0.394, 0.413, 0.431, 0.450, 0.469, 0.488, 0.508, 0.527, 0.547, 0.566, 0.585, 0.605, 0.624, 0.643, 0.662, 0.681, 0.700, 0.718, 0.736, 0.754, 0.772, 0.789, 0.805, 0.822, 0.837, 0.852, 0.867, 0.881, 0.894, 0.907, 0.919, 0.931, 0.941, 0.951, 0.960, 0.968, 0.975, 0.982, 0.987, 0.992, 0.995, 0.998, 0.999, 1.000, 1.000, 0.998, 0.996, 0.992, 0.987, 0.981, 0.975, 0.967, 0.958, 0.948, 0.936, 0.924, 0.911, 0.896, 0.881, 0.864, 0.847, 0.829, 0.809, 0.789, 0.768, 0.746, 0.723, 0.699, 0.675, 0.650, 0.625, 0.598, 0.572, 0.545, 0.517, 0.490, 0.462, 0.433, 0.405, 0.377, 0.349, 0.322, 0.294, 0.267, 0.241, 0.215, 0.190, 0.166, 0.143, 0.122, 0.101, 0.082, 0.065, 0.049, 0.035, 0.023, 0.014, 0.007, 0.002, 0.000, 0.000, 0.000, 0.000, 0.000, 0.000, 0.000, 0.000, 0.000, 0.000, 0.000, 0.000, 0.000, 0.000, 0.000, 0.000, 0.000, 0.000, 0.000, 0.000, 0.000, 0.000, 0.000, 0.000, 0.000, 0.000, 0.000, 0.000, 0.000, 0.000, 0.000, 0.000, 0.000, 0.000, 0.000} {
    -- ({2.2 * \sx * \n / 216 - 1.1 * \sx}, {1.4 * \sy * \y})
  };
  
  \colorlet{custom}{dark!50!white};
  \draw[custom, line width=1] (-1.1 * \sx, 0) 
  \foreach \y [count=\n] in {0.000, 0.000, 0.000, 0.000, 0.000, 0.000, 0.000, 0.000, 0.000, 0.000, 0.000, 0.000, 0.000, 0.000, 0.000, 0.000, 0.000, 0.000, 0.000, 0.000, 0.000, 0.000, 0.000, 0.000, 0.001, 0.001, 0.001, 0.002, 0.002, 0.003, 0.004, 0.005, 0.006, 0.007, 0.009, 0.010, 0.012, 0.014, 0.017, 0.019, 0.022, 0.025, 0.029, 0.032, 0.036, 0.041, 0.045, 0.050, 0.055, 0.061, 0.067, 0.073, 0.080, 0.087, 0.094, 0.102, 0.110, 0.119, 0.128, 0.137, 0.147, 0.157, 0.167, 0.178, 0.189, 0.200, 0.212, 0.224, 0.237, 0.250, 0.263, 0.276, 0.290, 0.304, 0.319, 0.333, 0.348, 0.363, 0.379, 0.394, 0.410, 0.426, 0.443, 0.459, 0.475, 0.492, 0.509, 0.525, 0.542, 0.559, 0.576, 0.593, 0.610, 0.627, 0.643, 0.660, 0.677, 0.693, 0.709, 0.725, 0.741, 0.757, 0.772, 0.787, 0.802, 0.816, 0.830, 0.844, 0.857, 0.870, 0.882, 0.894, 0.905, 0.916, 0.927, 0.936, 0.945, 0.954, 0.962, 0.969, 0.975, 0.981, 0.986, 0.990, 0.994, 0.997, 0.999, 1.000, 1.000, 0.999, 0.998, 0.996, 0.993, 0.989, 0.984, 0.978, 0.972, 0.964, 0.956, 0.946, 0.936, 0.925, 0.913, 0.900, 0.886, 0.871, 0.856, 0.840, 0.822, 0.804, 0.786, 0.766, 0.746, 0.725, 0.703, 0.681, 0.658, 0.635, 0.611, 0.587, 0.562, 0.537, 0.511, 0.486, 0.460, 0.434, 0.408, 0.381, 0.355, 0.330, 0.304, 0.279, 0.254, 0.229, 0.205, 0.182, 0.160, 0.139, 0.118, 0.099, 0.081, 0.065, 0.050, 0.037, 0.025, 0.016, 0.008, 0.003, 0.000, 0.000, 0.000, 0.000, 0.000, 0.000, 0.000, 0.000, 0.000, 0.000, 0.000, 0.000, 0.000, 0.000, 0.000, 0.000, 0.000, 0.000, 0.000, 0.000, 0.000, 0.000, 0.000, 0.000, 0.000, 0.000, 0.000, 0.000} {
    -- ({2.2 * \sx * \n / 216 - 1.1 * \sx}, {1.4 * \sy * \y})
  };

  \colorlet{custom}{dark!60!white};
  \draw[custom, line width=1] (-1.1 * \sx, 0) 
  \foreach \y [count=\n] in {0.000, 0.000, 0.000, 0.000, 0.000, 0.000, 0.000, 0.000, 0.000, 0.000, 0.000, 0.000, 0.000, 0.000, 0.000, 0.000, 0.000, 0.000, 0.000, 0.001, 0.001, 0.001, 0.002, 0.003, 0.004, 0.005, 0.007, 0.008, 0.010, 0.012, 0.015, 0.017, 0.021, 0.024, 0.028, 0.032, 0.036, 0.041, 0.046, 0.052, 0.058, 0.064, 0.071, 0.078, 0.085, 0.093, 0.101, 0.110, 0.119, 0.128, 0.138, 0.148, 0.159, 0.170, 0.181, 0.193, 0.205, 0.218, 0.230, 0.243, 0.257, 0.270, 0.284, 0.298, 0.313, 0.328, 0.343, 0.358, 0.373, 0.389, 0.405, 0.420, 0.436, 0.453, 0.469, 0.485, 0.502, 0.518, 0.534, 0.551, 0.567, 0.584, 0.600, 0.616, 0.633, 0.649, 0.665, 0.681, 0.696, 0.712, 0.727, 0.742, 0.757, 0.771, 0.785, 0.799, 0.813, 0.826, 0.839, 0.852, 0.864, 0.875, 0.887, 0.897, 0.908, 0.918, 0.927, 0.936, 0.944, 0.952, 0.959, 0.966, 0.972, 0.978, 0.983, 0.987, 0.991, 0.994, 0.996, 0.998, 0.999, 1.000, 1.000, 0.999, 0.998, 0.996, 0.993, 0.990, 0.986, 0.981, 0.975, 0.969, 0.963, 0.955, 0.948, 0.939, 0.930, 0.920, 0.909, 0.898, 0.887, 0.875, 0.862, 0.849, 0.835, 0.820, 0.806, 0.790, 0.775, 0.758, 0.742, 0.725, 0.707, 0.689, 0.671, 0.653, 0.634, 0.615, 0.596, 0.577, 0.557, 0.538, 0.518, 0.498, 0.478, 0.459, 0.439, 0.419, 0.399, 0.380, 0.360, 0.341, 0.322, 0.303, 0.285, 0.267, 0.249, 0.231, 0.214, 0.198, 0.182, 0.166, 0.151, 0.137, 0.123, 0.110, 0.098, 0.086, 0.075, 0.065, 0.055, 0.046, 0.038, 0.031, 0.025, 0.019, 0.014, 0.010, 0.007, 0.004, 0.002, 0.001, 0.000, 0.000, 0.000, 0.000, 0.000, 0.000, 0.000, 0.000, 0.000, 0.000, 0.000, 0.000, 0.000, 0.000} {
    -- ({2.2 * \sx * \n / 216 - 1.1 * \sx}, {1.4 * \sy * \y})
  };

  \colorlet{custom}{dark!70!white};
  \draw[custom, line width=1] (-1.1 * \sx, 0) 
  \foreach \y [count=\n] in {0.000, 0.000, 0.000, 0.000, 0.000, 0.000, 0.000, 0.000, 0.000, 0.000, 0.000, 0.000, 0.000, 0.000, 0.000, 0.000, 0.000, 0.000, 0.000, 0.000, 0.000, 0.000, 0.000, 0.000, 0.000, 0.000, 0.000, 0.000, 0.000, 0.000, 0.000, 0.000, 0.000, 0.000, 0.000, 0.000, 0.000, 0.000, 0.000, 0.000, 0.001, 0.002, 0.004, 0.006, 0.008, 0.011, 0.015, 0.019, 0.024, 0.030, 0.036, 0.043, 0.050, 0.059, 0.067, 0.077, 0.087, 0.097, 0.109, 0.120, 0.133, 0.146, 0.159, 0.173, 0.187, 0.202, 0.217, 0.233, 0.249, 0.266, 0.283, 0.300, 0.317, 0.335, 0.353, 0.371, 0.389, 0.408, 0.427, 0.445, 0.464, 0.483, 0.502, 0.521, 0.540, 0.558, 0.577, 0.596, 0.614, 0.633, 0.651, 0.669, 0.686, 0.704, 0.721, 0.737, 0.754, 0.770, 0.786, 0.801, 0.816, 0.830, 0.844, 0.858, 0.870, 0.883, 0.895, 0.906, 0.917, 0.927, 0.936, 0.945, 0.954, 0.961, 0.968, 0.975, 0.980, 0.985, 0.989, 0.993, 0.996, 0.998, 0.999, 1.000, 1.000, 0.999, 0.998, 0.995, 0.992, 0.989, 0.984, 0.979, 0.974, 0.967, 0.960, 0.952, 0.943, 0.934, 0.924, 0.914, 0.902, 0.890, 0.878, 0.865, 0.851, 0.837, 0.822, 0.807, 0.791, 0.775, 0.759, 0.741, 0.724, 0.706, 0.688, 0.669, 0.650, 0.631, 0.611, 0.592, 0.572, 0.552, 0.531, 0.511, 0.491, 0.470, 0.450, 0.430, 0.409, 0.389, 0.369, 0.349, 0.329, 0.310, 0.291, 0.272, 0.253, 0.235, 0.218, 0.200, 0.184, 0.168, 0.152, 0.137, 0.123, 0.109, 0.096, 0.084, 0.072, 0.062, 0.052, 0.043, 0.034, 0.027, 0.021, 0.015, 0.011, 0.007, 0.004, 0.002, 0.001, 0.000, 0.000, 0.000, 0.000, 0.000, 0.000, 0.000, 0.000, 0.000, 0.000, 0.000, 0.000, 0.000, 0.000, 0.000} {
    -- ({2.2 * \sx * \n / 216 - 1.1 * \sx}, {1.4 * \sy * \y})
  };  
  
  \colorlet{custom}{dark!80!white};
  \draw[custom, line width=1] (-1.1 * \sx, 0) 
  \foreach \y [count=\n] in {0.000, 0.000, 0.000, 0.000, 0.000, 0.000, 0.000, 0.000, 0.000, 0.000, 0.000, 0.000, 0.000, 0.000, 0.000, 0.001, 0.001, 0.002, 0.003, 0.004, 0.005, 0.006, 0.008, 0.010, 0.012, 0.015, 0.018, 0.021, 0.024, 0.028, 0.032, 0.036, 0.041, 0.046, 0.051, 0.057, 0.063, 0.069, 0.076, 0.082, 0.090, 0.097, 0.105, 0.113, 0.121, 0.130, 0.139, 0.148, 0.158, 0.168, 0.178, 0.188, 0.199, 0.210, 0.221, 0.233, 0.244, 0.256, 0.268, 0.280, 0.293, 0.305, 0.318, 0.331, 0.344, 0.358, 0.371, 0.385, 0.399, 0.412, 0.426, 0.440, 0.454, 0.468, 0.483, 0.497, 0.511, 0.525, 0.540, 0.554, 0.568, 0.583, 0.597, 0.611, 0.625, 0.639, 0.653, 0.667, 0.681, 0.694, 0.708, 0.721, 0.734, 0.747, 0.760, 0.772, 0.785, 0.797, 0.809, 0.820, 0.832, 0.843, 0.854, 0.864, 0.874, 0.884, 0.894, 0.903, 0.912, 0.921, 0.929, 0.936, 0.944, 0.951, 0.957, 0.963, 0.969, 0.974, 0.979, 0.983, 0.987, 0.991, 0.993, 0.996, 0.998, 0.999, 1.000, 1.000, 1.000, 0.999, 0.998, 0.996, 0.993, 0.990, 0.987, 0.983, 0.978, 0.973, 0.967, 0.960, 0.953, 0.946, 0.937, 0.929, 0.919, 0.909, 0.899, 0.888, 0.876, 0.864, 0.851, 0.837, 0.823, 0.809, 0.794, 0.778, 0.762, 0.746, 0.729, 0.711, 0.693, 0.674, 0.656, 0.636, 0.616, 0.596, 0.576, 0.555, 0.534, 0.512, 0.490, 0.468, 0.446, 0.424, 0.401, 0.378, 0.356, 0.333, 0.310, 0.287, 0.265, 0.243, 0.221, 0.199, 0.178, 0.157, 0.136, 0.117, 0.098, 0.080, 0.063, 0.048, 0.033, 0.021, 0.011, 0.003, 0.000, 0.000, 0.000, 0.000, 0.000, 0.000, 0.000, 0.000, 0.000, 0.000, 0.000, 0.000, 0.000, 0.000, 0.000, 0.000, 0.000, 0.000, 0.000, 0.000} {
    -- ({2.2 * \sx * \n / 216 - 1.1 * \sx}, {1.4 * \sy * \y})
  };  
  
  \colorlet{custom}{dark!90!white};
  \draw[custom, line width=1] (-1.1 * \sx, 0) 
  \foreach \y [count=\n] in {0.000, 0.000, 0.000, 0.000, 0.000, 0.000, 0.000, 0.000, 0.000, 0.000, 0.000, 0.000, 0.000, 0.000, 0.000, 0.000, 0.000, 0.000, 0.000, 0.000, 0.000, 0.000, 0.000, 0.000, 0.000, 0.000, 0.000, 0.000, 0.000, 0.000, 0.000, 0.000, 0.001, 0.001, 0.002, 0.003, 0.004, 0.006, 0.008, 0.010, 0.012, 0.015, 0.018, 0.022, 0.026, 0.031, 0.036, 0.041, 0.047, 0.053, 0.060, 0.067, 0.075, 0.083, 0.092, 0.101, 0.111, 0.121, 0.132, 0.143, 0.155, 0.167, 0.179, 0.192, 0.205, 0.219, 0.233, 0.248, 0.263, 0.278, 0.293, 0.309, 0.325, 0.342, 0.358, 0.375, 0.392, 0.410, 0.427, 0.445, 0.463, 0.481, 0.499, 0.517, 0.535, 0.553, 0.571, 0.589, 0.606, 0.624, 0.642, 0.659, 0.677, 0.694, 0.711, 0.727, 0.744, 0.760, 0.776, 0.791, 0.806, 0.821, 0.835, 0.849, 0.862, 0.875, 0.887, 0.899, 0.910, 0.921, 0.931, 0.940, 0.949, 0.957, 0.965, 0.972, 0.978, 0.983, 0.988, 0.992, 0.995, 0.997, 0.999, 1.000, 1.000, 0.999, 0.998, 0.996, 0.992, 0.989, 0.984, 0.978, 0.972, 0.965, 0.957, 0.948, 0.939, 0.929, 0.918, 0.906, 0.893, 0.880, 0.866, 0.852, 0.836, 0.820, 0.804, 0.787, 0.769, 0.751, 0.732, 0.712, 0.692, 0.672, 0.651, 0.630, 0.609, 0.587, 0.565, 0.543, 0.520, 0.498, 0.475, 0.452, 0.430, 0.407, 0.384, 0.362, 0.340, 0.317, 0.296, 0.274, 0.253, 0.233, 0.213, 0.193, 0.175, 0.156, 0.139, 0.123, 0.107, 0.092, 0.078, 0.065, 0.053, 0.043, 0.033, 0.025, 0.018, 0.012, 0.007, 0.004, 0.001, 0.000, 0.000, 0.000, 0.000, 0.000, 0.000, 0.000, 0.000, 0.000, 0.000, 0.000, 0.000, 0.000, 0.000, 0.000, 0.000, 0.000, 0.000, 0.000, 0.000, 0.000, 0.000, 0.000} {
    -- ({2.2 * \sx * \n / 216 - 1.1 * \sx}, {1.4 * \sy * \y})
  };
  
  \colorlet{custom}{dark!100!white};
  \draw[custom, line width=1] (-1.1 * \sx, 0) 
  \foreach \y [count=\n] in {0.000, 0.000, 0.000, 0.000, 0.000, 0.000, 0.000, 0.000, 0.000, 0.000, 0.000, 0.000, 0.000, 0.000, 0.000, 0.000, 0.000, 0.000, 0.000, 0.000, 0.000, 0.000, 0.000, 0.000, 0.000, 0.000, 0.000, 0.000, 0.000, 0.000, 0.000, 0.000, 0.001, 0.002, 0.004, 0.007, 0.011, 0.015, 0.020, 0.026, 0.032, 0.039, 0.047, 0.055, 0.064, 0.073, 0.083, 0.093, 0.104, 0.115, 0.127, 0.139, 0.152, 0.165, 0.178, 0.192, 0.205, 0.220, 0.234, 0.249, 0.264, 0.280, 0.295, 0.311, 0.327, 0.343, 0.359, 0.375, 0.392, 0.408, 0.425, 0.441, 0.458, 0.475, 0.491, 0.508, 0.524, 0.541, 0.557, 0.574, 0.590, 0.606, 0.622, 0.638, 0.653, 0.669, 0.684, 0.699, 0.714, 0.729, 0.743, 0.757, 0.771, 0.784, 0.798, 0.811, 0.823, 0.835, 0.847, 0.859, 0.870, 0.881, 0.891, 0.901, 0.910, 0.919, 0.928, 0.936, 0.944, 0.951, 0.958, 0.964, 0.970, 0.976, 0.980, 0.985, 0.988, 0.992, 0.994, 0.997, 0.998, 0.999, 1.000, 1.000, 0.999, 0.998, 0.997, 0.994, 0.991, 0.988, 0.984, 0.979, 0.974, 0.969, 0.962, 0.955, 0.948, 0.940, 0.931, 0.922, 0.913, 0.902, 0.891, 0.880, 0.868, 0.856, 0.842, 0.829, 0.815, 0.800, 0.785, 0.770, 0.753, 0.737, 0.720, 0.702, 0.685, 0.666, 0.647, 0.628, 0.609, 0.589, 0.569, 0.548, 0.528, 0.507, 0.485, 0.464, 0.442, 0.420, 0.399, 0.376, 0.354, 0.332, 0.310, 0.288, 0.266, 0.245, 0.223, 0.202, 0.181, 0.161, 0.141, 0.121, 0.103, 0.085, 0.068, 0.052, 0.037, 0.024, 0.013, 0.005, 0.000, 0.000, 0.000, 0.000, 0.000, 0.000, 0.000, 0.000, 0.000, 0.000, 0.000, 0.000, 0.000, 0.000, 0.000, 0.000, 0.000, 0.000, 0.000, 0.000, 0.000, 0.000, 0.000, 0.000} {
    -- ({2.2 * \sx * \n / 216 - 1.1 * \sx}, {1.4 * \sy * \y})
  };
  
  \draw [<->, >=stealth, line width=1] (-1.1 * \sx, 1.5 * \sy) -- (-1.1 * \sx, 0) -- (1.1 * \sx, 0);
  \node at (0, -0.5) { Temperature (Arbitrary Units) };
  \node[rotate=90] at (-1.4 * \sx, 0.75 * \sy) { Forcing Function };
  \node at (-1.2 * \sx, 0) { $0$ };
  \node at (-1.2 * \sx, 1.4 * \sy) { $1$ };
  
  \node at (0, 1.7 * \sy) { Modified Survival Model };

\end{tikzpicture}
} 
\caption{The (a) the standard survival model and (b) modified survival model with 
a soft thresholding warping function yield substantially different inferences for
the forcing function behavior.  Without any thresholding the standard survival model 
needs to squeeze the forcing function in order to limit the the growth of the 
cumulative hazard function and hence the decay of the survival model.  On the other 
hand the warping function of the modified survival model accounts for this behavior 
directly, allowing the posterior inferences to concentrate on more reasonable 
behaviors.}
\label{fig:inferred_forcing} 
\end{figure}
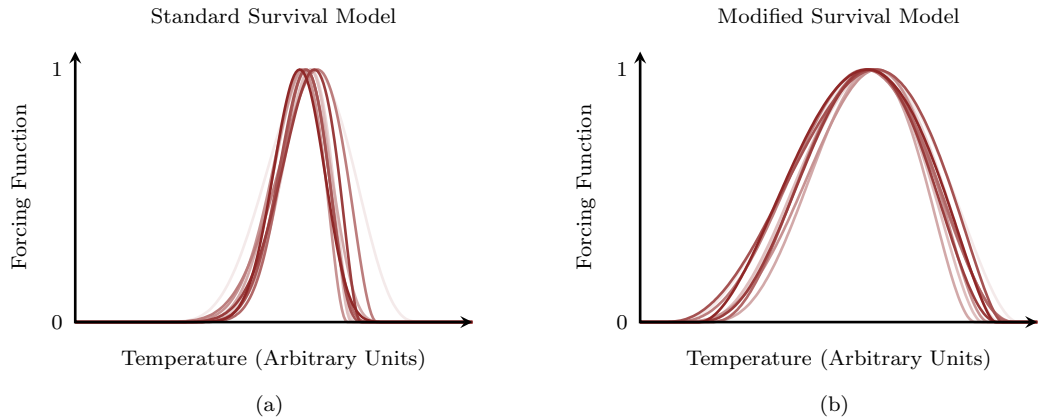

In order to accommodate the observed veraison data as well as possible the standard
survival model has to slow the decay of the survival function.  To do this the 
posterior distribution squeezes the forcing function as much is allowed by the 
constraints of the prior model so that the hotter spring and summer temperatures 
contribute relatively negligible forcings (Figure \ref{fig:inferred_forcing}).  
Without having to maintain this contortion the inferred forcing function behavior for 
the modified survival model concentrates on behaviors consistent with previous analyses
\citep{WolkovichEtAl:2022}.  An adequate model is critical to recovering meaningful
forcing function inferences.

\section{Conclusion}

When the instantaneous hazard function models an explicit, physical phenomena it may 
not be flexible enough for the resulting survival model to capture complex behaviors 
such as thresholding.  Those behaviors, however, can often be accommodated by modifying 
the standard survival model construction with a warping function that moderates the 
precise relationship between the cumulative hazard function and the final survival 
function.

So long as the derivative of this warping function can be readily evaluated then such 
modified survival models will be as straightforward, or difficult, to implement as the 
standard survival model.  Consequently modified survival models are a straightforward 
way to expand the utility of survival models in practice.

\section{Acknowledgements}

I thank Elizabeth Wolkovich for welcoming me into the collaboration that motivated
this work, wrangling of the data, invaluable domain expertise, and helpful comments 
on this manuscript.  Additionally I am grateful to I\~{n}aki Garc\'{i}a de Cort\'{a}zar-Atauri 
for generously sharing his domain expertise and helpful comments on this manuscript, 
Faith Jones for motivating the use of thresholding behavior, and Geoff Legault for 
helpful discussions.

\setcounter{section}{0}
\renewcommand{\thesection}{\Alph{section}}

\section{Appendix}

This appendix covers two topics that go beyond the nominal scope of the manuscript.
First I will show how modified survival functions can be derived from probability 
density functions on the cumulative hazard function.  Second I will go into more 
detail about the forcing functions used in the phenology analysis.

\subsection{Implicit Modified Survival Models} \label{sec:implicit_modified}

In Section \ref{sec:threshold_survival} I showed that a modified survival model with 
a soft thresholding warping function resulted in an event probability density function 
that simplified to a logistic probability density function on the accumulated 
hazard function,
\begin{align*}
\pi(t)
&=
\lambda (t) \, 
\frac{ \mathrm{d} g }{ \mathrm{d} \Lambda} \left( \Lambda(t) \right) \,
\exp \left( - g \left( \Lambda(t) \right) \right)
\\
&=
\lambda (t) \,
\frac{1}{\alpha}  \,
\frac{\exp \left( \frac{ \Lambda(t) - \Lambda_{0} }{ \alpha } \right)}
{ \left(  1 + \exp \left( \frac{ \Lambda(t) - \Lambda_{0} }{ \alpha } \right) \right)^{2} }
\\
&=
\lambda (t) \, \mathrm{logistic} (\Lambda(t); \Lambda_{0}, \alpha ).
\end{align*}

Conveniently we can also reverse this construction, transforming a probability
density function over the cumulative hazard into a modified survival model for
a particular warping function.  More formally any probability density function 
over cumulative hazards $\omega( \Lambda )$ induces the event probability density 
function
\begin{equation*}
\pi(t) 
= \frac{ \mathrm{d} \Lambda }{ \mathrm{d} t }(t) \, \omega(\Lambda(t))
= \lambda(t) \, \omega(\Lambda(t)).
\end{equation*}

In turn this event probability density function defines the survival function
\begin{align*}
S(t)
&=
\int_{t}^{\infty} \mathrm{d} t \, \pi(t)
\\
&=
\int_{t}^{\infty} \mathrm{d} t \, \lambda(t) \, \omega(\Lambda(t)).
\end{align*}
Under the change of variables $l = \Lambda(t)$ this becomes
\begin{align*}
S(t)
&=
\int_{t}^{\infty} \mathrm{d} t \, \lambda(t) \, \omega(\Lambda(t)) 
\\
&=
\int_{\Lambda(t)}^{\infty} \mathrm{d} l \, \omega(l)
\\
&=
\Omega_{c}(\Lambda(t)),
\end{align*}
where $\Omega_{c}$ is the complementary cumulative distribution function 
corresponding to the probability density function $\omega$.

At this point we can apply the natural logarithm and exponential functions to 
give
\begin{align*}
S(t)
&=
\Omega_{c}(\Lambda(t))
\\
&=
\exp \big( \log \big( \Omega_{c} \big( \Lambda(t) \big) \big) \big)
\\
&=
\exp \big( \log \circ \, \Omega_{c} \big( \Lambda(t) \big) \big)
\\
&=
\exp \big( - g\big( \Lambda(t) \big) \big)
\end{align*}
with 
\begin{equation*}
g(l) = - \log \circ \, \Omega_{c} (l).
\end{equation*}

Now all complementary cumulative distribution functions are by definition 
monotonically non-increasing.  Moreover the natural logarithm is monotonically 
increasing.  Consequently the composition $g = - \log \circ \, \Omega_{c}$ 
is always monotonically non-decreasing and defines a valid warping function.
In other words the cumulative hazard model $\omega(\Lambda(t))$ implicitly 
defines a modified survival model for the event time $t$.

\begin{figure}
\centering
\begin{tikzpicture}[scale=1.0]

  \pgfmathsetmacro{\sx}{6 / 18}
  \pgfmathsetmacro{\sy}{4 / 35}
  
  \draw[white] (-4.25, -3) rectangle (4.25, 3);
  
  \draw[gray80, line width=1, dashed] (\sx * 10 - 3, -2) -- +(0, 4);
  \node[gray80] at (\sx * 10 - 3, 2.15) { $\Lambda_{0}$ };
  
  \draw[dark, line width=2] (-3, -2) 
  \foreach \x/\y in {0.000/-0.000, 0.100/-0.000, 0.200/-0.000, 0.300/-0.000, 0.400/-0.000, 0.500/-0.000, 0.600/-0.000, 0.700/-0.000, 0.800/-0.000, 0.900/-0.000, 1.000/-0.000, 1.100/-0.000, 1.200/-0.000, 1.300/-0.000, 1.400/-0.000, 1.500/-0.000, 1.600/-0.000, 1.700/-0.000, 1.800/0.000, 1.900/0.000, 2.000/0.000, 2.100/0.000, 2.200/0.000, 2.300/0.000, 2.400/0.000, 2.500/0.000, 2.600/0.000, 2.700/0.000, 2.800/0.000, 2.900/0.000, 3.000/0.000, 3.100/0.000, 3.200/0.000, 3.300/0.000, 3.400/0.000, 3.500/0.000, 3.600/0.000, 3.700/0.000, 3.800/0.000, 3.900/0.000, 4.000/0.000, 4.100/0.000, 4.200/0.000, 4.300/0.000, 4.400/0.000, 4.500/0.000, 4.600/0.000, 4.700/0.000, 4.800/0.000, 4.900/0.000, 5.000/0.000, 5.100/0.000, 5.200/0.000, 5.300/0.000, 5.400/0.000, 5.500/0.000, 5.600/0.000, 5.700/0.000, 5.800/0.000, 5.900/0.000, 6.000/0.000, 6.100/0.000, 6.200/0.000, 6.300/0.000, 6.400/0.000, 6.500/0.000, 6.600/0.000, 6.700/0.000, 6.800/0.001, 6.900/0.001, 7.000/0.001, 7.100/0.002, 7.200/0.003, 7.300/0.003, 7.400/0.005, 7.500/0.006, 7.600/0.008, 7.700/0.011, 7.800/0.014, 7.900/0.018, 8.000/0.023, 8.100/0.029, 8.200/0.037, 8.300/0.046, 8.400/0.056, 8.500/0.069, 8.600/0.084, 8.700/0.102, 8.800/0.122, 8.900/0.146, 9.000/0.173, 9.100/0.203, 9.200/0.238, 9.300/0.277, 9.400/0.321, 9.500/0.369, 9.600/0.422, 9.700/0.481, 9.800/0.546, 9.900/0.617, 10.000/0.693, 10.100/0.776, 10.200/0.866, 10.300/0.962, 10.400/1.065, 10.500/1.176, 10.600/1.294, 10.700/1.419, 10.800/1.552, 10.900/1.692, 11.000/1.841, 11.100/1.998, 11.200/2.162, 11.300/2.335, 11.400/2.516, 11.500/2.706, 11.600/2.904, 11.700/3.111, 11.800/3.326, 11.900/3.550, 12.000/3.783, 12.100/4.025, 12.200/4.276, 12.300/4.535, 12.400/4.804, 12.500/5.082, 12.600/5.368, 12.700/5.664, 12.800/5.970, 12.900/6.284, 13.000/6.608, 13.100/6.941, 13.200/7.283, 13.300/7.635, 13.400/7.996, 13.500/8.366, 13.600/8.746, 13.700/9.135, 13.800/9.534, 13.900/9.942, 14.000/10.360, 14.100/10.787, 14.200/11.224, 14.300/11.671, 14.400/12.127, 14.500/12.592, 14.600/13.068, 14.700/13.553, 14.800/14.047, 14.900/14.551, 15.000/15.065, 15.100/15.588, 15.200/16.122, 15.300/16.665, 15.400/17.217, 15.500/17.779, 15.600/18.351, 15.700/18.933, 15.800/19.525, 15.900/20.126, 16.000/20.737, 16.100/21.357, 16.200/21.988, 16.300/22.628, 16.400/23.278, 16.500/23.938, 16.600/24.608, 16.700/25.287, 16.800/25.976, 16.900/26.675, 17.000/27.384, 17.100/28.103, 17.200/28.831, 17.300/29.570, 17.400/30.318, 17.500/31.077, 17.600/31.846, 17.700/32.626, 17.800/33.405, 17.900/34.172, 18.000/34.945} {
    -- ({\sx * \x - 3}, {\sy * \y - 2})
  };
 
  \draw [<->, >=stealth, line width=1] (-3, 2) -- (-3, -2) -- (3, -2);
  \node at (0, -2.5) { $\Lambda$ };
  \node[rotate=90] at (-3.5, 0) { Implied Warping Function };

\end{tikzpicture}
\caption{Modeling the cumulative hazard function with a normal probability 
density function results in a modified survival model for the event times.  
The implied warping function exhibits thresholding behavior similar to the 
warping function introduced in Section \ref{sec:threshold_survival}.  Indeed 
most unimodal models for the cumulative hazard function will result in similar 
thresholding behavior.}
\label{fig:implied_warping} 
\end{figure}
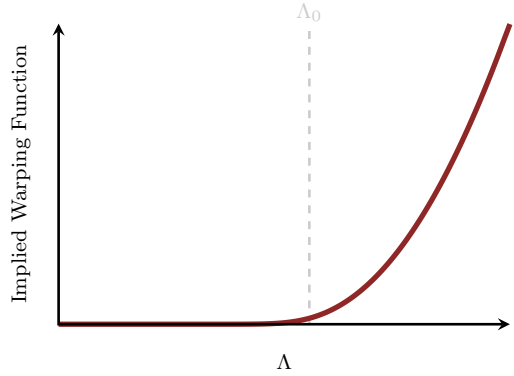

For example taking $\omega(\Lambda(t)) = \text{normal}(\Lambda(t); \Psi_{0}, \alpha)$ 
results in a modified survival model with the warping function 
(Figure \ref{fig:implied_warping})
\begin{align*}
g(l) 
&= 
- \log \big( 1 - \Phi(l; \Psi_{0}, \alpha) \big)
\\
&=
- \log \left( \frac{1}{2} 
  \left(1 - \mathrm{erf} \left( \frac{l - \Psi_{0}}{ \sqrt{2} \, \alpha } \right) \right) 
  \right)
\\
&=
\log 2 
- \log \left( 
  1 - \mathrm{erf} \left( \frac{l - \Psi_{0}}{ \sqrt{2} \, \alpha } \right) 
\right).
\end{align*}

Note that up to a factor of $\lambda(t)$ the cumulative hazard model 
$\omega(\Lambda(t)) = \text{normal}(\Lambda(t); \Psi_{0}, \alpha)$ is 
equivalent to the conventional phenology model introduced in Section 
\ref{sec:conventional_pheno}.  If the temperature forcings, and hence 
$\lambda(t)$, do not strongly vary from day to day then this missing 
factor will be negligible and we can interpret the conventional model 
as an approximation of this particular modified survival model.

This reserved construction is particularly useful when retrodictive checks 
suggest that an initial event model is inadequate.  For example heavy-tailed 
or even asymmetric event times can be accommodated by assuming heavy-tailed or 
asymmetric probability density functions directly on the cumulative hazard 
function without having to engineer an appropriate warping function.

That said we have to be careful with identifiability problems.  If we model 
the cumulative hazard with any location-scale family of probability density 
functions satisfying
\begin{equation*}
\omega( \gamma \cdot l \mid \Lambda_{0}, \tau, \eta) 
= 
\gamma^{-1} \cdot \omega( l \mid \gamma^{-1} \cdot \Lambda_{0}, \gamma^{-1} \cdot \tau, \eta)
\end{equation*}
then we will always have a structural degeneracy in the resulting event
probability density function,
\begin{align*}
\pi(t; \gamma, \Lambda_{0}, \tau, \eta) 
&= 
\gamma \cdot \lambda(t) \, \omega(\gamma \cdot \Lambda \mid \Lambda_{0}, \tau, \eta)
\\
\pi(t; \gamma, \Lambda_{0}, \tau, \eta) 
&=
\psi(t) \, \omega(\Psi(t) \mid \gamma^{-1} \cdot \Lambda_{0}, \gamma^{-1} \tau, \eta)
\\
\pi(t; \gamma, \Lambda_{0}, \tau, \eta) 
&=
\pi(t; 1, \gamma^{-1} \cdot \Lambda_{0}, \gamma^{-1} \cdot \tau, \eta).
\end{align*}

\subsection{An Everywhere-Differentiable Forcing Model} \label{sec:diff_forcing}

In this section I first present the mathematical properties of the Wang-Engel
forcing functions, both good and bad, before constructing a new family of forcing 
functions with enough flexibility to ensure differentiability at the temperature 
boundaries.

\subsubsection{Investigating The Wang-Engel Forcing Functions}

For $T_{\min} < T, T_{\mathrm{opt}} < T_{\max}$ the Wang-Engel function model is defined 
by the parametric family of functions
\begin{align*}
f(T) 
&=
2 \, \left( \frac{ T - T_{\min} }{ T_{\mathrm{opt}} - T_{\min} } \right)^{a}
- \left( \frac{ T - T_{\min} }{ T_{\mathrm{opt}} - T_{\min} } \right)^{2 \, a}
\\
&=
\left( \frac{ T - T_{\min} }{ T_{\mathrm{opt}} - T_{\min} } \right)^{a} \,
\left[ 2 - \left( \frac{ T - T_{\min} }{ T_{\mathrm{opt}} - T_{\min} } \right)^{a} \right]
\end{align*}
where
\begin{equation*}
a = \frac{ \log 2 }{ \log \frac{ T_{\max} - T_{\min} }{ T_{\mathrm{opt}} - T_{\min} } }.
\end{equation*}
For $T < T_{\min}$ and $T > T_{\max}$ the forcing functions are set to zero so that for any 
input temperature we have
\begin{equation*}
f(T) =
\left\{
\begin{array}{rr}
0, & T < T_{\min} \\
\left( \frac{ T - T_{\min} }{ T_{\mathrm{opt}} - T_{\min} } \right)^{a} \,
\left[ 2 - \left( \frac{ T - T_{\min} }{ T_{\mathrm{opt}} - T_{\min} } \right)^{a} \right]
& T_{\min} \le T \le T_{\max} \\
0, & T > T_{\max}
\end{array}
\right.
\end{equation*}

To simplify the analysis of these functions let's transform the input temperature into 
the unitless variable
\begin{equation*}
z = \frac{ T - T_{\min} }{ T_{\mathrm{opt}} - T_{\min} }.
\end{equation*}
With this new variable the forcing functions become
\begin{equation*}
f(z) =
\left\{
\begin{array}{rr}
0, & z < 0 \\
z^{a} \, \left( 2 - z^{a} \right)
& 0 \le z \le z_{\max} \\
0, & z > z_{\max}
\end{array}
\right.
\end{equation*}
where
\begin{equation*}
a = \frac{ \log 2 }{ \log z_{\max} }
\end{equation*}
and
\begin{equation*}
z_{\max} = \frac{ T_{\max} - T_{\min} }{ T_{\mathrm{opt}} - T_{\min} } > 1.
\end{equation*}
One immediate insight this reparameterization provides is that because the Wang-Engel functions 
depend on only $z$, and $z$ is invariant to linear translations and scalings of the temperatures, 
then the Wang-Engel forcing functions are also invariant to these transformations.  In
particular they have the same form for any choice of temperature units.

Inspecting this unitless form we can also see that $f(z = 0) = 0$ and hence 
$f(T = T_{\min}) = 0$.  The behavior at the upper boundary requires a bit more work to
extract; first we have
\begin{align*}
\lim_{ z \rightarrow z_{\max} } z^{a}
&=
\lim_{ z \rightarrow z_{\max} } z^{ \frac{ \log 2 }{ \log z_{\max} } }
\\
&=
\lim_{ z \rightarrow z_{\max} } \exp \left( \frac{ \log 2 }{ \log z_{\max} } \, \log z \right)
\\
&=
\exp \left( \frac{ \log 2 }{ \log z_{\max} } \, \log z_{\max} \right)
\\
&=
\exp \left( \log 2 \right)
\\
&=
2
\end{align*}
so that
\begin{equation*}
f(z = z_{\max}) = 2 \, (2 - 2) = 0
\end{equation*}
and, correspondingly, $f(T = T_{\max}) = 0$.  Consequently the Wang-Engel forcing
functions are all continuous at the boundaries $z = 0$ and $z = z_{\max}$, or 
equivalently $T = T_{\min}$ and $T = T_{\max}$.

The first-order derivative of the forcing functions within the interval $0 < z < z_{\max}$
are
\begin{align*}
\frac{\mathrm{d} f}{ \mathrm{d} z} (z) 
&=
a \, z^{a - 1} \, \left( 2 - z^{a} \right)
+ z^{a} \, \left(0 - a \, z^{a - 1} \right)
\\
&=
a \, z^{a - 1} \, \left( 2 - z^{a} \right)
- a \, z^{a - 1} \, \left( z^{a} \right)
\\
&=
a \, z^{a - 1} \, \left( 2 - z^{a} - z^{a} \right)
\\
&=
2 \, a \, z^{a - 1} \, \left(1 - z^{a} \right).
\end{align*}
Consequently the forcing functions are maximized at
\begin{align*}
0 
&= \frac{\mathrm{d} f}{ \mathrm{d} z} (z^{*})
\\
0
&=
2 \, a \, (z^{*})^{a - 1} \, \left(1 - (z^{*})^{a} \right)
\\
0
&=
1 - (z^{*})^{a}
\\
(z^{*})^{a} &= 1
\\
z^{*} &= 1,
\end{align*}
or equivalently
\begin{equation*}
T^{*} = (T_{\mathrm{opt}} - T_{\min}) \, z^{*} + T_{\min} = T_{\mathrm{opt}}.
\end{equation*}
The value attained at this maximum is
\begin{equation*}
f(z^{*}) = 1^{a} \, (2 - 1^{a}) = 1,
\end{equation*}
so that $0 \le f(T) \le 1$ for all input temperatures.

Outside of the interval $0 < z < z_{\max}$ the forcing functions are constant and hence 
the first-order derivatives vanish.  Unfortunately this behavior is not consistent with 
what we see within the interval.  At the upper boundary we have
\begin{equation*}
\lim_{z_{\max} \leftarrow z} \frac{\mathrm{d} f}{ \mathrm{d} z}(z)
=
0
\end{equation*}
but also
\begin{equation*}
\lim_{z \rightarrow z_{\max}} \frac{\mathrm{d} f}{ \mathrm{d} z}(z)
=
2 \, a \, \frac{2}{z_{\max}} \, \left(1 - 2 \right)
=
-\frac{4 \, a}{z_{\max}}.
\end{equation*}
Because the limit from below is strictly negative it conflicts with the limit from 
above; in other words the forcing functions are not differentiable at $z = z_{\max}$, 
or equivalently at $T = T_{\max}$.

The behavior at the lower boundary is a bit more complicated.  There we have
\begin{equation*}
\lim_{z \rightarrow 0} \frac{\mathrm{d} f}{ \mathrm{d} z}(z)
=
0
\end{equation*}
and
\begin{equation*}
\lim_{0 \leftarrow z} \frac{\mathrm{d} f}{ \mathrm{d} z}(z)
=
\lim_{0 \leftarrow z}
2 \, a \, z^{a - 1} \, \left(1 - z^{a} \right)
=
2 \, a \, z^{a - 1}.
\end{equation*}
If $a \ge 1$ then the limits from above and below will match and the forcing 
functions will be differentiable at $z = 0$, but if $a < 1$ then the derivative 
will diverge towards $-\infty$ from above.  This occurs when 
\begin{align*}
a &< 1
\\
\frac{ \log 2 }{ \log z_{\max} } &< 1
\\
\log 2 &< \log z_{\max}
\\
2 &< z_{\max}
\\
2 &< \frac{ T_{\max} - T_{\min} }{ T_{\mathrm{opt}} - T_{\min} }
\\
T_{\mathrm{opt}} - T_{\min} &< \frac{ T_{\max} - T_{\min} }{2}.
\end{align*}
In words the derivatives at $T_{\min}$ are discontinuous when $T_{\mathrm{opt}}$ 
is closer to $T_{\min}$than it is to $T_{\max}$.

\subsubsection{Constructing Differentiable Forcing Functions}

In order to avoid these cusps we need to build a family of forcing functions from the 
ground up while maintaining the qualitative features of the Wang-Engel forcing 
functions.  Specifically we want the forcing to be differentiable, vanish below $T_{\min}$ 
and above $T_{\max}$, and achieving the maximum value of $1$ at $T_{\mathrm{opt}}$.

To ensure invariance to linear transformations of the input temperature I will follow
the strategy of \cite{YinEtAl:1995} and base the new forcing functions around the 
unitless temperature
\begin{equation*}
x = \frac{T - T_{\min}}{T_{\max} - T_{\min}}.
\end{equation*}
Note that while superficially similar this is fundamentally different to the rescaled
temperature in the Wang-Engel model.  In particular we have $0 < x < 1$ regardless 
of the values of $T_{\min}$, $T_{\mathrm{opt}}$, and $T_{\max}$.

A convenient family of candidate functions within this unit interval is given by the
beta family of probability density functions which, up to normalization, are given by
\begin{equation*}
f(x) = C \, x^{\alpha - 1} \, (1 - x)^{\beta - 1}.
\end{equation*}
To ensure that these functional behaviors are unimodal and peak somewhere above $x = 0$
and below $x = 1$ we have to restrict both $\alpha > 1$ and $\beta > 1$.

The parameters $T_{\min}$ and $T_{\max}$ are encoded in the definition of the 
rescaled temperature $x$.  In order to enforce a peak at the optimal temperature 
$T_{\mathrm{opt}}$ we need to tune $\alpha$ or $\beta$ so that
\begin{equation*}
x_{\mathrm{opt}} = \frac{T_{\mathrm{opt}} - T_{\min}}{T_{\max} - T_{\min}}.
\end{equation*}
This requires
\begin{align*}
0 
&=
\frac{ \mathrm{d} f}{ \mathrm{d} x} (x_{\mathrm{opt}}) 
\\
0
&=
C \, x_{\mathrm{opt}}^{\alpha - 2} \, (1 - x_{\mathrm{opt}})^{\beta - 2} \,
\big[ (\alpha - 1) \, (1 - x_{\mathrm{opt}}) - (\beta - 1) \, x_{\mathrm{opt}} \big]
\\
0
&=
(\alpha - 1) \, (1 - x_{\mathrm{opt}}) - (\beta - 1) \, x_{\mathrm{opt}}
\\
0
&=
(\alpha - 1) - (\alpha + \beta - 2 ) \, x_{\mathrm{opt}}
\\
(\alpha + \beta - 2 ) \, x_{\mathrm{opt}}
&=
\alpha - 1
\\
x_{\mathrm{opt}}
&=
\frac{ \alpha - 1 }{ \alpha + \beta - 2 }.
\end{align*}
Another constraint on $\alpha$ and $\beta$ is enforced by the unit normalization
of the forcing functions,
\begin{align*}
1 
&= 
f(x_{\mathrm{opt}})
\\
1
&=
C \, x_{\mathrm{opt}}^{\alpha - 1} \, (1 - x_{\mathrm{opt}})^{\beta - 1}
\\
1
&=
C \, \left( \frac{ \alpha - 1 }{ \alpha + \beta - 2 } \right)^{\alpha - 1} \, 
\left( 1 - \frac{ \alpha - 1 }{ \alpha + \beta - 2 } \right)^{\beta - 1}
\\
1
&=
C \, \left( \frac{ \alpha - 1 }{ \alpha + \beta - 2 } \right)^{\alpha - 1} \, 
\left( \frac{ \beta - 1 }{ \alpha + \beta - 2 } \right)^{\beta - 1}
\\
1
&=
C \, \frac{ ( \alpha - 1 )^{\alpha - 1} \, ( \beta - 1 )^{\beta - 1} }
{ ( \alpha + \beta - 2 )^{\alpha + \beta - 2} }
\\
C
&=
\frac{ ( \alpha + \beta - 2 )^{\alpha + \beta - 2}}
{ ( \alpha - 1 )^{\alpha - 1} \, ( \beta - 1 )^{\beta - 1} }
\\
C
&=
\frac{ ( \alpha + \beta - 2 )^{\alpha - 1}}
{ ( \alpha - 1 )^{\alpha - 1} }
\,
\frac{ ( \alpha + \beta - 2 )^{\beta - 1}}
{ ( \beta - 1 )^{\beta - 1} }
\\
C
&=
\left( \frac{ \alpha + \beta - 2 }{ \alpha - 1 } \right)^{\alpha - 1} \,
\left( \frac{ \alpha + \beta - 2 }{ \beta - 1 } \right)^{\beta - 1}.
\end{align*}
Using the definition of $x_{\mathrm{opt}}$ this reduces to
\begin{align*}
C
&=
\left( \frac{ \alpha + \beta - 2 }{ \alpha - 1 } \right)^{\alpha - 1} \,
\left( \frac{ \alpha + \beta - 2 }{ \beta - 1 } \right)^{\beta - 1}
\\
&=
\left( \frac{ \alpha + \beta - 2 }{ \alpha - 1 } \right)^{\alpha - 1} \,
\left( \frac{ \alpha + \beta - 2 }{ \alpha - 1 } \frac{ \alpha - 1 }{ \beta - 1 }  \right)^{\beta - 1}
\\
&=
\left( \frac{1}{x_{\mathrm{opt}}} \right)^{\alpha - 1} \,
\left( \frac{1}{x_{\mathrm{opt}}} \frac{ \alpha - 1 }{ \beta - 1 }  \right)^{\beta - 1}
\end{align*}

Consequently all of the qualitative properties of the Wang-Engel forcing functions 
will manifest in the family
\begin{align*}
f(x) 
&=
\left( \frac{ \alpha + \beta - 2 }{ \alpha - 1 } \right)^{\alpha - 1} \,
\left( \frac{ \alpha + \beta - 2 }{ \beta - 1 } \right)^{\beta - 1} \,
x^{\alpha - 1} \, (1 - x)^{\beta - 1}
\\
&=
\left( \frac{1}{x_{\mathrm{opt}}} \right)^{\alpha - 1} \,
\left( \frac{1}{x_{\mathrm{opt}}} \frac{ \alpha - 1 }{ \beta - 1 }  \right)^{\beta - 1}
x^{\alpha - 1} \, (1 - x)^{\beta - 1}
\\
&=
\left( \frac{x}{x_{\mathrm{opt}}} \right)^{\alpha - 1} \, 
\left(  \frac{ \alpha - 1 }{ \beta - 1 } \frac{1 - x}{x_{\mathrm{opt}}} \right)^{\beta - 1}
\end{align*}
where
\begin{equation*}
x_{\mathrm{opt}}
=
\frac{ \alpha - 1 }{ \alpha + \beta - 2 }.
\end{equation*}
This is, up to normalization, equivalent to the beta family of \cite{YinEtAl:1995} .

Differentiability of these forcing functions, however, is not guaranteed for all 
choices of $\alpha$ or $\beta$.  The first-order derivatives are given by
\begin{align*}
\frac{ \mathrm{d} f}{ \mathrm{d} x} (x) 
&=
C \, (\alpha - 1) \, x^{\alpha - 2} \, (1 - x)^{\beta - 1}
- C \, (\beta - 1) \, x^{\alpha - 1} \, (1 - x)^{\beta - 2}
\\
&=
C \, x^{\alpha - 2} \, (1 - x)^{\beta - 2} \,
\bigg[ (\alpha - 1) \, (1 - x) - (\beta - 1) \, x \bigg].
\end{align*}
At the left boundary $x = 0$ we have
\begin{equation*}
\frac{ \mathrm{d} f}{ \mathrm{d} x} (0) 
=
C \, 0^{\alpha - 2} \, 1^{\beta - 2} \,
\big[ \alpha - 1 \big];
\end{equation*}
and the derivatives will diverge if $\alpha \le 2$.  Likewise at the right boundary
$x = 1$ we have
\begin{equation*}
\frac{ \mathrm{d} f}{ \mathrm{d} x} (1) 
=
C \, 1^{\alpha - 2} \, 0^{\beta - 2} \,
\big[ \beta - 1 \big]
\end{equation*}
and the derivatives will diverge if $\beta \le 2$.  

Restricting to $\alpha > 2$ and $\beta > 2$ ensures that the first-order derivatives 
vanish at the boundaries, matching the derivatives of the constant zeros below and 
above the boundaries.  In other words this restriction ensures forcing functions that 
are differentiable everywhere.  If $\alpha$ and $\beta$ are even larger than $2$ then 
higher-order derivatives will also vanish, ensuring an even smoother transition across
$x = 0$ and $x = 1$.

How best to enforce this condition depends on the behavior of $x_{\mathrm{opt}}$.  For
example if $T_{\mathrm{opt}}$ is closer to $T_{\max}$ than $T_{\min}$ then
$x_{\mathrm{opt}} > \frac{1}{2}$.  In this case we can replace $\beta$ with 
\begin{align*}
x_{\mathrm{opt}}
&=
\frac{ \alpha - 1 }{ \alpha + \beta - 2 }
\\
(\alpha - 2) \, x_{\mathrm{opt}} + \beta \,  x_{\mathrm{opt}}
&=
\alpha - 1
\\
\beta \,  x_{\mathrm{opt}}
&=
(\alpha - 1) - (\alpha - 2) \, x_{\mathrm{opt}}
\\
\beta
&=
\frac{ (\alpha - 1) - (\alpha - 2) \, x_{\mathrm{opt}} }{ x_{\mathrm{opt}} }.
\end{align*}
To ensure that $\beta > 2$ we need 
\begin{align*}
2 &< \beta
\\
2 
&<
\frac{ (\alpha - 1) - (\alpha - 2) \, x_{\mathrm{opt}} }{ x_{\mathrm{opt}} }
\\
2 \, x_{\mathrm{opt}} 
&<
(\alpha - 1) - (\alpha - 2) \, x_{\mathrm{opt}} 
\\
0
&<
\alpha - 1 - \alpha \, x_{\mathrm{opt}} 
\\
1
&<
\alpha \, (1 - x_{\mathrm{opt}})
\\
\alpha
&>
\frac{1}{1 - x_{\mathrm{opt}}}.
\end{align*}
This can be guaranteed introducing a new parameter $\delta > 0$ and defining
\begin{align*}
\alpha = (\delta + 1) \, \frac{1}{1 - x_{\mathrm{opt}}}.
\end{align*}

With these choices we have
\begin{align*}
\alpha - 1
&=
(\delta + 1) \, \frac{1}{1 - x_{\mathrm{opt}}} - 1
\\
&=
\frac{\delta + 1 - 1 + x_{\mathrm{opt}}}{1 - x_{\mathrm{opt}}} 
\\
&=
\frac{\delta + x_{\mathrm{opt}}}{1 - x_{\mathrm{opt}}}
\end{align*}
and
\begin{align*}
\beta - 1
&=
\frac{ (\alpha - 1) - (\alpha - 2) \, x_{\mathrm{opt}} }{ x_{\mathrm{opt}} } - 1
\\
&=
\frac{ \alpha - 1 - \alpha \, x_{\mathrm{opt}} + 2 \, x_{\mathrm{opt}} - x_{\mathrm{opt}} }
{ x_{\mathrm{opt}} }
\\
&=
\frac{ \alpha - 1 - \alpha \, x_{\mathrm{opt}} + x_{\mathrm{opt}} }{ x_{\mathrm{opt}} }
\\
&=
\frac{ (\alpha - 1) - (\alpha - 1) \, x_{\mathrm{opt}} }{ x_{\mathrm{opt}} }
\\
&=
(\alpha - 1) \frac{ 1 - x_{\mathrm{opt}} }{ x_{\mathrm{opt}} }.
\end{align*}
This allows us to reduce the forcing function to
\begin{align*}
f(x)
&=
\left( \frac{x}{x_{\mathrm{opt}}} \right)^{\alpha - 1} \, 
\left(  \frac{ \alpha - 1 }{ \beta - 1 } \frac{1 - x}{x_{\mathrm{opt}}} \right)^{\beta - 1}
\\
&=
\left( \frac{x}{x_{\mathrm{opt}}} \right)^{\alpha - 1} \, 
\left(  \frac{x_{\mathrm{opt}}}{1 - x_{\mathrm{opt}}} \frac{1 - x}{x_{\mathrm{opt}}} \right)^{\beta - 1}
\\
&=
\left( \frac{x}{x_{\mathrm{opt}}} \right)^{\alpha - 1} \, 
\left( \frac{1 - x}{1 - x_{\mathrm{opt}}} \right)^{\beta - 1}
\\
&=
\left( \frac{x}{x_{\mathrm{opt}}} \right)^{\alpha - 1} \, 
\left( \frac{1 - x}{1 - x_{\mathrm{opt}}} \right)^{ 
(\alpha - 1) \frac{ 1 - x_{\mathrm{opt}} }{ x_{\mathrm{opt}} } }
\\
&=
\left(
\left( \frac{x}{x_{\mathrm{opt}}} \right) \, 
\left( \frac{1 - x}{1 - x_{\mathrm{opt}}} \right)^{ \frac{ 1 - x_{\mathrm{opt}} }{ x_{\mathrm{opt}} } } 
\right)^{\alpha - 1}
\end{align*}
One advantage of this particular form is that it facilitates numerical analysis of the forcing 
functions.  In particular we can see that as $x_{\mathrm{opt}}$ approaches $1$ the term 
$(1 - x) / (1 - x_{\mathrm{opt}})$ diverges.  At the same time the exponent 
$(1 - x_{\mathrm{opt}} ) / x_{\mathrm{opt}}$ goes to zero in this limit so that the overall power
approaches zero.  

Ensuring that intermediate divergences don't spoil this limiting behavior is key to a robust 
numerical implementation.  For example if $x < x_{\mathrm{opt}}$ then we might compute 
$(1 - x_{\mathrm{opt}}) / (1 - x)$ as the intermediate quantity before raising it to the power
$-(1 - x_{\mathrm{opt}} ) / x_{\mathrm{opt}}$.  Instead of having to take the power of an 
overflowing intermediate term we would take the power of an underflowing intermediate term
which is better suited towards the vanishing output.  Alternatively if a numerically stable
implementation of the function $y \log y$ is available then we might evaluate $\log f(x)$
first and then exponentiate to give the $f(x)$.

In the complementary situation where $T_{\mathrm{opt}}$ is closer to $T_{\min}$ than $T_{\max}$,
and $x_{\mathrm{opt}} \le 0.5$ we can instead replace $\alpha$ with 
\begin{align*}
x_{\mathrm{opt}}
&=
\frac{ \alpha - 1 }{ \alpha + \beta - 2 }
\\
\alpha \, x_{\mathrm{opt}} + (\beta - 2) \,  x_{\mathrm{opt}}
&=
\alpha - 1
\\
(\beta - 2) \,  x_{\mathrm{opt}} + 1
&=
\alpha \, (1 -  x_{\mathrm{opt}})
\\
\frac{ 1 + (\beta - 2) \,  x_{\mathrm{opt}}}{1 -  x_{\mathrm{opt}}}
&=
\alpha.
\end{align*}
To ensure that $\alpha > 2$ we need 
\begin{align*}
2 &< \alpha
\\
2 
&<
\frac{ 1 + (\beta - 2) \,  x_{\mathrm{opt}}}{1 -  x_{\mathrm{opt}}}
\\
2 -  2 \, x_{\mathrm{opt}} 
&<
1 + (\beta - 2) \,  x_{\mathrm{opt}}
\\
1 -  2 \, x_{\mathrm{opt}} 
&<
\beta \, x_{\mathrm{opt}} - 2 \, x_{\mathrm{opt}}
\\
1
&<
\beta \, x_{\mathrm{opt}}
\\
\frac{1}{ x_{\mathrm{opt}} } 
&<
\beta.
\end{align*}
As above this can inequality is guaranteed by introducing a new parameter $\delta > 0$ and 
taking
\begin{align*}
\beta = (\delta + 1) \, \frac{1}{x_{\mathrm{opt}}}.
\end{align*}

These choices give
\begin{align*}
\beta - 1
&=
(\delta + 1) \, \frac{1}{x_{\mathrm{opt}}} - 1
\\
&=
\frac{\delta + 1 - x_{\mathrm{opt}}}{x_{\mathrm{opt}}}
\end{align*}
and
\begin{align*}
\alpha - 1
&=
\frac{ 1 + (\beta - 2) \,  x_{\mathrm{opt}}}{1 -  x_{\mathrm{opt}}} - 1
\\
&=
\frac{ 1 + \beta \, x_{\mathrm{opt}} - 2 \,  x_{\mathrm{opt}} - 1 +  x_{\mathrm{opt}}}
{1 -  x_{\mathrm{opt}}}
\\
&=
\frac{ \beta \, x_{\mathrm{opt}} - x_{\mathrm{opt}}}{1 - x_{\mathrm{opt}}}
\\
&=
(\beta - 1) \frac{x_{\mathrm{opt}}}{1 - x_{\mathrm{opt}}}.
\end{align*}
Consequently
\begin{align*}
f(x)
&=
\left( \frac{x}{x_{\mathrm{opt}}} \right)^{\alpha - 1} \, 
\left(  \frac{ \alpha - 1 }{ \beta - 1 } \frac{1 - x}{x_{\mathrm{opt}}} \right)^{\beta - 1}
\\
&=
\left( \frac{x}{x_{\mathrm{opt}}} \right)^{\alpha - 1} \, 
\left(  \frac{x_{\mathrm{opt}}}{1 - x_{\mathrm{opt}}} \frac{1 - x}{x_{\mathrm{opt}}} \right)^{\beta - 1}
\\
&=
\left( \frac{x}{x_{\mathrm{opt}}} \right)^{\alpha - 1} \, 
\left( \frac{1 - x}{1 - x_{\mathrm{opt}}} \right)^{\beta - 1}
\\
&=
\left( \frac{x}{x_{\mathrm{opt}}} \right)^{ (\beta - 1) \frac{x_{\mathrm{opt}}}{1 - x_{\mathrm{opt}}} } \, 
\left( \frac{1 - x}{1 - x_{\mathrm{opt}}} \right)^{\beta - 1}
\\
&=
\left( 
\left( \frac{x}{x_{\mathrm{opt}}} \right)^{ \frac{x_{\mathrm{opt}}}{1 - x_{\mathrm{opt}}} } \, 
\left( \frac{1 - x}{1 - x_{\mathrm{opt}}} \right)
\right)^{\beta - 1}
\end{align*}
Reducing the forcing functions this far once again clearly identifies potential numerical 
problems.  Here as $x_{\mathrm{opt}}$ approaches $0$ the term $x / x_{\mathrm{opt}}$ diverges
but the exponent $x_{\mathrm{opt}} / (1 - x_{\mathrm{opt}} )$ goes to zero.  Ultimately the
vanishing exponent moderates the diverging argument so that the power decays to zero, but we 
have to be careful to avoid the diverging intermediate term so that it doesn't result in 
floating point overflow in a numerical implementation of the forcing functions.

Combining these cases gives the desired family of forcing functions,
\begin{equation*}
f(x) =
\left\{
\begin{array}{rr}
0, & x < 0 \\
\left( 
\left( \frac{x}{x_{\mathrm{opt}}} \right)^{ \eta } \, 
\left( \frac{1 - x}{1 - x_{\mathrm{opt}}} \right)^{ \kappa }
\right)^{\gamma}, & 0 \le x \le 1 \\
0, & x > 1
\end{array}
\right.
\end{equation*}
with
\begin{align*}
\eta &= 1
\\
\kappa &= 
\frac{1 - x_{\mathrm{opt}}}{x_{\mathrm{opt}}}
\\
\gamma &= \frac{\delta + x_{\mathrm{opt}}}{1 - x_{\mathrm{opt}}}
\end{align*}
for $x_{\mathrm{opt}} > 0.5$ and
\begin{align*}
\eta &= \frac{x_{\mathrm{opt}}}{1 - x_{\mathrm{opt}}}
\\
\kappa &= 1
\\
\gamma &= \frac{\delta + 1 - x_{\mathrm{opt}}}{x_{\mathrm{opt}}}
\end{align*}
for $x_{\mathrm{opt}} \le 0.5$.

\begin{figure}
\centering
\begin{tikzpicture}[scale=1]

  \pgfmathsetmacro{\sx}{4}
  \pgfmathsetmacro{\sy}{4}

  \draw[gray90, dashed, line width=2] (-1.1 * \sx, \sy) -- (1.1 * \sx, \sy);
  \node[black, align=center] at (1.2 * \sx, \sy) { $1$ };
  
  \draw[gray90, dashed, line width=2] (-\sx, 0) -- +(0, 1.2 * \sy);
  \node[black, align=center] at (-\sx, -0.5) { $T_{\min}$ };
  
  \draw[gray90, dashed, line width=2] (\sx, 0) -- +(0, 1.2 * \sy);
  \node[black, align=center] at (\sx, -0.5) { $T_{\max}$ };

  \colorlet{custom}{dark!100!white};
  
  \draw[custom, line width=2] (\sx * -1.050, \sy * 0.000) 
    \foreach \x/\y in {-1.000/0.000, -0.980/0.000, -0.960/0.000, -0.939/0.000, -0.919/0.000, 
                       -0.899/0.000, -0.879/0.000, -0.859/0.000, -0.838/0.000, -0.818/0.001, 
                       -0.798/0.001, -0.778/0.001, -0.758/0.002, -0.737/0.003, -0.717/0.005, 
                       -0.697/0.006, -0.677/0.008, -0.657/0.011, -0.636/0.014, -0.616/0.018, 
                       -0.596/0.023, -0.576/0.028, -0.556/0.034, -0.535/0.041, -0.515/0.050, 
                       -0.495/0.059, -0.475/0.069, -0.455/0.081, -0.434/0.093, -0.414/0.107, 
                       -0.394/0.123, -0.374/0.139, -0.354/0.157, -0.333/0.177, -0.313/0.198, 
                       -0.293/0.220, -0.273/0.243, -0.253/0.268, -0.232/0.294, -0.212/0.321, 
                       -0.192/0.349, -0.172/0.379, -0.152/0.409, -0.131/0.440, -0.111/0.472, 
                       -0.091/0.505, -0.071/0.538, -0.051/0.571, -0.030/0.604, -0.010/0.637, 
                       0.010/0.670, 0.030/0.703, 0.051/0.735, 0.071/0.766, 0.091/0.796, 
                       0.111/0.825, 0.131/0.852, 0.152/0.878, 0.172/0.902, 0.192/0.923, 
                       0.212/0.943, 0.232/0.960, 0.253/0.974, 0.273/0.985, 0.293/0.993, 
                       0.313/0.998, 0.333/1.000, 0.354/0.998, 0.374/0.993, 0.394/0.984, 
                       0.414/0.972, 0.434/0.955, 0.455/0.936, 0.475/0.912, 0.495/0.885, 
                       0.515/0.855, 0.535/0.821, 0.556/0.784, 0.576/0.745, 0.596/0.703, 
                       0.616/0.658, 0.636/0.612, 0.657/0.564, 0.677/0.515, 0.697/0.465, 
                       0.717/0.415, 0.737/0.366, 0.758/0.317, 0.778/0.270, 0.798/0.225, 
                       0.818/0.183, 0.838/0.144, 0.859/0.109, 0.879/0.078, 0.899/0.052, 
                       0.919/0.032, 0.939/0.016, 0.960/0.006, 0.980/0.001, 1.000/0.000} {
    -- (\sx * \x, \sy * \y)
  } -- (\sx * 1.050, \sy * 0.000);
  
  \colorlet{custom}{dark!60!white};
  
  \draw[custom, line width=2] (\sx * -1.050, \sy * 0.000) 
    \foreach \x/\y in {-1.000/0.000, -0.980/0.001, -0.960/0.006, -0.939/0.016, -0.919/0.032, 
                       -0.899/0.052, -0.879/0.078, -0.859/0.109, -0.838/0.144, -0.818/0.183, 
                       -0.798/0.225, -0.778/0.270, -0.758/0.317, -0.737/0.366, -0.717/0.415, 
                       -0.697/0.465, -0.677/0.515, -0.657/0.564, -0.636/0.612, -0.616/0.658, 
                       -0.596/0.703, -0.576/0.745, -0.556/0.784, -0.535/0.821, -0.515/0.855, 
                       -0.495/0.885, -0.475/0.912, -0.455/0.936, -0.434/0.955, -0.414/0.972, 
                       -0.394/0.984, -0.374/0.993, -0.354/0.998, -0.333/1.000, -0.313/0.998, 
                       -0.293/0.993, -0.273/0.985, -0.253/0.974, -0.232/0.960, -0.212/0.943, 
                       -0.192/0.923, -0.172/0.902, -0.152/0.878, -0.131/0.852, -0.111/0.825, 
                       -0.091/0.796, -0.071/0.766, -0.051/0.735, -0.030/0.703, -0.010/0.670, 
                       0.010/0.637, 0.030/0.604, 0.051/0.571, 0.071/0.538, 0.091/0.505, 
                       0.111/0.472, 0.131/0.440, 0.152/0.409, 0.172/0.379, 0.192/0.349, 
                       0.212/0.321, 0.232/0.294, 0.253/0.268, 0.273/0.243, 0.293/0.220, 
                       0.313/0.198, 0.333/0.177, 0.354/0.157, 0.374/0.139, 0.394/0.123, 
                       0.414/0.107, 0.434/0.093, 0.455/0.081, 0.475/0.069, 0.495/0.059, 
                       0.515/0.050, 0.535/0.041, 0.556/0.034, 0.576/0.028, 0.596/0.023, 
                       0.616/0.018, 0.636/0.014, 0.657/0.011, 0.677/0.008, 0.697/0.006, 
                       0.717/0.005, 0.737/0.003, 0.758/0.002, 0.778/0.001, 0.798/0.001, 
                       0.818/0.001, 0.838/0.000, 0.859/0.000, 0.879/0.000, 0.899/0.000, 
                       0.919/0.000, 0.939/0.000, 0.960/0.000, 0.980/0.000, 1.000/0.000} {
    -- (\sx * \x, \sy * \y)
  } -- (\sx * 1.050, \sy * 0.000);

  \colorlet{custom}{dark!100!white};
  \draw[custom, dashed, line width=2] (\sx * 0.3333333, 0) -- +(0, 1.2 * \sy);
  \node[custom, align=center] at (\sx * 0.3333333, -0.5) { $T_{\mathrm{opt}}$ };

  \colorlet{custom}{dark!60!white};
  \draw[custom, dashed, line width=2] (-\sx * 0.3333333, 0) -- +(0, 1.2 * \sy);
  \node[custom, align=center] at (-\sx * 0.3333333, -0.5) { $T_{\max} - T_{\mathrm{opt}} + T_{\min}$ };

  \draw [<->, >=stealth, line width=1] (-1.1 * \sx, 0) -- (1.1 * \sx, 0);
  \node[] at (0, -1.25) { Temperature (Arbitrary Units) };

\end{tikzpicture}
\caption{Unlike the Wang-Engel functions the generalized forcing functions are exactly 
symmetric with respect to the optimal temperature: for fixed $\delta$, $T_{\min}$, and 
$T_{\max}$ the functions given by the maxima $T_\mathrm{opt}$ and 
$T_{\max} - T_{\mathrm{opt}} + T_{\min}$ are perfect reflections of each other.
}
\label{fig:new_symmetry} 
\end{figure}
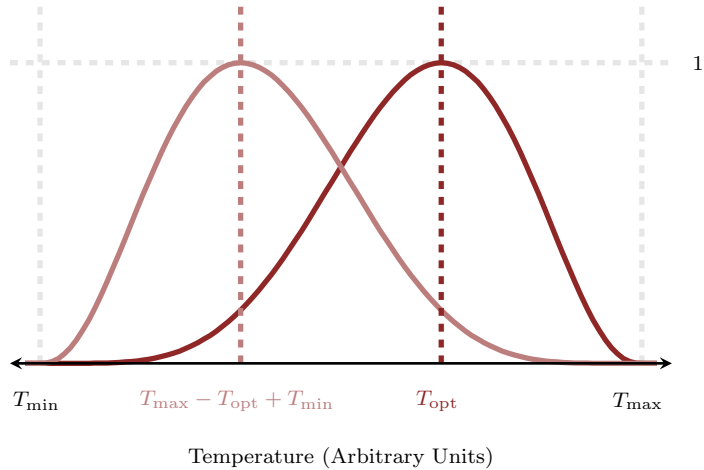

In terms of the nominal temperatures this becomes 
(Figures \ref{fig:diff_forcing}, \ref{fig:new_symmetry})
\begin{equation*}
f(T) =
\left\{
\begin{array}{rr}
0, & T < T_{\min} \\
\left( 
\left( \frac{T - T_{\mathrm{min}}}{T_{\mathrm{opt}} - T_{\mathrm{min}}} \right)^{ \eta } \, 
\left( \frac{T_{\mathrm{max}} - T}{T_{\mathrm{max}} - T_{\mathrm{opt}}} \right)^{ \kappa }
\right)^{\gamma},
& T_{\min} \le T \le T_{\max} \\
0, & T > T_{\max}
\end{array}
\right.
\end{equation*}
with
\begin{align*}
\eta &= 1
\\
\kappa &= 
\frac{T_{\mathrm{max}} - T_{\mathrm{opt}}}{T_{\mathrm{opt}} - T_{\mathrm{min}}}
\\
\gamma &= 
\frac{\delta \, T_{\mathrm{max}} + T_{\mathrm{opt}} - (\delta + 1) \, T_{\mathrm{min}}}
{ T_{\mathrm{max}} - T_{\mathrm{opt}} }
\end{align*}
for $T_{\mathrm{opt}} > \frac{1}{2}( T_{\min} + T_{\max} )$ and 
\begin{align*}
\eta &= \frac{T_{\mathrm{opt}} - T_{\mathrm{min}}}{T_{\mathrm{max}} - T_{\mathrm{opt}}}
\\
\kappa &= 1
\\
\gamma &= 
\frac{ (\delta + 1) \, T_{\mathrm{max}} - T_{\mathrm{opt}} - \delta \, T_{\mathrm{min}} }
{ T_{\mathrm{opt}} - T_{\mathrm{min}} }
\end{align*}
for $ T_{\mathrm{opt}} \le \frac{1}{2} ( T_{\min} + T_{\max} )$.

\bibliography{threshold_survival}

\begin{thebibliography}{18}

\bibitem[\protect\citeauthoryear{Betancourt}{2018}]{Betancourt:2018b}
\begin{barticle}[author]
\bauthor{\bsnm{Betancourt},~\bfnm{Michael}\binits{M.}}
(\byear{2018}).
\btitle{A Conceptual Introduction to Hamiltonian Monte Carlo}.
\bjournal{ArXiv e-prints}
\bvolume{1701.02434}.
\end{barticle}
\endbibitem

\bibitem[\protect\citeauthoryear{Betancourt}{2020}]{Betancourt:2020a}
\begin{barticle}[author]
\bauthor{\bsnm{Betancourt},~\bfnm{Michael}\binits{M.}}
(\byear{2020}).
\btitle{Towards A Principled Bayesian Workflow}.
\bjournal{\url{https://betanalpha.github.io/assets/case_studies/principled_bayesian_workflow.html}}
\bvolume{commit aeab31509b8e37ff05b0828f87a3018b1799b401}.
\end{barticle}
\endbibitem

\bibitem[\protect\citeauthoryear{Betancourt}{2022}]{Betancourt:2022a}
\begin{barticle}[author]
\bauthor{\bsnm{Betancourt},~\bfnm{Michael}\binits{M.}}
(\byear{2022}).
\btitle{Outwit, Outlast, Outmodel}.
\bjournal{\url{https://betanalpha.github.io/assets/case_studies/survival_modeling.html}}
\bvolume{commit 0b667a51ac437694b76894acb88ea4c2ec361963}.
\end{barticle}
\endbibitem

\bibitem[\protect\citeauthoryear{Brown and Heathcote}{2005}]{BrownEtAl:2005}
\begin{barticle}[author]
\bauthor{\bsnm{Brown},~\bfnm{Scott}\binits{S.}} \AND
  \bauthor{\bsnm{Heathcote},~\bfnm{Andrew}\binits{A.}}
(\byear{2005}).
\btitle{A ballistic model of choice response time.}
\bjournal{Psychological review}
\bvolume{112}
\bpages{117}.
\end{barticle}
\endbibitem

\bibitem[\protect\citeauthoryear{Coombe and Dry}{1992}]{CoombeEtAl:1992}
\begin{bbook}[author]
\beditor{\bsnm{Coombe},~\bfnm{Bryan}\binits{B.}} \AND
  \beditor{\bsnm{Dry},~\bfnm{Peter}\binits{P.}}, eds.
(\byear{1992}).
\btitle{Viticulture Volume 2 -- Practices}.
\bpublisher{winetitles}, \baddress{Adelaide, SA, Australia}.
\end{bbook}
\endbibitem

\bibitem[\protect\citeauthoryear{{Cornelius, Christine and Petermeier, Hannes
  and Estrella, Nicole and Menzel, Annette}}{2011}]{CorneliusEtAl:2011}
\begin{barticle}[author]
\bauthor{\bsnm{{Cornelius, Christine and Petermeier, Hannes and Estrella,
  Nicole and Menzel, Annette}}}
(\byear{2011}).
\btitle{{A comparison of methods to estimate seasonal phenological development
  from BBCH scale recording}}.
\bjournal{{International journal of biometeorology}}
\bvolume{55}
\bpages{867--877}.
\end{barticle}
\endbibitem

\bibitem[\protect\citeauthoryear{Cox and Oakes}{1984}]{CoxEtAl:1984}
\begin{bbook}[author]
\bauthor{\bsnm{Cox},~\bfnm{D.~R.}\binits{D.~R.}} \AND
  \bauthor{\bsnm{Oakes},~\bfnm{D.}\binits{D.}}
(\byear{1984}).
\btitle{Analysis of survival data}.
\bseries{Monographs on Statistics and Applied Probability}.
\bpublisher{Chapman \& Hall, London}.
\end{bbook}
\endbibitem

\bibitem[\protect\citeauthoryear{Hosmer, Lemeshow and
  May}{2008}]{HosmerEtAl:2008}
\begin{bbook}[author]
\bauthor{\bsnm{Hosmer},~\bfnm{D.~W.}\binits{D.~W.}},
  \bauthor{\bsnm{Lemeshow},~\bfnm{S.}\binits{S.}} \AND
  \bauthor{\bsnm{May},~\bfnm{S.}\binits{S.}}
(\byear{2008}).
\btitle{Applied Survival Analysis: Regression Modeling of Time-to-Event Data}.
\bseries{Wiley Series in Probability and Statistics}.
\bpublisher{Wiley}.
\end{bbook}
\endbibitem

\bibitem[\protect\citeauthoryear{Ibrahim, Chen and
  Sinha}{2001}]{IbrahimEtAl:2001}
\begin{bbook}[author]
\bauthor{\bsnm{Ibrahim},~\bfnm{Joseph~G.}\binits{J.~G.}},
  \bauthor{\bsnm{Chen},~\bfnm{Ming-Hui}\binits{M.-H.}} \AND
  \bauthor{\bsnm{Sinha},~\bfnm{Debajyoti}\binits{D.}}
(\byear{2001}).
\btitle{Bayesian Survival Analysis}.
\bpublisher{Springer New York}.
\end{bbook}
\endbibitem

\bibitem[\protect\citeauthoryear{Lambers, Chapin~III and
  Pons}{2008}]{LambersEtAl:2008}
\begin{bincollection}[author]
\bauthor{\bsnm{Lambers},~\bfnm{H.}\binits{H.}},
  \bauthor{\bsnm{Chapin~III},~\bfnm{F.~Stuart}\binits{F.~S.}} \AND
  \bauthor{\bsnm{Pons},~\bfnm{T.~L.}\binits{T.~L.}}
(\byear{2008}).
\btitle{Life cycles: environmental influences and adaptations}.
In \bbooktitle{Plant Physiological Ecology}
\bedition{2nd} ed.
(\beditor{\bfnm{H.}\binits{H.}~\bsnm{Lambers}},
  \beditor{\bfnm{F.~Stuart}\binits{F.~S.}~\bsnm{Chapin~III}} \AND
  \beditor{\bfnm{T.~L.}\binits{T.~L.}~\bsnm{Pons}}, eds.)
\bpages{375-402}.
\bpublisher{Spinger}, \baddress{New York}.
\end{bincollection}
\endbibitem

\bibitem[\protect\citeauthoryear{Lee and Whitmore}{2006}]{LeeEtAl:2006}
\begin{barticle}[author]
\bauthor{\bsnm{Lee},~\bfnm{Mei-Ling~Ting}\binits{M.-L.~T.}} \AND
  \bauthor{\bsnm{Whitmore},~\bfnm{G.~A.}\binits{G.~A.}}
(\byear{2006}).
\btitle{Threshold regression for survival analysis: modeling event times by a
  stochastic process reaching a boundary}.
\bjournal{Statist. Sci.}
\bvolume{21}
\bpages{501--513}.
\end{barticle}
\endbibitem

\bibitem[\protect\citeauthoryear{Lorenz et~al.}{1994}]{LorenzEtAl:1994}
\begin{barticle}[author]
\bauthor{\bsnm{Lorenz},~\bfnm{DH}\binits{D.}},
  \bauthor{\bsnm{Eichhorn},~\bfnm{KW}\binits{K.}},
  \bauthor{\bsnm{Bleiholder},~\bfnm{H}\binits{H.}},
  \bauthor{\bsnm{Klose},~\bfnm{R}\binits{R.}},
  \bauthor{\bsnm{Meier},~\bfnm{U}\binits{U.}} \AND
  \bauthor{\bsnm{Weber},~\bfnm{E}\binits{E.}}
(\byear{1994}).
\btitle{{Ph{\"a}nologische Entwicklungsstadien der Weinrebe (Vitis vinifera L.
  ssp. vinifera). Codierung und Beschreibung nach der erweiterten BBCH-Skala}}.
\bjournal{Wein-Wissenschaft}
\bvolume{49}
\bpages{66--70}.
\end{barticle}
\endbibitem

\bibitem[\protect\citeauthoryear{Tank et~al.}{2002}]{TankEtAl:2002}
\begin{barticle}[author]
\bauthor{\bsnm{Tank},~\bfnm{AMG~Klein}\binits{A.~K.}},
  \bauthor{\bsnm{Wijngaard},~\bfnm{JB}\binits{J.}},
  \bauthor{\bsnm{K{\"o}nnen},~\bfnm{GP}\binits{G.}},
  \bauthor{\bsnm{B{\"o}hm},~\bfnm{R}\binits{R.}},
  \bauthor{\bsnm{Demar{\'e}e},~\bfnm{G}\binits{G.}},
  \bauthor{\bsnm{Gocheva},~\bfnm{A}\binits{A.}},
  \bauthor{\bsnm{Mileta},~\bfnm{M}\binits{M.}},
  \bauthor{\bsnm{Pashiardis},~\bfnm{S}\binits{S.}},
  \bauthor{\bsnm{Heejkrlik},~\bfnm{L}\binits{L.}},
  \bauthor{\bsnm{Kern-Hansen},~\bfnm{C}\binits{C.}} \betal{et~al.}
(\byear{2002}).
\btitle{Daily surface air temperature and precipitation dataset 1901--1999 for
  European Climate Assessment (ECA)}.
\bjournal{Int. J. Climatol}
\bvolume{22}
\bpages{1441--1453}.
\end{barticle}
\endbibitem

\bibitem[\protect\citeauthoryear{{Stan Development Team}}{2019a}]{Stan:2019a}
\begin{bmisc}[author]
\bauthor{\bsnm{{Stan Development Team}}}
(\byear{2019}a).
\btitle{{S}tan: {A} {C}++ Library for Probability and Sampling, Version 2.19}.
\bhowpublished{http://mc-stan.org/}.
\end{bmisc}
\endbibitem

\bibitem[\protect\citeauthoryear{{{Stan Development Team}}}{2019b}]{RStan:2019}
\begin{bmisc}[author]
\bauthor{\bsnm{{{Stan Development Team}}}}
(\byear{2019}b).
\btitle{{{RStan}: the {R} interface to {Stan}}}.
\end{bmisc}
\endbibitem

\bibitem[\protect\citeauthoryear{Wang and Engel}{1998}]{WangEtAl:1998}
\begin{barticle}[author]
\bauthor{\bsnm{Wang},~\bfnm{Enli}\binits{E.}} \AND
  \bauthor{\bsnm{Engel},~\bfnm{Thomas}\binits{T.}}
(\byear{1998}).
\btitle{Simulation of phenological development of wheat crops}.
\bjournal{Agricultural Systems}
\bvolume{58}
\bpages{1-24}.
\end{barticle}
\endbibitem

\bibitem[\protect\citeauthoryear{{Wolkovich, Elizabeth M and Garc\'{i}a de
  Cort\'{a}zar-Atauri, I\~{n}aki}}{2022}]{WolkovichEtAl:2022}
\begin{bmisc}[author]
\bauthor{\bsnm{{Wolkovich, Elizabeth M and Garc\'{i}a de Cort\'{a}zar-Atauri,
  I\~{n}aki}}}
(\byear{2022}).
\bhowpublished{Personal Communication}.
\end{bmisc}
\endbibitem

\bibitem[\protect\citeauthoryear{Yin et~al.}{1995}]{YinEtAl:1995}
\begin{barticle}[author]
\bauthor{\bsnm{Yin},~\bfnm{Xinyou}\binits{X.}},
  \bauthor{\bsnm{Kropff},~\bfnm{Martin~J.}\binits{M.~J.}},
  \bauthor{\bsnm{McLaren},~\bfnm{Graham}\binits{G.}} \AND
  \bauthor{\bsnm{Visperas},~\bfnm{Romeo~M.}\binits{R.~M.}}
(\byear{1995}).
\btitle{A nonlinear model for crop development as a function of temperature}.
\bjournal{Agricultural and Forest Meteorology}
\bvolume{77}
\bpages{1-16}.
\end{barticle}
\endbibitem

\end{thebibliography}
\bibliographystyle{imsart-nameyear}

\end{document}